\documentclass[floats,floatfix,showpacs,amssymb,physrev,twocolumn,superscriptaddress,reprint,
nofootinbib, longbibliography]{revtex4-2}

\usepackage{amssymb,amsmath,verbatim,mathtools,needspace,enumitem,etoolbox,graphicx,physics,microtype,afterpage,xspace}
\graphicspath{{figures/}}

\usepackage[dvipsnames, usenames]{xcolor}
\usepackage{standalone}
\definecolor{linkcolor}{rgb}{0.0,0.3,0.5}
\usepackage[unicode, colorlinks=true, linkcolor=linkcolor, citecolor=linkcolor, filecolor=linkcolor,urlcolor=linkcolor, pdfusetitle]{hyperref}
\usepackage[all]{hypcap}
\usepackage[T1]{fontenc}
\usepackage{tikz}
\usepackage{tabularx}
\usepackage{xcolor}
\usepackage{natbib}
\definecolor{britishracinggreen}{rgb}{0.0, 0.26, 0.15}
\definecolor{cadmiumgreen}{rgb}{0.0, 0.42, 0.24}
\definecolor{forestgreen}{rgb}{0.13, 0.55, 0.13}
\allowdisplaybreaks
\usepackage[english]{babel}
\usepackage{eufrak}
\usepackage{aas_macros}

\def\sqr#1#2{{\vcenter{\vbox{\hrule height.#2pt\hbox{\vrule
width.#2pt height#1pt \kern#1pt\vrule width.#2pt}\hrule height.#2pt}}}}

\makeatletter

\@addtoreset{equation}{section}
\makeatother

\begin{document}

\title{Third post-Newtonian effective-one-body Hamiltonian\\
in scalar-tensor and Einstein-scalar-Gauss-Bonnet gravity}

\author{F\'elix-Louis Juli\'e}
\email{felix-louis.julie@aei.mpg.de}
\affiliation{Max Planck Institute for Gravitational Physics (Albert Einstein Institute),
\\ Am M\"uhlenberg 1, 14476 Potsdam, Germany}

\author{Vishal Baibhav}
\affiliation{Center for Interdisciplinary Exploration and Research in Astrophysics (CIERA),
\\ Northwestern University, 1800 Sherman Ave, Evanston, IL 60201, USA}

\author{Emanuele Berti}
\affiliation{Department of Physics and Astronomy, Johns Hopkins University, 3400 N. Charles Street, Baltimore, MD 21218, USA}

\author{Alessandra Buonanno}
\affiliation{Max Planck Institute for Gravitational Physics (Albert Einstein Institute),
\\ Am M\"uhlenberg 1, 14476 Potsdam, Germany}

\date{\today}

\begin{abstract}
We build an effective-one-body (EOB) Hamiltonian at third post-Newtonian (3PN) order  in scalar-tensor (ST) and Einstein-scalar-Gauss-Bonnet (ESGB) theories of gravity.
The latter is an extension of general relativity that predicts scalar hair for black holes.
We start from the known two-body Lagrangian at 3PN order, and use order-reduction methods to construct its ordinary Hamiltonian counterpart.
 We then reduce the conservative two-body dynamics to the (nongeodesic) motion of a test particle in an effective metric by means of canonical transformations.
The resulting EOB Hamiltonian is a modification 
of the general relativistic Hamiltonian, 
and already at 3PN order, it must account for nonlocal-in-time tail contributions. We include the latter beyond circular orbits and up to sixth order in the binary's orbital eccentricity.
We finally calculate the orbital frequency at the innermost stable circular orbit (ISCO) of binary black holes in the shift-symmetric ESGB model.
Our work extends F.L. Juli\'e and N. Deruelle~[Phys. Rev. D 95, 124054 (2017)],
and it is an essential step toward the accurate modeling of gravitational waveforms beyond general relativity.

\end{abstract}

\maketitle

\section{Introduction}
\label{sec:I}

The observations of gravitational waves (GWs) from coalescing binary systems composed of black holes (BHs) and neutron stars (NSs)~\cite{LIGOScientific:2016aoc,TheLIGOScientific:2017qsa,LIGOScientific:2020aai,LIGOScientific:2021qlt,LIGOScientific:2021djp} with the LIGO and Virgo detectors~\cite{LIGOScientific:2014pky,VIRGO:2014yos} offer the unique opportunity to unveil the nature of these compact objects and to test Einstein's theory of general relativity (GR) in the highly dynamical strong-field regime~\cite{TheLIGOScientific:2016src,Abbott:2018lct,LIGOScientific:2019fpa,LIGOScientific:2020tif,LIGOScientific:2021sio}.  The GW signals are at first ``chirps'' produced during the long inspiral phase, where the two bodies steadily and adiabatically come closer to each other, losing energy because of GW emission. The inspiral is followed by a short plunge and merger stage, where nonlinearities prevail, and then by the so-called ``ringdown'' phase for binary BHs~\cite{Kokkotas:1999bd,Berti:2009kk}, or by more complex pre- and postmerger signals (depending on the equation of state of the NS and on the properties of the BH) for binaries comprising at least one NS~\cite{Baiotti:2016qnr,Kyutoku:2021icp}.

Tests of GR for the different stages of the binary coalescence have been developed within theory-independent and theory-specific frameworks. In theory-independent tests, the underlying GW signal is assumed to be well-described by GR, and beyond-GR parameters are included in the waveform models to describe small deviations from GR (a nonexhaustive list includes Refs.~\cite{Blanchet:1994ez,Arun:2006hn,Yunes:2009ke,Li:2011cg,Agathos:2013upa,Barausse:2016eii,Cardoso:2019mqo,McManus:2019ulj,Khalil:2019wyy,Maselli:2019mjd,Carullo:2019flw,Ghosh:2021mrv,Bhagwat:2021kfa,Mehta:2022pcn}). By contrast, studies that analyze directly the data with waveform models constructed in beyond-GR theories of gravity are part of the theory-specific framework (see, e.g., Refs.~\cite{Will:2014kxa,Berti:2015itd,Yunes:2016jcc}). 

So far, the majority of the tests of GR with GW signals has been carried out following the theory-independent approach. However, in this framework the parametrizations are nonunique, the beyond-GR degrees of freedom can be degenerate with each other, and they are not necessarily guaranteed to represent the (infinite) landscape of beyond-GR theories. Thus, it is relevant to develop, both analytically (see, e.g., Refs.~\cite{Damour:1992we,Damour:1993hw,Yagi:2011xp,Mirshekari:2013vb,Lang:2013fna,Sennett:2016klh,Bernard:2018hta,Bernard:2018ivi,Bernard:2019yfz,Julie:2017pkb,Julie:2017ucp,Julie:2017rpw,Cardenas:2017chu,Julie:2018lfp,Julie:2019sab,Khalil:2018aaj,Shiralilou:2021mfl, Julie:2022huo}) and numerically (see, e.g., Refs.~\cite{Healy:2011ef,Barausse:2012da,Berti:2013gfa,Shibata:2013pra,Okounkova:2017yby,Witek:2018dmd,Okounkova:2019dfo,Julie:2020vov,Witek:2020uzz,Okounkova:2020rqw,Silva:2020omi,East:2020hgw,East:2021bqk,Figueras:2021abd,Corman:2022xqg,Doneva:2022byd,Elley:2022ept,Hegade:2022hlf}), waveform models in specific beyond-GR theories of gravity. Eventually, as already done for GR waveforms~\cite{Ossokine:2020kjp,Pratten:2020ceb,Gamba:2021ydi}, the combination of analytical and numerical-relativity (NR) results will produce accurate beyond-GR inspiral-merger-ringdown (IMR) waveform models, which will be used to probe gravity with the LIGO-Virgo-KAGRA interferometers, and with future detectors on the ground (Einstein Telescope and Cosmic Explorer)~\cite{Punturo:2010zz,Reitze:2019iox} and in space (LISA)~\cite{LISA:2017pwj}.  Importantly, next-decade facilities promise signal-to-noise ratios one or two orders of magnitude higher than what is achievable with current and near-future observations on the ground, thus allowing for exquisite tests of GR~\cite{Perkins:2020tra}.

Among the simplest modifications of GR, scalar-tensor (ST) theories add one massless scalar degree of freedom, which couples universally to matter.
They were introduced by Jordan, Fierz, Thiry, Brans and Dicke~\cite{Goenner:2012cq} and put in a
modern perspective in Refs.~\cite{Will:1989sk,Nordtvedt:1970uv,Damour:1992we}.
The corresponding two-body dynamics has been computed within the post-Newtonian (PN) formalism~\cite{Damour:1992we,Mirshekari:2013vb,Bernard:2018hta,Bernard:2018ivi,Lang:2013fna,Sennett:2016klh,Bernard:2022noq}.  
Interestingly, compact objects in ST theories can undergo a phase transition associated with the spontaneous symmetry breaking of the scalar field near the compact object in the  presence of large curvature or relativistic matter~\cite{Damour:1993hw}. For NSs, this phase transition leads to a rapid growth of the scalar charge (``spontaneous scalarization''). An analogous nonperturbative phenomenon (``dynamical scalarization'') was found in binary NS and NS-BH simulations in NR~\cite{Barausse:2012da,Shibata:2013pra,Palenzuela:2013hsa}. Various methods to describe these nonperturbative effects in waveform models have been proposed~\cite{Sampson:2013jpa,Khalil:2019wyy,Khalil:2022sii}.

However in ST theories, vacuum BH solutions are the same as in GR. By contrast, Einstein-scalar-Gauss-Bonnet (ESGB) theories have attracted particular attention because they have the interesting property that (i) for certain functional forms of the coupling constant, BH solutions in ESGB gravity are different from the solutions of GR, but admit the ordinary Kerr solutions as a special limit; and (ii) there is the possibility of ``spontaneous scalarization''~\cite{Silva:2017uqg,Doneva:2017bvd,Antoniou:2017acq} (i.e., BHs can ``grow hair'').
These observations opened up a much richer phenomenology for binary BHs~\cite{Doneva:2022ewd}. Recent progress in gravitational waveform modeling within ESGB gravity includes the calculation of inspiral waveforms using PN theory~\cite{Julie:2019sab,Shiralilou:2021mfl,Julie:2022huo} 
and the first calculation of quasinormal mode frequencies of rotating ESGB BHs at quadratic order in a small-spin expansion~\cite{Blazquez-Salcedo:2016enn,Pierini:2021jxd,Pierini:2022eim}.
The numerical calculation of merger-ringdown waveforms in ESGB gravity has also made remarkable progress, at first using a small-coupling approximation to numerically solve the field equations in an ``effective field theory'' approach~\cite{Witek:2018dmd,Witek:2020uzz,Okounkova:2020rqw,Silva:2020omi,Elley:2022ept}, and then by showing that numerical evolutions are possible in the full theory, although hyperbolicity can break down in some regions of the parameter space~\cite{East:2020hgw,Julie:2020vov,East:2021bqk,Corman:2022xqg,Hegade:2022hlf}.

An important step to build semianalytic IMR waveforms
is to construct an accurate analytic description of the two-body conservative inspiral dynamics. We achieve this here within the effective-one-body (EOB) formalism~\cite{Buonanno:1998gg,Buonanno:2000ef,Damour:2000we}.
The EOB approach builds IMR waveforms by combining analytical predictions for the inspiral, notably PN results, with perturbative calculations for the ringdown, and physically motivated ansatzes for the plunge-merger stage. 
The EOB waveforms are then informed and made highly accurate by calibration to NR simulations (see, e.g., Refs.~\cite{Ossokine:2020kjp,Gamba:2021ydi}). One key ingredient of the EOB formalism is the conservative EOB Hamiltonian.
The latter, for nonspinning compact objects and in GR, is built by mapping the two-body dynamics into that of an effective body moving in a deformed Schwarzschild spacetime, whose deformation parameter is the symmetric mass ratio $\nu = \mu/M$, where $\mu=m_A\,m_B/M$ is the binary's reduced mass, $m_A$ and $m_B$ are the component masses, and $M=m_A+m_B$ is the total mass~\cite{Buonanno:1998gg,Buonanno:2000ef}. Previous work extended the EOB Hamiltonian to ST and Einstein-Maxwell-scalar theories at 2PN and 1PN, respectively~\cite{Julie:2017pkb,Julie:2017ucp,Julie:2017rpw,Khalil:2018aaj,Julie:2018lfp}.
In this paper we
build upon Ref.~\cite{Julie:2017pkb}, and take advantage of recent progress in PN calculations in ST and ESGB theories~\cite{Bernard:2018hta,Bernard:2018ivi,Julie:2019sab}, to construct an EOB Hamiltonian at 3PN order for NSs and BHs in ST and ESGB theories. 

This paper is organized as follows. In Sec.~\ref{sec:II}, starting from the two-body 3PN Lagrangian in ST and ESGB theories, we derive, using order-reduction methods, the two-body Hamiltonian at 3PN order in the Einstein frame. In Sec.~\ref{sec:III}, we construct a canonical transformation that maps the two-body Hamiltonian into the EOB Hamiltonian, including nonlocal--in-time terms due to tail effects, which are already present at 3PN order in ST and ESGB theories. More specifically, we compute such tails for generic orbits in an expansion in the orbital eccentricity parameter.
In Sec.~\ref{sec:IV} we specify our EOB Hamiltonian to BH binaries in the shift-symmetric ESGB model, and we calculate the orbital frequency at the ISCO.
In Sec.~\ref{sec:V} we summarize our main conclusions and future research directions. Various technical details are relegated to the appendixes. In Appendix~\ref{app:EinsteinVsJordan} we develop a dictionary to relate quantities in the Einstein and Jordan frames. In Appendix~\ref{app:3PNlagrangian} we list the expression of the 3PN Lagrangian. In Appendix~\ref{app:contactTransfo} we discuss contact transformations of the two-body Lagrangian. In Appendix~\ref{app:twoBodHamiltonian} we give the two-body Hamiltonian. Finally, in Appendix~\ref{app:generatingFunctionCoeffs} we list the coefficients of the canonical transformations. Throughout this paper we use geometrical units $(G=c=1)$.

\section{The two-body Hamiltonian\label{sec:2bodyHamiltonian}}
\label{sec:II}

\subsection{ST and ESGB gravity}

We consider the theory described by the Einstein-frame action~\cite{Julie:2019sab, Damour:1992we}
\begin{align}
I  = \frac{1}{16 \pi}  \int\! d^{4}x & \sqrt{-g} \big(R - 2g^{\mu\nu} \partial_{\mu}\varphi \partial_{\nu}\varphi +\ell^2 f(\varphi)\mathcal G\big)\nonumber\\
&+I_{\rm m}[\Psi,\mathcal A^2(\varphi)g_{\mu\nu}]\,,
\label{eq:action}
\end{align}
where $R$ is the Ricci scalar, $g = {\rm det} \, g_{\mu\nu}$ is the metric
determinant, and
$\mathcal G=R^{\mu\nu\rho\sigma}R_{\mu\nu\rho\sigma} - 4R^{\mu\nu}R_{\mu\nu} + R^2$ is the Gauss-Bonnet scalar, with $R^{\mu}{}_{\nu\rho\sigma}$ and $R_{\mu\nu}$ the Riemann and Ricci
tensors, respectively.
The integral of the Gauss-Bonnet scalar over a four-dimensional spacetime $\int d^4x\sqrt{-g}\,\mathcal G$ is a boundary term \cite{Myers:1987yn}.
Matter fields $\Psi$ are minimally coupled to the Jordan metric $\tilde g_{\mu\nu}=\mathcal A^2(\varphi)g_{\mu\nu}$.
The dimensionless functions $\mathcal A$ and $f$ and the constant quantity $\ell$ (with dimensions of length) specify the theory.
We recover ST theories when either $\ell=0$ or $f$ is a constant, and GR when moreover $\mathcal A$ (and $\varphi$) are constant. 

When dealing with compact bodies, we adopt the phenomenological treatment initiated in Refs.~\cite{1975ApJ...196L..59E,Damour:1992we} in ST theories,
and describe them as point particles:
\begin{align}
I_{\rm m}\to I_{\rm m}^{\rm pp}[g_{\mu\nu},\varphi, \{x_A^\mu\}]=-\sum_A\int m_A(\varphi)\, d s_A\,,
\label{eq:actionSkel}
\end{align}
 where $d s_A=\sqrt{-g_{\mu\nu}dx_A^\mu dx_A^\nu}$ and $x_A^\mu[s_A]$ is the worldline of particle $A$. 
The constant GR mass is replaced by a function $m_A(\varphi)$ that depends on the internal structure of body $A$ and on the value of the scalar field at $x_A^\mu(s_A)$.
For an explicit calculation of the mass $m_A(\varphi)$ of an ESGB BH, see Refs.~\cite{Julie:2019sab,Julie:2022huo}; see also Refs.~\cite{Damour:1992we,Damour:1993hw,Zaglauer:1992bp} for NSs in ST theories.

From now on, we will refer to the theory with action \eqref{eq:action} as ``ESGB gravity,'' but we note that the action includes ST gravity as a special case.

\subsection{The two-body Lagrangian at 3PN\label{subsec:2-bodyLag3PN}}

In this paper, we focus on the conservative dynamics of compact binaries on bound orbits. When the relative orbital velocity is small and the gravitational field is weak, the motion can be studied in the PN framework.\footnote{We denote by $n$PN the relative $\mathcal O(v^{2n})\sim\mathcal O(M/r)^n$ corrections to Newtonian gravity, with $v$ the system's relative orbital velocity, $r$ the orbital separation, and $M$ the total mass.}
To do so, the field equations of the theory \eqref{eq:action} with \eqref{eq:actionSkel} are solved iteratively around a flat metric $g_{\mu\nu}=\eta_{\mu\nu}+\delta g_{\mu\nu}$ and a constant scalar background $\varphi=\varphi_0+\delta\varphi$, where $\varphi_0$ is imposed by the binary's cosmological environment. In particular, the functions $m_A(\varphi)$ and $m_B(\varphi)$, describing bodies $A$ and $B$, can be expanded at 3PN by introducing
\begin{subequations}
\begin{align}
\alpha_A^0&=\frac{d\ln m_A}{d\varphi}(\varphi_0)\,,\\
\beta_A^0&=\frac{d\alpha_A}{d\varphi}(\varphi_0)\,,\\
{\beta'}_A^0&=\frac{d\beta_A}{d\varphi}(\varphi_0)\,,\\
{\beta''}_A^0&=\frac{d{\beta'}_A}{d\varphi}(\varphi_0)\,,
\end{align}\label{eq:sensis}%
\end{subequations}
and their counterparts for body $B$, where from now on the superscript $0$ denotes a quantity evaluated at $\varphi=\varphi_0$.

The ST two-body Lagrangian was derived at 1PN by Damour and Esposito-Far\`ese  \cite{Damour:1992we}, at 2PN by Mirshekari and Will  \cite{Mirshekari:2013vb},
and at 3PN by Bernard  \cite{Bernard:2018hta,Bernard:2018ivi}. It was then generalized by Juli\'e and Berti, who derived its ESGB corrections in Ref.~\cite{Julie:2019sab}.
However, the results in Refs.~\cite{Mirshekari:2013vb,Bernard:2018hta,Bernard:2018ivi} are presented using a different, ``Jordan-frame'' formulation of ST theories based on a set of Brans-Dicke-inspired parameters.
To recover the conventions of the present paper, we must proceed as follows:
\begin{enumerate}
\item Translate the parameters in Refs.~\cite{Mirshekari:2013vb,Bernard:2018hta,Bernard:2018ivi} in terms of the quantities \eqref{eq:sensis}. The conversion is detailed in Appendix~\ref{app:EinsteinVsJordan}.
\item \label{item2EinsteinVsJordan} Observe that Refs.~\cite{Mirshekari:2013vb,Bernard:2018hta,Bernard:2018ivi} use a coordinate system $\{\tilde x^\mu\}$ such that the Jordan metric $\tilde g_{\mu\nu}=\mathcal A^2(\varphi)g_{\mu\nu}$ is Minkowski at infinity, $\tilde g_{\mu\nu}\to\eta_{\mu\nu}$. By contrast, we use here coordinates $\{x^\mu\}$ such that $g_{\mu\nu}\to\eta_{\mu\nu}$.
This means that
\begin{align}
\tilde x^\mu=\mathcal A_0 x^\mu
\end{align}
with $\mathcal A_0=\mathcal A(\varphi_0)$, so that the orbital radius $\tilde r$ and body accelerations $\tilde{\mathbf a}_A$ entering Refs.~\cite{Mirshekari:2013vb,Bernard:2018hta,Bernard:2018ivi} translate as $\tilde r=\mathcal A_0 r$ and $\tilde{\mathbf a}_A=\mathbf a_A/\mathcal A_0 $ in our conventions.
\item Since also $\tilde t=\mathcal A_0 t$, the two-body Lagrangians $\tilde L$ given in Refs.~\cite{Mirshekari:2013vb,Bernard:2018hta,Bernard:2018ivi} must be rescaled as $ L=\mathcal A_0\tilde L$.
\end{enumerate}

We denote by $\mathbf x_A$ the spatial position of body $A$, and introduce the notations $r=|\mathbf x_A-\mathbf x_B|$, $\mathbf n=(\mathbf x_A-\mathbf x_B)/r$, $\mathbf v_A=\dot{\mathbf x}_A=d\mathbf x_A/dt$ and $\mathbf a_A=\dot{\mathbf v}_A$. The ESGB two-body Lagrangian is then, in harmonic coordinates such that $\partial_\mu\left(\sqrt{-g}g^{\mu\nu}\right)=0$:
\begin{align}
L&=-m_A^0-m_B^0+L_{0\rm PN}+L_{1\rm PN}+L_{2\rm PN}\nonumber\\
&+L_{3\rm PN}+\mathcal O(v^{10})\,,\label{eq:twoBodyLstructure}
\end{align}
where the contributions up to 2PN were 
 presented in Refs.~\cite{Julie:2017pkb,Julie:2017ucp} and are recalled in Appendix~\ref{app:3PNlagrangian}. We decompose the new 3PN contribution as
\begin{align}
L_{3\rm PN}&=\sum_{i=0}^4 L_{3\rm PN}^{(i)}+L_{3\rm PN}^{\rm tail}+\Delta L_{3\rm PN}^{\rm ESGB}\,,\label{eq:3PNlagrangianStructure}
\end{align}
where the lengthy expressions of the terms $L_{3\rm PN}^{(i)}$ are also given in Appendix~\ref{app:3PNlagrangian}.
 They depend on the logarithms $\ln(r/r_A)$ and $\ln(r/r_B)$, where $r_A$ and $r_B$ are regularization lengths that we shall eliminate later.

However, one of us noticed, while performing the conversion from the Jordan to the Einstein frame, that some terms in $L_{3\rm PN}^{(i)}$, originated from the results of Ref.~\cite{Bernard:2018ivi}, must be revised. The Einstein-frame two-body dynamics is indeed described by the action \eqref{eq:action} with matter explicitly accounted for by Eq.~\eqref{eq:actionSkel}.
The associated PN Lagrangian should thus not depend on $\mathcal A$ and its derivatives at infinity.
Yet,
the prefactors of the first lines in Eqs.~\eqref{eq:2bodyL3PN3} and \eqref{eq:2bodyL3PN4} are inversely proportional to $\tilde\alpha=(1+\alpha_A^0\alpha_B^0)/(1+\alpha_0^2)$, where $\alpha_0=(d\ln\mathcal A/d\varphi)_{\varphi_0}$, cf. Appendix~\ref{app:EinsteinVsJordan}.
This issue will be addressed in an upcoming publication~\cite{LBandFLJ}.
For now, we note that adjusting $\tilde\alpha$ will not affect the structure of our results.

 The ESGB correction beyond ST reads~\cite{Julie:2019sab} 
\begin{align}
\!\Delta L_{3\rm PN}^{\rm ESGB}&=\frac{\ell^2 f'(\varphi_0)}{M^2}\left(\frac{M}{r}\right)^2\frac{m_A^0m_B^0}{r^2}\nonumber\\
&\times\left[m_A^0(\alpha_B^0+2\alpha_A^0)+m_B^0(\alpha_A^0+2\alpha_B^0)\right]\,,\label{eq:2bodyLagrangianESGB}
\end{align}
with $f'(\varphi_0)=(df/d\varphi)_{\varphi_0}$ and $M=m_A^0+m_B^0$. It is numerically of the same order of magnitude as a 3PN term whenever $\ell^2 f'(\varphi_0)\lesssim M^2$. It turns out that this condition is satisfied by the nonperturbative ESGB BH solutions studied in Ref.~\cite{Julie:2022huo}.

Finally, $L_{3\rm PN}$ depends on a nonlocal-in-time ``tail'' contribution which we converted from the Jordan-frame expression of Ref.~\cite{Bernard:2018ivi},
\begin{align}
L_{3\rm PN}^{\rm tail}=\frac{2M\mathcal A_0^2}{3}\ddot D^i(t)\left(\underset{2r(t)}{\rm PF}\int_{-\infty}^{+\infty}\!\frac{d\tau}{|\tau|}\ddot D^i(t+\tau)\right)\,,\label{eq:LagrangianTail}
\end{align}
which is driven by the acceleration of the scalar dipole $D^i=m_A^0\alpha_A^0 x_A^i+m_B^0\alpha_B^0 x_B^i$.
This tail term is absent in GR. 
By the same arguments we made earlier, the tail term should be independent of $\mathcal A_0=\mathcal A(\varphi_0)$.
This issue will also be addressed in Ref.~\cite{LBandFLJ}, and for now, we note that replacing $\mathcal A(\varphi_0)$ by a different constant will not change the structure of our final results.
Here,  ``PF'' denotes the Hadamard partie finie, and we follow the conventions of Refs.~\cite{Damour:2014jta,Bernard:2018hta,Bernard:2018ivi}: given a regular function $f(t)$ vanishing sufficiently fast at infinity and a constant $s$, we have
\begin{align}
\underset{2s}{\rm PF}\!\int_{\Bbb R}\frac{d\tau}{|\tau|} f(t+\tau)=\!\int_{\Bbb R+}\!\!\!\!d\tau\ln\left(\frac{\tau}{2s}\right)\left(\dot f(t-\tau)-\dot f(t+\tau)\right)\,.\label{eq:defPF}
\end{align}

The two-body Lagrangian \eqref{eq:twoBodyLstructure} depends on the theory-dependent combination $\ell^2 f'(\varphi_0)$ entering Eq.~\eqref{eq:2bodyLagrangianESGB}, and on ten body-dependent parameters: the masses of each body and their logarithmic derivatives \eqref{eq:sensis} at infinity.
It is also useful to introduce the following quantities, ordered by the PN level at which they appear, from 0PN to 3PN:
\begin{widetext}
\begin{subequations}
\begin{align}
G_{AB}&=1+\alpha_A^0\alpha_B^0\,,\\
\bar\gamma_{AB}&= -\frac{2\alpha_A^0\alpha_B^0}{1+\alpha_A^0\alpha_B^0}\,,\quad\bar\beta_A=\frac{1}{2}\frac{\beta_A^0(\alpha_B^0)^2}{(1+\alpha_A^0\alpha_B^0)^2}\,,\label{eq:combinations1PN}\\
\delta_A&=\frac{(\alpha_A^0)^2}{(1+\alpha_A^0\alpha_B^0)^2}\,,\quad\epsilon_A=\frac{{\beta'}_A^0(\alpha_B^0)^3}{(1+\alpha_A^0\alpha_B^0)^3}\,,\quad\zeta_{AB}=\frac{\beta_A^0\beta_B^0\alpha_A^0\alpha_B^0}{(1+\alpha_A^0\alpha_B^0)^3}\,,\label{eq:combinations2PN}\\
\omega_A&=\frac{(\alpha_A^0)^2\beta_A^0(\beta_B^0)^2}{(1+\alpha_A^0\alpha_B^0)^4}\,,\quad
\kappa_A=\frac{{\beta''}_A^0(\alpha_B^0)^4}{8(1+\alpha_A^0\alpha_B^0)^4}\,,\quad
\xi_A=\frac{(\alpha_A^0)^2\alpha_B^0\beta_A^0{\beta'}_B^0}{(1+\alpha_A^0\alpha_B^0)^4}\,,\quad
\psi_A=\frac{\alpha_A^0\alpha_B^0\beta_A^0}{(1+\alpha_A^0\alpha_B^0)^3}\,,\label{eq:combinations3PN}
\end{align}\label{eq:combinations}%
\end{subequations}
\end{widetext}
and their $(A\leftrightarrow B)$ counterparts.
The quantities \eqref{eq:combinations3PN} are new to this paper, and we named the first three of them according to their (field theory) diagrammatic interpretation, as was initiated at 2PN by Damour and Esposito-Far\`ese in Ref.~\cite{Damour:1995kt}.
We recover the ST Lagrangian in the limit $\ell^2 f'(\varphi_0)=0$, such that \eqref{eq:2bodyLagrangianESGB} vanishes. We recover GR when moreover $m_A(\varphi)$ and $m_B(\varphi)$ are constants: then, Eqs.~\eqref{eq:sensis} and their $B$ counterparts are zero, so that $G_{AB}=1$ and \eqref{eq:combinations1PN}-\eqref{eq:combinations3PN} all vanish.

\subsection{The order-reduced Lagrangian}

The Lagrangian \eqref{eq:twoBodyLstructure} is written in harmonic coordinates, and it depends on the accelerations $\mathbf a_A$ and $\mathbf a_B$ of the bodies, both linearly via $L_{2\rm PN}$, $L_{3\rm PN}^{(1)}$ and $L_{3\rm PN}^{(2)}$ [cf. Appendix~\ref{app:3PNlagrangian}], and  quadratically via the tail contribution $L_{3\rm PN}^{\rm tail}$.
 To deal with an ordinary Lagrangian depending on positions and velocities only, we can replace the accelerations by their on-shell 1PN expressions, as we now prove.

Consider a degree of freedom $q(t)$ described by the action $I=\int dt L[q,\dot q,\ddot q]$, where
\begin{align}
L[q,\dot q,\ddot q]&=L_0(q,\dot q)+\epsilon L_1(q,\dot q)\nonumber\\
&+\epsilon^2[L_2(q,\dot q)+\ell_2(q,\dot q)\ddot q]\nonumber\\
&+\epsilon^3\Big[L_3(q,\dot q)+\ell_3(q,\dot q)\ddot q+ \ddot q\,\underset{2q}{\rm PF}\!\int_{\Bbb R}\frac{d\tau}{|\tau|} \ddot q(t+\tau)\Big]\nonumber\\
&+\mathcal O(\epsilon^4)\,,\label{eq:lagrangian1TM}
\end{align}
with $\epsilon\ll 1$ an expansion parameter.
The Lagrangian \eqref{eq:lagrangian1TM} depends on $\ddot q$ linearly at $\mathcal O(\epsilon^2)$ and $\mathcal O(\epsilon^3)$, and also quadratically via a nonlocal-in-time contribution at $\mathcal O(\epsilon^3)$.
The Euler-Lagrange variation of
\begin{align}
    F(q,\dot q)=L_0(q,\dot q)+\epsilon L_1(q,\dot q)
\end{align}
reads
\begin{align}
\frac{\delta F}{\delta q}&=\frac{\partial F}{\partial q}-\frac{d}{dt}\left(\frac{\partial F}{\partial {\dot{q}}}\right)\nonumber\\
&=\frac{\partial F}{\partial q}-\dot q\,\frac{\partial}{\partial q}\left(\frac{\partial F}{\partial\dot q}\right)-\ddot q\, \frac{\partial^2\! F}{\partial\dot q^2}\,,\label{eq:ELfromF}
\end{align}
where the second equality follows from the chain rule. Now introduce the notations
\begin{subequations}
\begin{align}
H_F(q,\dot q)&=\frac{\partial^2\! F}{\partial\dot q^2}\,,\\
\ddot q_F(q,\dot q)&=\frac{1}{H_F}\left[\frac{\partial F}{\partial q}-\dot q\,\frac{\partial}{\partial q}\left(\frac{\partial F}{\partial\dot q}\right)\right]\,,
\end{align}
\end{subequations}
where $H_F(q,\dot q)$ is the Hessian of $F(q,\dot q)$. Then Eq.~\eqref{eq:ELfromF} can be rewritten as the identity
\begin{align}
\ddot q=\ddot q_F-\frac{1}{H_F}\frac{\delta F}{\delta q}\,,\label{eq:contactTransfoAccelOnShellF}
\end{align}
which reduces to $\ddot q=\ddot q_F(q,\dot q)+\mathcal O(\epsilon^2)$ when the Euler-Lagrange equations of $F(q,\dot q)$ are satisfied,  $\delta F/\delta q=0$.
We can then insert \eqref{eq:contactTransfoAccelOnShellF} into \eqref{eq:lagrangian1TM} to find:
\begin{align}
L&=L_{\rm red}\nonumber\\
&-\frac{1}{H_F}\frac{\delta F}{\delta q}\Big[\epsilon^2\ell_2+\epsilon^3\Big(\ell_3+2\underset{2q}{\rm PF}\int_{\Bbb R}\frac{d\tau}{|\tau|}\ddot q_F\rvert_{t+\tau}\Big)\Big]\nonumber\\
&+\epsilon^3\frac{1}{H_F}\frac{\delta F}{\delta q}\underset{2q}{\rm PF}\!\int_{\Bbb R}\frac{d\tau}{|\tau|}\frac{1}{H_F}
\left. \frac{\delta F}{\delta q} \right|_{t+\tau}\nonumber\\
&+\mathcal O(\epsilon^4)\,,\label{eq:reducedLagrangianToyModel}%
\end{align}
where 
\begin{align}
L_{\rm red}[q,\dot q]=L[q,\dot q,\ddot q_F(q,\dot q)]
\end{align}
is an ordinary Lagrangian depending only on $q$ and $\dot q$, obtained by replacing the acceleration $\ddot q$ in $L[q,\dot q,\ddot q]$ by its on-shell expression deduced from $F(q,\dot q)$.\footnote{Using  Eq.~\eqref{eq:defPF} we have that
\begin{align*}
\underset{2q(t)}{\rm PF}\int_{\Bbb R}\frac{d\tau}{|\tau|}B(t+\tau)=\underset{2s}{\rm PF}\int_{\Bbb R}\frac{d\tau}{|\tau|}B(t+\tau)+2\ln\left(\frac{s}{q(t)}\right)B(t)\,,
\end{align*}
from which we deduce an identity that is useful to prove Eq.~\eqref{eq:reducedLagrangianToyModel}:
\begin{align*}
\int_{\Bbb R}dt A(t)\underset{2q(t)}{\rm PF}\int_{\Bbb R}\frac{d\tau}{|\tau|}B(t+\tau)=\int_{\Bbb R}dt B(t)\underset{2q(t)}{\rm PF}\int_{\Bbb R}\frac{d\tau}{|\tau|}A(t+\tau)\,.
\end{align*}
\label{footnotePFprop}}

The third line of Eq.~\eqref{eq:reducedLagrangianToyModel} is doubly zero: its $\mathcal O(\epsilon^3)$ contributions to the equations of motion are at least linear in $\delta F/\delta q$, which vanishes on shell. Thus it can be discarded.
As for the second line of Eq.~\eqref{eq:reducedLagrangianToyModel}, it can also be eliminated via a variable change $q\to q+\delta q[q,\dot q]$ with $\delta q=\mathcal O(\epsilon^2)$. Indeed, the Lagrangian then transforms, by definition, as
\begin{align}
L\to L+\frac{\delta F}{\delta q}\delta q+\mathcal O(\epsilon^4)\,,\label{eq:variableChangeToyModel}
\end{align}
modulo an irrelevant total time derivative, and we can choose
\begin{align}
\delta q[q,\dot q]=\frac{1}{H_F}\Big[\epsilon^2\ell_2+\epsilon^3\Big(\ell_3+ 2\underset{2q}{\rm PF}\!\int_{\Bbb R}\frac{d\tau}{|\tau|}\ddot q_F\rvert_{t+\tau}\Big)\Big]\,.\label{eq:contactTransfoQ}
\end{align}
This variable change belongs to the class of contact transformations introduced by Sch\"{a}fer and Damour in Refs.~\cite{Schafer:1984mr,Damour:1990jh},
which we generalized to include nonlocal-in-time terms for our purpose.

Now return to the two-body Lagrangian \eqref{eq:twoBodyLstructure}. 
We can replace the accelerations by their on-shell expressions deduced from $F=L_{0\rm PN}+L_{1\rm PN}$:
\begin{align}
L_{\rm red}=L[a_{A}^i\to (a_F)_A^{i},\,a_{B}^i\to (a_F)_B^{i}]+\mathcal O(v^{10})\,,\label{eq:lagRedTwoBody}
\end{align}
where $(a_F)_A^{i}$ is a function of the positions and velocities given in Appendix \ref{app:contactTransfo}.
Note that it is sufficient to replace the accelerations entering the Lagrangian at 3PN level by their 0PN expressions.
This procedure amounts to making an implicit 4D coordinate change via a contact transformation $\mathbf x_A\to\mathbf x_A+\delta \mathbf x_A$ resembling \eqref{eq:contactTransfoQ},
\begin{subequations}
\begin{align}
&\!\delta x_A^i=\sum_{B}(H_F^{-1})^{Ai}_{Bj}\Big[\frac{\partial L_{2\rm PN}}{\partial a_B^j}+\frac{\partial (L_{3\rm PN}^{(1)}+L_{3\rm PN}^{(2)})}{\partial a_B^j}\\
 &\quad+\frac{4M\mathcal A_0^2}{3} m_B^0\alpha^0_B\sum_C m_C^0\alpha^0_C\,\underset{2r}{\rm PF}\!\int_{\Bbb R}\frac{d\tau}{|\tau|}(a_F)_C^j\rvert_{t+\tau}\Big]\,,\nonumber\\
&\! (H_F)^{Ai}_{Bj}=\frac{\partial^2 F}{\partial v_A^i\partial v_B^j}\,,\label{eq:Hessian1PNdef}%
\end{align}\label{eq:Hessian1PNdefComplete}%
\end{subequations}
which we also give explicitly in Appendix \ref{app:contactTransfo}. We verified that applying the contact transformation \eqref{eq:appendixContactExplicit} to the two-body Lagrangian \eqref{eq:twoBodyLstructure} yields a result that matches, modulo total time derivatives and doubly zero terms, the order-reduced Lagrangian \eqref{eq:lagRedTwoBody}.

From now on we work with the order-reduced Lagrangian $L_{\rm red}$, and thus, in a coordinate system other than harmonic.

\subsection{The two-body Hamiltonian at 3PN\label{sec:twoBodyH3PN}}

From the order-reduced Lagrangian \eqref{eq:lagRedTwoBody}, we can infer an ordinary Hamiltonian via the Legendre transformation:
\begin{align}
H&=\mathbf p_A\cdot \mathbf v_A+\mathbf p_B\cdot \mathbf v_B-L_{\rm red}\,,\label{eq:Legendre}
\end{align}
with
\begin{align}
\mathbf p_A=\frac{\partial L_{\rm red}}{\partial \mathbf v_A}\,,\quad\mathbf p_B=\frac{\partial L_{\rm red}}{\partial \mathbf v_B}\,.\label{eq:momenta}
\end{align}

A technically useful remark is that,
when deriving a Hamiltonian from a $n$PN Lagrangian, it is sufficient to calculate \eqref{eq:momenta} at $(n-1)$PN order, since after inversion, the $n$PN corrections to $\mathbf v_A(\mathbf p_A,\mathbf p_B)$ and $\mathbf v_B(\mathbf p_A,\mathbf p_B)$ cancel out in Eq.~\eqref{eq:Legendre}. We thus need the momenta at 2PN order only.

In the center-of-mass frame such that $\mathbf p_A+\mathbf p_B=0$, the conjugate variables are $\mathbf r=\mathbf x_A-\mathbf x_B$ and $\mathbf p=\mathbf p_A=-\mathbf p_B$.
The motion being planar, we use polar coordinates $(r,\phi)$ with conjugate momenta $p_r=\mathbf n\cdot \mathbf p$ and $p_\phi=r (\mathbf n\times\mathbf p)_z$.
We introduce the reduced mass and mass ratios 
\begin{align}
\mu&=\frac{m_A^0m_B^0}{M}\,,\quad \nu=\frac{\mu}{M}\,,\quad m_-=\frac{m_A^0-m_B^0}{M}\,,\label{def:reducedMass}
\end{align}
together with the following dimensionless quantities:
\begin{align}
\hat r&=\frac{r}{M}\,,\quad  \hat t=\frac{t}{M}\,,\quad \hat\tau=\frac{\tau}{M}\,,\nonumber\\
\hat p_r&=\frac{p_r}{\mu}\,, \quad\hat p_\phi=\frac{p_\phi}{\mu M}\,,\quad \hat p^2=\hat p_r^2+\frac{\hat p_\phi^2}{\hat r^2}\,.
\end{align}
We denote by a subscript $+$ (respectively, $-$) the (anti) symmetrization of the quantities~\eqref{eq:combinations}, as in, e.g., $\bar\beta_+=\bar\beta_A+\bar\beta_B$ and $\bar\beta_-=\bar\beta_A-\bar\beta_B$ [note the factor of 2 compared to Refs.~\cite{Sennett:2016klh,Bernard:2018hta,Bernard:2018ivi}].
The two-body Hamiltonian is then:
\begin{align}
\hat H=\frac{H}{\mu}&=\frac{M}{\mu}+\hat H_{0\rm PN}+\hat H_{1\rm PN}+\hat H_{2\rm PN}\nonumber\\
&+\hat H_{3\rm PN}+\mathcal O(\hat p^{10})\,,\label{eq:2bodyHamiltonian}
\end{align}
where the contributions up to 2PN were first derived in Refs.~\cite{Julie:2017pkb,Julie:2017ucp} and are recalled in Appendix~\ref{app:twoBodHamiltonian}. The new 3PN contributions read
\begin{align}
\hat H_{3\rm PN}=\sum_{i=0}^4 \hat H_{3\rm PN}^{(i)}+\hat H_{3\rm PN}^{\rm tail}+\Delta \hat H_{3\rm PN}^{\rm ESGB}\,,\label{eq:2bodyHamiltonian3PN}
\end{align}
where the lengthy expressions of $\hat H_{3\rm PN}^{(i)}$ are also given in Appendix~\ref{app:twoBodHamiltonian}. They depend on $\ln_\pm=\ln(\hat r_A)\pm\ln(\hat r_B)$, where $\hat r_A=r_A/M$ and $\hat r_B=r_B/M$ are the dimensionless regularization lengths mentioned below Eq.~\eqref{eq:3PNlagrangianStructure}.

For the reason given above,
the tail and ESGB contributions are equal and opposite to their Lagrangian counterparts:
\begin{align}
\!\!\!\!\!\hat H_{3\rm PN}^{\rm tail}&=\frac{G_{AB}^4\, k_{\rm tail}}{\hat r^2}\,\underset{2\hat r}{\rm PF}\!\int_{\Bbb R}\frac{d\hat\tau}{|\hat\tau|}\frac{\cos\Delta\phi}{\hat r^2(\hat t+\hat \tau)}\,,\label{eq:2bodyHamiltonianTail}
\end{align}
where the cosine of $\Delta\phi=\phi(\hat t+\hat\tau)-\phi(\hat t)$  follows from the order-reduced, center-of-mass frame acceleration $\ddot D^i=-G_{AB} \nu(\alpha_A^0-\alpha_B^0) n^i/\hat r^2$ taken at $\hat t$ and $\hat t+\hat\tau$, and
\begin{align}
    k_{\rm tail}=-\frac{2\mathcal A_0^2}{3}\frac{(\alpha_A^0-\alpha_B^0)^2\nu}{(1+\alpha_A^0\alpha_B^0)^2}\,.\label{def:ktail}%
\end{align}
The ESGB correction beyond ST reads
\begin{align}
\Delta \hat H_{3\rm PN}^{\rm ESGB}&=\frac{G_{AB}^4\, k_{\rm ESGB}}{\hat r^4}\,,\label{eq:2bodyHamiltonianESGB}
\end{align}
with
\begin{align}
k_{\rm ESGB}&=-\frac{\ell^2 f'(\varphi_0)}{2M^2}\frac{3(\alpha_A^0+\alpha_B^0)+m_-(\alpha_A^0-\alpha_B^0)}{(1+\alpha_A^0\alpha_B^0)^4}\,.
\end{align}

Finally, to prepare for the calculations
of Sec.~\ref{sec:EOBhamiltonian} below when the eccentricity is nonzero, and thus $\hat r$ is not constant, we use the first identity in footnote~\ref{footnotePFprop} to get
\begin{align}
\hat H_{3\rm PN}^{\rm tail}&=G_{AB}^4\, k_{\rm tail}\\
\times &\left[\frac{1}{\hat r^2}\underset{2\hat s}{\rm PF}\!\int_{\Bbb R}\frac{d\hat\tau}{|\hat\tau|}\frac{\cos\Delta\phi}{\hat r^2(\hat t+\hat \tau)}-\frac{2\ln(\hat r/\hat s)}{\hat r^4}\right]\nonumber\,,
\end{align}
where $\hat s=s/M$ is an arbitrary constant [which will not appear in our final EOB Hamiltonian]. Following what was done in GR at 4PN order in Ref.~\cite{Damour:2014jta}, we then decompose the two-body Hamiltonian \eqref{eq:2bodyHamiltonian} into local-in-time and nonlocal-in-time parts,
\begin{align}
\hat H=\hat H^{\rm I}+\hat H^{\rm II}+\mathcal O(\hat p^{10})\,,\label{eq:decompositionH2body}
 \end{align}
with
\begin{subequations}
\begin{align}
\hat H^{\rm I}&=\frac{M}{\mu}+\hat H_{0\rm PN}+\hat H_{1\rm PN}+\hat H_{2\rm PN}\label{eq:decomposed2bodyHamiltonian1}\\
&+\sum_{i=0}^4 \hat H_{3\rm PN}^{(i)}-2G_{AB}^4 k_{\rm tail}\frac{\ln(\hat r/\hat s)}{\hat r^4}+\Delta \hat H_{3\rm PN}^{\rm ESGB}\,,\nonumber\\
\hat H^{\rm II}&=\frac{G_{AB}^4k_{\rm tail}}{\hat r^2}\,\underset{2\hat s}{\rm PF}\!\int_{\Bbb R}\frac{d\hat\tau}{|\hat\tau|}\frac{\cos\Delta\phi}{\hat r^2(\hat t+\hat \tau)}\,.\label{eq:decomposed2bodyHamiltonian2}
\end{align}\label{eq:decomposed2bodyHamiltonian}%
\end{subequations}

\section{The EOB mapping\label{sec:EOBhamiltonian}}
\label{sec:III}

\subsection{The EOB Hamiltonian at 3PN order\label{subsec:EOBhamiltonian}}

In Sec.~\ref{sec:2bodyHamiltonian}, we derived a two-body Hamiltonian
at 3PN order and in the center-of-mass frame.
We shall now use canonical transformations $(r,\phi,p_r,p_\phi)\to(R,\Phi,P_R,P_\Phi)$ to identify
it with an EOB Hamiltonian \cite{Buonanno:1998gg,Damour:2015isa,Damour:2016gwp}
\begin{align}
\hat H_{\rm EOB}=\frac{H_{\rm EOB}}{\mu}=\frac{M}{\mu}\sqrt{1+2\nu\left(\hat H_{\rm eff}-1\right)}\,,\label{eq:defHeob}
\end{align}
 where $\hat H_{\rm eff}$ is an effective Hamiltonian to be constructed. In this paper, 
we write the effective Hamiltonian in the same gauge as that used in GR at 2PN, 3PN and 4PN orders~\cite{Buonanno:1998gg,Damour:2000we,Damour:2015isa}, that is
\begin{align}
\hat H_{\rm eff}=\frac{H_{\rm eff}}{\mu}=\sqrt{A\left(1+{AD\, \hat P_R^2}+\frac{\hat P_\phi^2}{\hat R^2}\right)+Q}\,,\label{eq:effHam}
\end{align}
which depends on three potentials. Through 3PN order they can be expanded as
\begin{subequations}
\begin{align}
A&=1+\frac{a_1}{\hat R}+\frac{a_2}{\hat R^2}+\frac{a_3}{\hat R^3}+\frac{{ a_4}+{ a_{4,\ln}}\ln\hat R}{\hat R^4}\,,\\
D&=1+\frac{  d_1}{\hat R}+\frac{  d_2}{\hat R^2}+\frac{{  d_3}+{  d_{3,\ln}}\ln\hat R}{\hat R^3}\,,\\
Q&=\frac{({  q_{1}}+{  q_{1,\ln}}\ln\hat R)\,\hat P_R^4}{\hat R^2}+\frac{({  q_{2}}+{  q_{2,\ln}}\ln \hat R)\,\hat P_R^6}{\hat R}\,,\label{eq:Q}
\end{align}\label{eq:ADQ}%
\end{subequations}
where
\begin{align}
\hat R=\frac{R}{M}\,,\quad  \hat P_R=\frac{P_R}{\mu}\,, \quad\hat P_\Phi=\frac{P_\phi}{\mu M}\,.
\end{align}

When restricted to 2PN order, the EOB Hamiltonian above depends on five coefficients $(a_1,a_2,a_3)$ and $(d_1,d_2)$,  which were derived in Ref.~\cite{Julie:2017pkb} in ST theories. We will recall their expressions in Sec.~\ref{subseq:completePotentials} for completeness.
In this paper, we introduce the remaining eight coefficients to include 3PN contributions in ST-ESGB gravity.

The effective Hamiltonian \eqref{eq:effHam} describes the motion of a test particle with mass $\mu$ in an effective static, spherically symmetric metric [in Schwarzschild-Droste coordinates with $\theta=\pi/2$]
\begin{align}
ds^2_{\rm eff}=-A\,dt^2+\frac{dR^2}{AD}+R^2d\Phi^2\,,\label{eq:effectiveMetric}
\end{align}
but it is now deformed by a nongeodesic 3PN potential $Q$
which vanishes for circular orbits such that $\hat P_R=0$.
The reason is twofold:

\begin{enumerate}
\item
At 3PN order and already in GR, the two-body dynamics cannot be reduced to geodesic motion~\cite{Damour:2000we}.
Following Damour, Jaranowski and Sch\"{a}fer, we thus include a postgeodesic correction controlled by the coefficient $q_1$.

\item
In GR, the two-body Hamiltonian depends at 4PN order on a quadrupole-driven tail~\cite{Damour:2014jta,Bernard:2015njp},
which can yet be accounted for in a local-in-time EOB Hamiltonian by extending $Q$ to an infinite postgeodesic series~\cite{Damour:2015isa}
\begin{align}
Q=\sum_{n=2} Q_{2n}(\hat R)\hat P_R^{2n}\,,\label{eq:QinfiniteSeries}
\end{align}
together with $\ln\hat R$-dependent EOB potentials.
Our ansatz \eqref{eq:ADQ} adapts this strategy to include the dipole-driven tail entering already at 3PN order.
\end{enumerate}
Note that in practice, one must truncate $Q$. We choose to do so at $\mathcal O(\hat P_R^6)$, as otherwise it would diverge at infinity [cf. the $\hat R$-dependence in Eq.~\eqref{eq:Q}].
This will amount to including beyond-GR tails up to sixth order in the binary's orbital eccentricity in Sec.~\ref{sec:nonloc}.
For circular orbits such that $\hat P_R=0$, we have $Q=0$; in this simpler subcase than what is done here, $\hat H_{\rm eff}$ depends only on $A$, and the tails affect $a_4$ and $a_{4,\ln}$ only.

Finally, following Ref.~\cite{Damour:2015isa}, we find it  useful to split the potentials as
\begin{subequations}
\begin{align}
A&=A^{\rm I}+A^{\rm II}\,,\\
D&=D^{\rm I}+D^{\rm II}\,,\\
Q&=Q^{\rm I}+Q^{\rm II}\,,
\end{align}\label{eq:splitADQ}
\end{subequations}
where
\begin{subequations}
\begin{align}
A^{\rm I}&=1+\frac{a_1}{\hat R}+\frac{a_2}{\hat R^2}+\frac{a_3}{\hat R^3}+\frac{{ a_4^{\rm I}}+{ a_{4,\ln}^{\rm I}}\ln\hat R}{\hat R^4}\,,\\
D^{\rm I}&=1+\frac{  d_1}{\hat R}+\frac{  d_2}{\hat R^2}+\frac{{  d_3^{\rm I}}+{  d_{3,\ln}^{\rm I}}\ln\hat R}{\hat R^3}\,,\\
Q^{\rm I}&=\frac{({  q_{1}^{\rm I}}+{  q_{1,\ln}^{\rm I}}\ln\hat R)\,\hat P_R^4}{\hat R^2}+\frac{({  q_{2}^{\rm I}}+{  q_{2,\ln}^{\rm I}}\ln \hat R)\,\hat P_R^6}{\hat R}\,,
\end{align}\label{eq:localPotentials}%
\end{subequations}
and
 \begin{subequations}
\begin{align}
A^{\rm II}&=\frac{{ a_4^{\rm II}}+{ a_{4, \ln}^{\rm II}}\ln\hat R}{\hat R^4}\,,\\
D^{\rm II}&=\frac{{  d_3^{\rm II}}+{  d_{3,\ln}^{\rm II}}\ln\hat R}{\hat R^3}\,,\\
Q^{\rm II}&=\frac{({  q_{1}^{\rm II}}+{  q_{1,\ln}^{\rm II}}\ln\hat R)\,\hat P_R^4}{\hat R^2}+\frac{({  q_{2}^{\rm II}}+{  q_{2,\ln}^{\rm II}}\ln \hat R)\,\hat P_R^6}{\hat R}\,.
\end{align}\label{eq:NONlocalPotentials}%
\end{subequations}
The EOB Hamiltonian can then be decomposed in a similar fashion as its two-body counterpart,
\begin{subequations}
\begin{align}
\hat H_{\rm EOB}=\hat H_{\rm EOB}^{\rm I}+\hat H_{\rm EOB}^{\rm II}+\mathcal O(\hat P^{10})\,,
\end{align}
\end{subequations}
where $\hat H_{\rm EOB}^{\rm I}$ is obtained by formally setting $A^{\rm II}$, $D^{\rm II}$ and $Q^{\rm II}$ to zero in Eq.~\eqref{eq:defHeob} while, at first order,
\begin{align}
\hat H_{\rm EOB}^{\rm II}=\frac{1}{2}\big(A^{\rm II}+D^{\rm II}\hat P_R^2+Q^{\rm II}\big)\,.\label{eq:defHeobII}%
\end{align}

 \vfill\null 

\subsection{Local-in-time contributions\label{sec:local}}

Let us first focus on the local-in-time part $\hat H^{\rm I}$ of the two-body Hamiltonian, cf. Eq.~\eqref{eq:decomposed2bodyHamiltonian1}.
We perform a canonical transformation such that $ H^{\rm I}$ is a scalar and its action only changes by a boundary term:
\begin{align}
\int \big(P_R dR &+P_\Phi d\Phi+dF\big)=\int \big(p_r dr+p_\phi d\phi\big)\,,\label{eq:canoicalGeneric}
\end{align}
and thus
\begin{align}
dF
&=p_r dr+p_\phi d\phi-(P_R dR+P_\Phi d\Phi)\,.
\end{align}
For practical reasons, we will rather use $G(r,\phi,P_R,P_\Phi)=F+(P_R R+P_\Phi \Phi)-(P_R\, r +P_\Phi \phi)$ such that
\begin{align}
 dG&=dr(p_r-P_R)+d\phi (p_\phi-P_\Phi)\nonumber\\
 &+dP_R (R-r)+dP_\Phi (\Phi-\phi)\,,
\end{align}
which generates a canonical transformation introduced in Refs.~\cite{Julie:2017pkb,Julie:2017ucp},
\begin{subequations}
\begin{align}
\hat R(\hat r,\phi,\hat P_R,\hat P_\Phi)&=\hat r+\frac{\partial \hat G}{\partial \hat P_R}\,,\\
\Phi(\hat r,\phi,\hat P_R,\hat P_\Phi)&=\phi+\frac{\partial \hat G}{\partial \hat P_\Phi}\,,\\
\hat p_r(\hat r,\phi,\hat P_R,\hat P_\Phi)&=\hat P_R+\frac{\partial \hat G}{\partial \hat r}\,,\\
\hat p_\phi(\hat r,\phi,\hat P_R,\hat P_\Phi)&=\hat P_\Phi+\frac{\partial \hat G}{\partial\phi}\,.
\end{align}\label{eq:canoTransfos}%
\end{subequations}
We choose the ansatz:
\begin{align}
\hat G=\frac{G}{M\mu}=\hat r\hat P_R \sum_{i,j,k} \left(\gamma_{ijk}+\gamma_{ijk}^{\ln}\ln \hat r\right)\frac{\mathcal P^{2i}\hat P_R^{2j}}{\hat r^k}\label{eq:generatingFunctionCoefficients}%
\end{align}
with 
\begin{align}
\mathcal P^2=\hat P_R^2+\frac{\hat P_\Phi^2}{\hat r^2}\,,
\end{align}
which yields coordinate changes between 1PN and 3PN levels when the positive integers $i$, $j$ and $k$ satisfy $1\leqslant i+j+k \leqslant 3$.
Our ansatz does not depend on $\phi$ to preserve isotropy, and thus $p_\phi=P_\Phi$.
Moreover, for circular orbits such that $p_r=P_R=0$, we have $\Phi=\phi$.

From Eqs.~\eqref{eq:canoTransfos} we can express both $\hat H^{\rm I}$ and $\hat H_{\rm EOB}^{\rm I}$ in the same mixed coordinate system $(r,\phi,P_R,P_\Phi)$. We then solve, order-by-order, the equation
\begin{align}
\hat H^{\rm I}(\hat r,\phi,\hat P_R,\hat P_\Phi)=\hat H_{\rm EOB}^{\rm I}(\hat r,\phi,\hat P_R,\hat P_\Phi)
\end{align}
to fix the coefficients of the potentials \eqref{eq:localPotentials} and of the generating function \eqref{eq:generatingFunctionCoefficients}.
The solution is unique, and the new 3PN coefficients are:
\begin{widetext}
\begin{subequations}
\begin{align}
a_4^{\rm I}&=4 G_{AB}^4 k_{\rm tail} \ln \hat s+2G_{AB}^4 k_{\rm ESGB}
+\frac{G_{AB}^4 \nu  \bar{\gamma }_{AB} \big(11 \left(\bar{\gamma }_{AB}+2\right)^2-2 \delta _+-2 \delta _- m_-\big)}{4 \tilde\alpha  \left(\bar{\gamma }_{AB}+2\right)}\nonumber\\
&+\frac{G_{AB}^4}{12}\Big[8 \delta _+ \bar{\gamma }_{AB}-6 \epsilon _+ \bar{\gamma }_{AB}-60 \bar{\gamma }_{AB}^3-78 \bar{\gamma }_{AB}^2-24 \bar{\gamma }_{AB}-\bar{\beta }_+ \left(-47 \bar{\gamma }_{AB}^2+28 \bar{\gamma }_{AB}+2 \delta _++2 \delta _- m_-+28\right)\nonumber\\
&+m_- \bar{\beta }_- \left(-47 \bar{\gamma }_{AB}^2+28 \bar{\gamma }_{AB}-36 \bar{\beta }_++2 \delta _++28\right)+8 \delta _- m_- \bar{\gamma }_{AB}+6 m_- \epsilon _- \bar{\gamma }_{AB}+18 \bar{\beta }_+^2+\delta _- \left(6 m_-^2-4\right) \bar{\beta }_-\nonumber\\
&+3 \left(5 m_-^2+1\right) \bar{\beta }_-^2+4 \delta _++4 \kappa _++4 \delta _- m_--4 \kappa _- m_-\Big]\nonumber\\
&+\frac{G_{AB}^4\nu}{1152}\Big[-192 \bar{\beta }_+ \left(2 \bar{\gamma }_{AB}^2+92 \bar{\gamma }_{AB}+4 \delta _++5\right)+126 \pi ^2 \delta _+ \bar{\gamma }_{AB}+1056 \delta _+ \bar{\gamma }_{AB}-6912 \zeta _{AB} \bar{\gamma }_{AB}+1152 \epsilon _+ \bar{\gamma }_{AB}+63 \pi ^2 \bar{\gamma }_{AB}^3\nonumber\\
&-432 \bar{\gamma }_{AB}^3-180 \pi ^2 \bar{\gamma }_{AB}^2+15296 \bar{\gamma }_{AB}^2-1350 \pi ^2 \bar{\gamma }_{AB}+37184 \bar{\gamma }_{AB}+96 \bar{\beta }_- \left(m_- \left(4 \bar{\gamma }_{AB}^2+64 \bar{\gamma }_{AB}-36 \bar{\beta }_++8 \delta _++19\right)+16 \delta _-\right)\nonumber\\
&+288 \delta _- m_- \bar{\gamma }_{AB}+2880 \bar{\beta }_+^2+288 \left(m_-^2-3\right) \bar{\beta }_-^2+3456 \zeta _{AB}+252 \pi ^2 \delta _+-5888 \delta _+-1152 \kappa _++192 \delta _- m_-+384 \kappa _- m_-\nonumber\\
&+576 m_- \xi _-+576 m_- w_-+768 m_- \psi _--576 m_- \epsilon _-+576 \xi _++576 w_+-1536 \psi _++576 \epsilon _+-1476 \pi ^2+36096\Big]\,,
\label{eq:a4I}\\
d_3^{\rm I}&=\frac{G_{AB}^3}{12}\Big[-12 \bar{\beta }_+ \left(3 \bar{\gamma }_{AB}+8\right)+2 \delta _+ \left(3 \bar{\gamma }_{AB}+8\right)-9 \bar{\gamma }_{AB}^3-52 \bar{\gamma }_{AB}^2-64 \bar{\gamma }_{AB}+36 m_- \bar{\beta }_- \bar{\gamma }_{AB}+6 \delta _- m_- \bar{\gamma }_{AB}\nonumber\\
&+96 m_- \bar{\beta }_-+16 \delta _- m_--8 m_- \epsilon _-+8 \epsilon _+\Big]
-\frac{G_{AB}^3\nu}{12}\Big[12 \bar{\beta }_+ \left(5 \bar{\gamma }_{AB}-9\right)+8 \delta _+ \left(3 \bar{\gamma }_{AB}+7\right)-36 \bar{\gamma }_{AB}^3-308 \bar{\gamma }_{AB}^2\nonumber\\
&-692 \bar{\gamma }_{AB}-72 m_- \bar{\beta }_- \bar{\gamma }_{AB}-99 m_- \bar{\beta }_--108 \zeta _{AB}+12 \delta _- m_-+6 m_- \epsilon _-+10 \epsilon _+-624\Big]\nonumber\\
&+G_{AB}^3\nu^2\Big[-\bar{\gamma }_{AB}^2-10 \bar{\gamma }_{AB}+9 \bar{\beta }_++6 \zeta _{AB}-2 \delta _+-\epsilon _+-6\Big]\,,\\
q_{1}^{\rm I}&=\frac{G_{AB}^2\nu}{6}\Big[15 \bar{\gamma }_{AB}^2+52 \bar{\gamma }_{AB}+2 \bar{\beta }_+-2 m_- \bar{\beta }_--2 \delta _+-2 \delta _- m_-+48\Big]+G_{AB}^2\nu^2\Big[-4 \bar{\gamma }_{AB}+\bar{\beta }_+-m_- \bar{\beta }_--6\Big]\,,\\
q_{2}^{\rm I}&=0\,,
\end{align}
\end{subequations}
\end{widetext}
with logarithmic counterparts
\begin{subequations}
\begin{align}
a_{4,\ln}^{\rm I}&=-4G_{AB}^4 k_{\rm tail}\,,\\
d_{3,\ln}^{\rm I}&=0\,,\\
q_{1,\ln}^{\rm I}&=0\,,\\
q_{2,\ln}^{\rm I}&=0\,.
\end{align}
\end{subequations}
The coefficients of the canonical transformation \eqref{eq:generatingFunctionCoefficients} are given in Appendix~\ref{app:generatingFunctionCoeffs} for completeness.

We explicitly checked that in the GR limit, $\hat H^{\rm I}$ can also be identified to the 3PN (ADM) Hamiltonian of Ref.~\cite{Damour:2014jta} via a canonical transformation whose coefficients are given in Appendix~\ref{app:generatingFunctionCoeffs}.

\subsection{Nonlocal-in-time contributions\label{sec:nonloc}}

Let us now turn to the nonlocal-in-time 3PN Hamiltonian $\hat H^{\rm II}$, which (we recall) reads
\begin{align}
    \hat H^{\rm II}&=\frac{G_{AB}^4k_{\rm tail}}{\hat r^2}\,\underset{2\hat s}{\rm PF}\!\int_{\Bbb R}\frac{d\hat\tau}{|\hat\tau|}\frac{\cos\Delta\phi}{\hat r^2(\hat t+\hat \tau)}\,,\label{eq:decomposed2bodyHamiltonian2Recall}
\end{align}
with $\Delta\phi=\phi(\hat t+\hat\tau)-\phi(\hat t)$.
We wish to identify it, modulo canonical transformations, to a local-in-time, ordinary EOB counterpart $\hat H^{\rm II}_{\rm EOB}$ depending on positions and momenta only.
To do so, we can Taylor expand $\hat r(\hat t+\hat \tau)$ and $\phi(\hat t+\hat\tau)$ around $\hat\tau=0$, and treat $\hat H^{\rm II}$ as a local-in-time function of $\hat r(\hat t)$ and $\phi(\hat t)$, and their arbitrarily high-order time derivatives.
The Newtonian equations of motion can then be used to order reduce these derivatives, as we now prove.

Consider a pair of phase-space variables $q(t)$ and $p(t)$ described by the action $I=\int dt (p\dot q-H)$ with
\begin{align}
H&=H_0(q,p)+\epsilon^3  \Delta H(q,\dot q,\ddot q,\cdots;p,\dot p,\ddot p,\cdots)\nonumber\\
&+\mathcal O(\epsilon^4) \,,\label{eq:toyModelHamiltonHamiltonian}
\end{align}
where $\epsilon\ll 1$, and where $ \Delta H$ depends on arbitrarily high-order time derivatives of $q$ and $p$.
The Euler-Lagrange variations of 
\begin{align}
    L_0=p\dot q-H_0(q,p)
\end{align}
 with respect to $p$ and $q$ yield, respectively,
\begin{subequations}
\begin{align}
\dot q&=\dot q_0+\frac{\delta L_0}{\delta p}\,,\label{eq:toyModelHamiltonVariationP}\\
\dot p&=\dot p_0-\frac{\delta L_0}{\delta q}\,,
\end{align}\label{eq:toyModelHamiltonVariation}%
\end{subequations}
where we have introduced the notation
\begin{subequations}
\begin{align}
\dot q_0(q,p)&=\frac{\partial H_0}{\partial p}\,,\\
\dot p_0(q,p)&=-\frac{\partial H_0}{\partial q}\,.
\end{align}
\end{subequations}
As usual, the system \eqref{eq:toyModelHamiltonVariation} reduces to the Hamilton equations $\dot q=\dot q_0(q,p)+\mathcal O(\epsilon^3)$ and $\dot p=\dot p_0(q,p)+\mathcal O(\epsilon^3)$ when $\delta L_0/\delta p=\delta L_0/\delta q=0$.

We can then insert \eqref{eq:toyModelHamiltonVariation} into \eqref{eq:toyModelHamiltonHamiltonian} and expand at first order only in $\delta L_0/\delta p$ and $\delta L_0/\delta q$ (and their time derivatives), since higher-order contributions are doubly zero. We then find, modulo total time derivatives:
\begin{align}
H&=H_0+\epsilon^3\Delta H\left(q,\dot q_0,\left(\dot q_0\right)^{\boldsymbol{\cdot}},\cdots;p,\dot p_0,\left(\dot p_0\right)^{\boldsymbol{\cdot}},\cdots\right)\nonumber\\
&+\epsilon^3 \frac{\delta L_0}{\delta p}\frac{\delta \Delta H}{\delta\dot q}-\epsilon^3 \frac{\delta L_0}{\delta q}\frac{\delta \Delta H}{\delta\dot p}+\mathcal O(\epsilon^4)\,,\label{eq:toyModelHamiltonianIntermediaire}
\end{align}
where we have defined the Euler-Lagrange variations of $\Delta H$ with respect to $\dot q$ and $\dot p$:
\begin{subequations}
\begin{align}
\frac{\delta \Delta H}{\delta\dot q}&=\frac{\partial \Delta H}{\partial\dot q}-\left(\frac{\partial\Delta H}{\partial \ddot q}\right)^{\!\!\boldsymbol{\cdot}}+\cdots\\
\frac{\delta \Delta H}{\delta\dot p}&=\frac{\partial \Delta H}{\partial\dot p}-\left(\frac{\partial\Delta H}{\partial \ddot p}\right)^{\!\!\boldsymbol{\cdot}}+\cdots
\end{align}
\end{subequations}
Now we denote the order-reduced (over) accelerations, obtained recursively from the Hamilton equations of $H_0(q,p)$, by
\begin{align}
\ddot q_0(q,p)&=\frac{\partial \dot q_0}{\partial q}\dot q_0+\frac{\partial \dot q_0}{\partial p}\dot p_0\,,\nonumber\\
\dddot q_{\!0}(q,p)&=\frac{\partial \ddot q_0}{\partial q}\dot q_0+\frac{\partial \ddot q_0}{\partial p}\dot p_0\,,\nonumber\\
&\cdots\nonumber\\
q_0^{(n+1)}(q,p)&=\frac{\partial q_0^{(n)}}{\partial q}\dot q_0+\frac{\partial q_0^{(n)}}{\partial p}\dot p_0\,,
\end{align}
with $n\geqslant 1$ and, similarly,
\begin{align}
p_0^{(n+1)}(q,p)=\frac{\partial p_0^{(n)}}{\partial q}\dot q_0+\frac{\partial p_0^{(n)}}{\partial p}\dot p_0\,.
\end{align}
We then have, using Eqs.~\eqref{eq:toyModelHamiltonVariation} again,
\begin{subequations}
\begin{align}
(\dot q_0)^{(n)}&=q_0^{(n+1)}(q,p)\\
&+\sum_{i=0}^{n-1}\left(\frac{\partial q_0^{(n-i)}}{\partial q}\frac{\delta L_0}{\delta p}-\frac{\partial q_0^{(n-i)}}{\partial p}\frac{\delta L_0}{\delta q}\right)^{\!\!(i)} ,\nonumber\\
(\dot p_0)^{(n)}&=p_0^{(n+1)}(q,p)\\
&+\sum_{i=0}^{n-1}\left(\frac{\partial p_0^{(n-i)}}{\partial q}\frac{\delta L_0}{\delta p}-\frac{\partial p_0^{(n-i)}}{\partial p}\frac{\delta L_0}{\delta q}\right)^{\!\!(i)} ,\nonumber
\end{align}
\end{subequations}
which we can plug into the second term in the right-hand side of Eq.~\eqref{eq:toyModelHamiltonianIntermediaire}. Expanding the result at first order in $\delta L_0/\delta p$ and $\delta L_0/\delta q$ (and their time derivatives) and integrating by parts finally yields:
\begin{align}
H&=H_{\rm red}\nonumber\\
&+\epsilon^3\frac{\delta L_0}{\delta p}\bigg(\frac{\delta\Delta H}{\delta\dot q}+\sum_{n=1}^{\infty} \frac{\delta \Delta H}{\delta q^{(n+1)}}\frac{\partial q_0^{(n)}}{\partial q}\nonumber\\
&\qquad\qquad\qquad\quad +\sum_{n=1}^{\infty} \frac{\delta \Delta H}{\delta p^{(n+1)}}\frac{\partial p_0^{(n)}}{\partial q}\bigg)_{\!\rm red}\nonumber\\
&-\epsilon^3\frac{\delta L_0}{\delta q}\bigg(\frac{\delta\Delta H}{\delta\dot p}+ \sum_{n=1}^\infty \frac{\delta\Delta H}{\delta q^{(n+1)}}\frac{\partial q_0^{(n)}}{\partial p}\nonumber\\
&\qquad\qquad\qquad\quad+ \sum_{n=1}^\infty \frac{\delta\Delta H}{\delta p^{(n+1)}}\frac{\partial p_0^{(n)}}{\partial p}\bigg)_{\!\rm red}\nonumber\\
&+\mathcal O(\epsilon^4)\,,\label{eq:ToyModelHamiltonFinal}
\end{align}
modulo doubly zero terms and total time derivatives. The subscript ``$\rm red$'' indicates an order-reduced quantity, as in
\begin{align}
H_{\rm red}(q,p)&=H_0(q,p)\\
&+\epsilon^3\Delta H\big(q,\dot q_0(q,p),\ddot q_0(q,p),\cdots;\nonumber\\
&\qquad\qquad\, p,\dot p_0(q,p),\ddot p_0(q,p),\cdots\big)\,.\nonumber
\end{align}

It is now elementary to eliminate the second to fifth lines of Eq.~\eqref{eq:ToyModelHamiltonFinal}: under a phase-space contact transformation $(q,p)\to (q+\delta q,p+\delta p)$ with $\delta q=\mathcal O(\epsilon^3)$ and $\delta p=\mathcal O(\epsilon^3)$, the Hamiltonian transforms as
\begin{align}
(H-p\dot q)&\to (H-p\dot q)\nonumber\\
&-\frac{\delta L_0}{\delta p}\delta p-\frac{\delta L_0}{\delta q}\delta q+\mathcal O(\epsilon^6)
\end{align}
modulo an irrelevant total time derivative, and we can choose to identify $\delta p(q,p)$ and $\delta q(q,p)$ with the long coefficients of $\delta L_0/\delta p$ and $\delta L_0/\delta q$ in Eq.~\eqref{eq:ToyModelHamiltonFinal},  respectively.
This toy model is an adaptation of Refs.~\cite{Schafer:1984mr,Damour:1990jh, Damour:1999cr}, which we have extended to Hamiltonians depending on arbitrarily high-order time derivatives of $q$ and $p$ for our purpose.

Indeed, return now to the nonlocal-in-time 3PN Hamiltonian $\hat H^{\rm II}$. 
We shall see that there exists a set of phase-space variables other than polar in which the steps above are elegantly carried out.
This is the route of Refs.~\cite{Damour:2015isa,Damour:2016abl}, which we adapt here to the ST-ESGB case.

In polar coordinates $(r,\phi, p_r, p_\phi)$, the Keplerian trajectory can be parametrized by the semimajor axis $\hat a=a/M$ and the eccentricity $e$ as \cite{Deruelle:2018ltn}
\begin{subequations}
\begin{align}
\hat r&=\hat a(1-e\cos\eta)\,,\label{eq:KeplerianOrbits1}\\
\tan\frac{\phi}{2}&=\sqrt{\frac{1+e}{1-e}}\tan\frac{\eta}{2}\,.
\end{align}\label{eq:KeplerianOrbits}%
\end{subequations}
We set $\hat t=0$ at the periastron without loss of generality, and define the eccentric anomaly $\eta$ as
\begin{equation}
\hat\Omega\, \hat t=\eta-e\sin\eta\,,\label{eq:defEccentricAnomaly}
\end{equation}
where
\begin{align}
\hat\Omega(\hat a)&=M\Omega=\sqrt{\frac{G_{AB}}{\hat a^3}}
\end{align}
is the mean orbital frequency.

Now, we observe that $\hat a$ and $e$ can be treated as functions of the Delaunay action-angles $(\mathcal L,\mathcal G,l,g)$.\footnote{
For completeness: the Delaunay action variables are defined as~\cite{Morbidelli}
\begin{align*}
\mathcal L&=\frac{1}{2\pi}\left(\oint\hat p_r(E,J,\hat r) d\hat r+\oint \hat p_\phi(J) d\phi\right)\,,\\
\mathcal G&=\frac{1}{2\pi}\oint \hat p_\phi(J) d\phi\,,
\end{align*}
which are calculated on an orbital cycle with constant $\hat p_\phi=J$ and $\hat H_0=E<0$.
The conjugate angles are then defined as
\begin{align*}
l=\frac{\partial \hat S}{\partial \mathcal L}\,,\quad g=\frac{\partial \hat S}{\partial \mathcal G}\,,\quad \text{with}\ \hat S=\int \hat p_r(\mathcal L,\mathcal G,\hat r) d\hat r+\hat p_\phi(\mathcal G) d\phi\,.
\end{align*}
}
Indeed, it is a textbook exercise to show that, on Keplerian orbits~\cite{Morbidelli},
\begin{subequations}
\begin{align}
\mathcal L&=\sqrt{G_{AB}\hat a}\,,\\
\mathcal G&=\sqrt{G_{AB}\hat a(1-e^2)}\,,
\end{align}
\end{subequations}
which can be inverted as
\begin{subequations}
\begin{align}
\hat a(\mathcal L)&=\frac{\mathcal L^2}{G_{AB}}\,,\\
e(\mathcal L,\mathcal G)&=\sqrt{1-\frac{\mathcal G^2}{\mathcal L^2}}\,,
\end{align}\label{eq:aANDeGL}%
\end{subequations}
while the conjugate angles are the mean anomaly and argument of the periastron, respectively:
\begin{subequations}
\begin{align}
l&=\hat \Omega\, \hat t\,,\\
g&=\omega\,.
\end{align}
\end{subequations}
In these canonical variables, the 0PN equations of motion are particularly simple: $\hat H_0$ is the Delaunay Hamiltonian,
\begin{align}
\hat H_0(\mathcal L)&=-\frac{G_{AB}^2}{2\mathcal L^2}\,,\label{eq:DelaunayHamiltonianDef}
\end{align}
and thus
\begin{subequations}
\begin{align}
\frac{dl}{d\hat t}&=\frac{\partial \hat H_0}{\partial\mathcal L}=\hat\Omega(\mathcal L)\,,\\
\frac{dg}{d\hat t}&=\frac{\partial \hat H_0}{\partial\mathcal G}=0\,,\\
\frac{d\mathcal L}{d\hat t}&=-\frac{\partial \hat H_0}{\partial l}=0\,,\\
\frac{d\mathcal G}{d\hat t}&=-\frac{\partial \hat H_0}{\partial g}=0\,.
\end{align}\label{eq:DelaunayEOMs}%
\end{subequations}

We can thus consider $\hat H^{\rm II}$, recalled in Eq.~\eqref{eq:decomposed2bodyHamiltonian2Recall}, as a nonlocal-in-time function $\hat H^{\rm II}[\mathcal L,\mathcal G,l]$ of the Delaunay variables, and use the relations above to turn it into a local-in-time, ordinary Hamiltonian as follows: when $e\ll 1$, invert Eq.~\eqref{eq:defEccentricAnomaly} iteratively as (recall that $l=\hat\Omega\,\hat t$)
\begin{align}
\eta&=l
+e \sin l
+\frac{1}{2}e^2  \sin 2 l+\frac{1}{8} e^3 (3 \sin 3 l- \sin l)\nonumber\\
&+\frac{1}{6} e^4 ( 2\sin 4 l- \sin 2l)\nonumber\\
&+\frac{1}{384} e^5 (2 \sin l-81  \sin 3l+125 \sin 5 l)\nonumber\\
&+\frac{1}{240} e^6 (5 \sin 2 l-64  \sin 4l+81  \sin 6l)\nonumber\\
&+\mathcal O(e^7)\,,\label{eq:etaP6}
\end{align}
and insert it into Eqs.~\eqref{eq:KeplerianOrbits} to deduce $\hat r(\hat a(\mathcal L),e(\mathcal L,\mathcal G),l)$ and $\phi(e(\mathcal L,\mathcal G),l)$, which enter $\hat H^{\rm II}$.
To evaluate them at $\hat t+\hat\tau$, Taylor expand $\mathcal L$, $\mathcal G$ and $l$ around $\hat t$, and order reduce their arbitrarily high-order time derivatives at time $\hat t$ using the 0PN equations of motion~\eqref{eq:DelaunayEOMs}. We recall that this step is equivalent to an implicit phase-space contact transformation, as clarified by our toy model above.
The result is very simple (more so than in polar coordinates) since only the first time derivative of $l$, $d l/d\hat t=\hat \Omega(\mathcal L)$, is nonzero on shell:
\begin{align}
l(\hat t+\hat \tau)=l(\hat t)+\hat\Omega(\mathcal L)\hat\tau\,,
\end{align}
while $\mathcal L$ and $\mathcal G$ are constants.

The order-reduced Hamiltonian then has the structure:
 \begin{align}
\hat H_{\rm red}^{\rm II}&=\frac{G_{AB}^4k_{\rm tail}}{\hat a(\mathcal L)^4}\,\underset{2\hat s}{\rm PF}\!\int_{\Bbb R}\frac{d\hat\tau}{|\hat\tau|}\Big(\cos\hat\Omega\hat\tau+\sum_{i=1}^6 e(\mathcal L,\mathcal G)^i I_i\Big)\nonumber\\
&+\mathcal O(e^7)\,,\label{eq:HIIKepler}
\end{align}
where
\begin{align}
I_i=\sum_{m,n}\big(a_{imn}\,{\cos}(m l)\,{\cos}(n\hat\Omega\hat\tau)&\nonumber\\
+b_{imn}\,{\sin}(m l)\,{\sin}(n\hat\Omega\hat\tau)&\big)\,.\label{eq:Ii}
\end{align}
Here $m$ and $n$ are non-negative integers, and the constants $a_{imn}$ and $b_{imn}$ are rational numbers.
The Hadamard partie finie can be computed via~\cite{Damour:2015isa}
\begin{subequations}
\begin{align}
\underset{2\hat s}{\rm PF}\!\int_{\Bbb R}\frac{d\hat\tau}{|\hat\tau|}{\cos}(n\hat\Omega\hat\tau)&=-2[\gamma_{\rm E}+{\ln}(2n\hat\Omega\hat s)]\,,\\
\underset{2\hat s}{\rm PF}\!\int_{\Bbb R}\frac{d\hat\tau}{|\hat\tau|}{\sin}(n\hat\Omega\hat\tau)&=0\,,\label{PFsin}
\end{align}\label{PF}%
\end{subequations}
where $\gamma_{\rm E}$ is Euler's constant. Equation~\eqref{PFsin} follows from the symmetry of the integrand, and it implies that the second line of $I_i$ can be discarded. The result is a local-in-time, ordinary Hamiltonian:
\vfill\eject
\vspace*{-0.8cm}
\begin{align}
\hat H_{\rm red}^{\rm II}(\mathcal L,\mathcal G,l)&=-\frac{2G_{AB}^4k_{\rm tail}}{\hat a^4(\mathcal L)}\left[\gamma_{\rm E}+\ln(2\hat\Omega(\mathcal L)\hat s)\right]\nonumber\\
&+\sum_{i=1}^6 \sum_{m,n}e(\mathcal L,\mathcal G)^i A_{imn}(\mathcal L){\cos} (m l)\,,\label{eq:H2reducedCleanedUp}
\end{align}
where
\begin{align}
A_{imn}(\mathcal L)=-\frac{2G_{AB}^4k_{\rm tail}\, a_{imn}}{\hat a(\mathcal L)^4} \left[\gamma_{\rm E}+\ln(2n\hat\Omega(\mathcal L)\hat s)\right]\,.
\end{align}

Another purpose of the Delaunay action-angles is that
$\hat H^{\rm II}_{\rm red}$ is a perturbation of the 0PN Hamiltonian \eqref{eq:DelaunayHamiltonianDef}, which depends only on $\mathcal L$ in these variables.
That way, the $l$ dependence in $\hat H_{\mathcal D}=\hat H_0(\mathcal L)+\hat H_{\rm red}^{\rm II}(\mathcal L,\mathcal G,l)$ can be eliminated through a canonical transformation resembling Eq.~\eqref{eq:canoTransfos},
\begin{subequations}
\begin{align}
l'(l,g,\mathcal L',\mathcal G')&=l+\frac{\partial \hat G}{\partial \mathcal L'}\,,\\
g'(l,g,\mathcal L',\mathcal G')&=g+\frac{\partial \hat G}{\partial \mathcal G'}\,,\\
\mathcal L(l,g,\mathcal L',\mathcal G')&=\mathcal L'+\frac{\partial \hat G}{\partial l}\,,\label{eq:canoTransfos3}\\
\mathcal G(l,g,\mathcal L',\mathcal G')&=\mathcal G'+\frac{\partial \hat G}{\partial g}\,.
\end{align}
\end{subequations}
Indeed, the ansatz (for $m\neq 0$)
\begin{align}
\hat G=-\sum_{i=1}^6 \sum_{m,n}e(\mathcal L',\mathcal G')^i A_{imn}(\mathcal L')\frac{\sin(m l)}{m\hat \Omega(\mathcal L')}
\end{align}
\vfill\null
\noindent
yields a 3PN coordinate change, such that [only \eqref{eq:canoTransfos3} matters]
\begin{align}
\hat H_{\mathcal D}(\mathcal L,\mathcal G, l)&=\hat H_0(\mathcal L')+\hat H^{\rm II}_{\rm red}(\mathcal L',\mathcal G',l)\nonumber\\
&+\frac{\partial \hat H_0}{\partial \mathcal L}\frac{\partial \hat G}{\partial l}+\mathcal O(\hat p^{10})\,.
\end{align}
Since $\partial \hat H_0/\partial \mathcal L=\hat\Omega$, the second line eliminates all $m\neq 0$ terms in $\hat H^{\rm II}_{\rm red}(\mathcal L',\mathcal G',l)$, cf. Eq.~\eqref{eq:H2reducedCleanedUp}.
We thus discard them in practice, and only keep the part $m=0$. The final canonically transformed Hamiltonian reads (dropping the primes for simplicity)
\begin{align}
\hat {H}_{\rm red}^{\rm II}& =\frac{G_{A B}^4k_{\rm tail}}{\hat a^4}\Big(
3 \ln \hat{a}-\ln G_{A B}-2 \left(\gamma _{\rm E}+\ln \left(2 \hat{s}\right)\right)\nonumber\\
&\qquad-e^2 \left(-9 \ln \hat{a}+3 \ln G_{A B}+6 \gamma _{\rm E}+6 \ln \hat{s}+14 \ln 2\right)\nonumber\\
&\qquad-\frac{3}{32} e^4 \left(-180 \ln \hat{a}+60 \ln G_{A B}+120 \gamma _{\rm E}+120 \ln\hat{s}\right.\nonumber\\
&\qquad\qquad\left.-251 \ln 2+243 \ln 6\right)\nonumber\\
&\qquad-\frac{1}{16} e^6 \left(-420 \ln \hat{a}+140 \ln G_{A B}+280 \gamma _{\rm E}\right.\nonumber\\
&\left.\qquad\qquad+280 \ln\hat{s}+2929 \ln 2-729 \ln 6\right)\nonumber\\
&\qquad+\mathcal O(e^8)
\Big)\,,\label{eq:HeobIIRealDelaunayPostCano}%
\end{align}
where we recall that $\hat a$ and $e$ are the functions \eqref{eq:aANDeGL} of $\mathcal L$ and $\mathcal G$.
 
The same steps can be applied to $\hat H_{\rm EOB}^{\rm II}$ of Eq.~\eqref{eq:defHeobII}. It is already local-in-time and ordinary, but we rewrite it in terms of $\hat a$ and $e$ using Eq.~\eqref{eq:KeplerianOrbits1} with $\hat r\to\hat R$, and $\hat P_R=d\hat R/d\hat t$.\footnote{At leading order in the eccentricity, $\hat P_R=\mathcal O(\epsilon)$. This means that $\hat H_{\rm EOB}^{\rm II}$, which we truncated at $\mathcal O(\hat P_R^6)$ [cf. below Eq.~\eqref{eq:QinfiniteSeries}], can be identified to the two-body Hamiltonian modulo canonical transformations up to $\mathcal O(\epsilon^6)$.}
We then use Eq.~\eqref{eq:etaP6} and invoke canonical transformations to discard $l$-dependent terms with the same form as in Eq.~\eqref{eq:H2reducedCleanedUp}. We find:
\begin{widetext}
\begin{align}
{\hat H}_{\rm EOB}^{\rm II}&=\frac{1}{\hat a^4}\Big(
\frac{1}{2} \left(a_{4,\ln}^{\rm II} \ln\hat a+a_{4}^{\rm II}\right)
+\frac{1}{8} e^2 \left(2 \ln\hat a \left(d_{3,\ln}^{\rm II} G_{A B}+6 a_{4,\ln}^{\rm II}\right)+2 d_3^{\rm II} G_{A B}+12 a_{4}^{\rm II}-7 a_{4,\ln}^{\rm II}\right)\nonumber\\
&\qquad+\frac{1}{64} e^4 \left(4 \ln\hat a \left( 3 q_{1,\ln}^{\rm II} G^2_{A B}+10 d_{3,\ln}^{\rm II}G_{A B}+45 a_{4,\ln}^{\rm II}\right)+12 q_1^{\rm II} G_{A B}^2+(40 d_3^{\rm II} -18 d_{3,\ln}^{\rm II})G_{A B}+180 a_{4}^{\rm II}-171 a_{4,\ln}^{\rm II}\right)\nonumber\\
&\qquad+\frac{1}{384} e^6 \left(60 \ln\hat a \left( q_{2,\ln}^{\rm II} G_{A B}^3+3 q_{1,\ln}^{\rm II} G_{A B}^2+7 d_{3,\ln}^{\rm II}G_{A B}+28 a_{4,\ln}^{\rm II}\right)+60 q_2^{\rm II} G_{A B}^3+(180 q_1^{\rm II} -66 q_{1,\ln}^{\rm II})G_{A B}^2 \right.\nonumber\\
&\qquad\left.+(420 d_3^{\rm II} -319 d_{3,\ln}^{\rm II} )G_{A B}+1680 a_{4}^{\rm II}-2046 a_{4,\ln}^{\rm II}\right)+\mathcal O(e^8)\Big)\,.\label{eq:HeobIIEffectDelaunayPostCano}
\end{align}%
\end{widetext}
\vspace*{-0.7cm}
Now assume that $\hat a$ and $e$ in Eqs.~\eqref{eq:HeobIIRealDelaunayPostCano} and \eqref{eq:HeobIIEffectDelaunayPostCano} are functions of the same action variables $(\mathcal L,\mathcal G)$. The identification $\hat {H}_{\rm red}^{\rm II}={\hat H}_{\rm EOB}^{\rm II}$ term-by-term then yields the unique solution:
\vfill\null
\vspace*{-0.6cm}
\begin{subequations}
\begin{align}
a_4^{\rm II}&=-4G_{AB}^4 k_{\rm tail}\big(\gamma_E+\frac{1}{2}\ln G_{AB}+\ln(2\hat s)\big)\,,\label{eq:a4II}\\
d_3^{\rm II}&=G_{AB}^3 k_{\rm tail}(21-32\ln 2)\,,\\
q_1^{\rm II}&=\frac{G_{AB}^2 k_{\rm tail}}{6} (93+1753 \ln 2-729 \ln 6)\,,\\
q_2^{\rm II}&=\frac{3G_{AB} k_{\rm tail}}{10} (37-5707 \ln 2+2187 \ln 6)\,,
\end{align}
\end{subequations}
with logarithmic counterparts
\begin{subequations}
\begin{align}
a_{4,\ln}^{\rm II}&=6G_{AB}^4 k_{\rm tail}\,,\\
d_{3,\ln}^{\rm II}&=0\,,\\
q_{1,\ln}^{\rm II}&=0\,,\\
q_{2,\ln}^{\rm II}&=0\,.
\end{align}
\end{subequations}
\\ \\
\subsection{Complete EOB potentials at 3PN order\label{subseq:completePotentials}}

In Eqs.~\eqref{eq:splitADQ} and below,
we split the EOB potentials into their parts $\rm I$ and $\rm II$, which we determined in Secs.~\ref{sec:local} and \ref{sec:nonloc}, respectively.
We recall that $\hat P_R=P_R/\mu$, and we introduce the notation
\begin{align}
u=\frac{G_{AB}M}{R}\,.\label{def:u}
\end{align}
Adding the results yields:
\begin{widetext}
\begin{subequations}
\begin{align}
A(u)&=1-2u+2(\langle\bar\beta\rangle-\bar\gamma_{AB})u^{2}+(2 \nu+\delta \bar a_{3})u^{3}+\big[\nu\big(94/3-41\pi^2/32\big)+\delta  \bar a_4+ \bar a_{4,\ln}\ln u\big]u^{4}\,, \label{eq:AeobPotentialFinal}\\
D(u)&=1-2\bar\gamma_{AB}\,u+(6\nu+\delta \bar d_{2})u^{2}+(52\nu-6\nu^2+\delta \bar d_3)u^{3}\,,\\
Q(u,\hat P_R)&=(8\nu-6\nu^2+\delta  \bar q_{1})\hat P_R^4u^{2}+ \bar q_{2}\hat P_R^6u\,,
\end{align}\label{eq:finalEOBpotentials}%
\end{subequations}
with
\begin{subequations}
\begin{align}
\delta\bar a_3&=\frac{1 }{12} \Big[\nu(-24 \zeta _{A B}-36\bar{\beta }_++4   \bar{\gamma }_{A B}^2+40 \bar{\gamma }_{A B}+8\delta _++4  \epsilon _+)-24\langle\bar\beta\rangle\left(1-2 \bar{\gamma }_{A B}\right)-35 \bar{\gamma }_{A B}^2\nonumber\\
&\qquad  -20 \bar{\gamma }_{A B}+4\langle\delta\rangle-4\langle\epsilon\rangle \Big]\,,\\
\delta\bar d_2&=\frac{1}{4} \Big[-3 \bar{\gamma }_{A B}^2-12 \bar{\gamma }_{A B}+4 \langle\delta \rangle-24 \langle\bar{\beta }\rangle+8 \nu(2 \bar{\gamma }_{A B} -\langle\bar{\beta }\rangle) \Big]\,,
\end{align}\label{eq:coeffsEOB2PN}%
\end{subequations}
and, at 3PN order,
\begin{subequations}
\begin{align}
\delta\bar a_4&=2 k_{\rm ESGB}-4k_{\rm tail} (\gamma _E+\ln 2)+\frac{\nu  \bar{\gamma }_{AB} \big(11 \left(\bar{\gamma }_{AB}+2\right)^2-4 \langle\delta\rangle\big)}{4 \tilde\alpha  \left(\bar{\gamma }_{AB}+2\right)}\nonumber\\
&+\frac{1}{12}\Big[2 \langle\bar\beta\rangle \left(47 \bar{\gamma }_{AB}^2-28 \bar{\gamma }_{AB}+6 \bar{\beta }_++4 \delta _+-12 \langle\delta\rangle-28\right)-60 \bar{\gamma }_{AB}^3-78 \bar{\gamma }_{AB}^2-24 \bar{\gamma }_{AB}+\langle\delta\rangle \left(16 \bar{\gamma }_{AB}+8\right)-12 \langle\epsilon\rangle \bar{\gamma }_{AB}\nonumber\\
&-4 \delta _- \bar{\beta }_--4 \delta _+ \bar{\beta }_++3 \bar{\beta }_-^2-3 \bar{\beta }_+^2+8 \langle\delta\rangle \bar{\beta }_++60 \langle\bar\beta\rangle^2+8 \langle\kappa\rangle\Big]\nonumber\\
&+\frac{\nu}{1152}\Big[-288 \bar{\beta }_+ \left(40 \bar{\gamma }_{AB}-3\right)+192 \langle\bar\beta\rangle \left(-4 \bar{\gamma }_{AB}^2-64 \bar{\gamma }_{AB}+30 \bar{\beta }_+-8 \delta _+-19\right)\nonumber\\
&+\delta _+ \left(126 \pi ^2 \bar{\gamma }_{AB}+768 \bar{\gamma }_{AB}+252 \pi ^2-6080\right)-6912 \zeta _{AB} \bar{\gamma }_{AB}+1152 \epsilon _+ \bar{\gamma }_{AB}+63 \pi ^2 \bar{\gamma }_{AB}^3-432 \bar{\gamma }_{AB}^3-180 \pi ^2 \bar{\gamma }_{AB}^2\nonumber\\
&+15296 \bar{\gamma }_{AB}^2-1350 \pi ^2 \bar{\gamma }_{AB}+37184 \bar{\gamma }_{AB}+192 \langle\delta\rangle\left(3 \bar{\gamma }_{AB}+2\right)+1536 \delta _- \bar{\beta }_--864 \bar{\beta }_-^2-288 \bar{\beta }_+^2+3456 \zeta _{AB}\nonumber\\
&+1152 \langle\bar\beta\rangle^2+1152 \langle\epsilon\rangle-768 \kappa _+-768 \langle\kappa\rangle-1536 \langle\psi\rangle+1152 \langle w\rangle-768 \psi _++1152 \langle\xi\rangle\Big]\,,\label{Eq:da4}\\
\delta \bar d_3&=k_{\rm tail}(21-32 \ln 2 )+\frac{1}{12} \Big[4 (\langle\delta\rangle-6 \langle\bar\beta\rangle) \left(3 \bar{\gamma }_{AB}+8\right)-9 \bar{\gamma }_{AB}^3-52 \bar{\gamma }_{AB}^2-64 \bar{\gamma }_{AB} +16 \langle\epsilon\rangle\Big]\nonumber\\
&+\frac{\nu}{12} \Big[4 \left(-\delta _+ \left(6 \bar{\gamma }_{AB}+11\right)+9 \bar{\gamma }_{AB}^3+77 \bar{\gamma }_{AB}^2+173 \bar{\gamma }_{AB}+27 \zeta _{AB}-6 \langle\delta\rangle+3 \langle\epsilon\rangle-4 \epsilon _+\right)\nonumber\\
&+3 \bar{\beta }_+ \left(4 \bar{\gamma }_{AB}+69\right)-18 \langle\bar\beta\rangle \left(8 \bar{\gamma }_{AB}+11\right)\Big]+\nu^2\Big[-\bar{\gamma }_{AB}^2-10 \bar{\gamma }_{AB}+9 \bar{\beta }_++6 \zeta _{AB}-2 \delta _+-\epsilon _+\Big]\,,\label{Eq:dd3}\\
\delta\bar q_1&=\frac{1}{6}k_{\rm tail}(93+1753 \ln 2-729 \ln 6)
+\frac{\nu}{6}\Big[\bar{\gamma }_{AB} \left(15 \bar{\gamma }_{AB}+52\right)+4( \langle\bar\beta\rangle- \langle\bar\delta\rangle)\Big]
+2 \nu ^2 \Big[\langle\bar\beta\rangle-2 \bar{\gamma }_{AB}\Big]\,,\label{Eq:dq1}\\
\bar q_2&=\frac{3}{10}k_{\rm tail}(37-5707 \ln 2+2187 \ln 6)\,,\label{Eq:q2}
\end{align}
with the logarithmic counterpart
\begin{align}
    \bar a_{4,\ln}&=-2k_{\rm tail}\,.
\end{align}\label{eq:coeffsEOB3PN}%
\end{subequations}
\end{widetext}
The coefficients depend on the mean values
\begin{subequations}
\begin{align}
\langle\bar\beta\rangle&=\frac{m_A^0\bar\beta_B+m_B^0 \bar\beta_A}{M}\,,\label{eq:bMean}\\
\langle\delta\rangle&=\frac{m_A^0\bar\delta_A+m_B^0 \delta_B}{M}\,,\\
\langle\epsilon\rangle&=\frac{m_A^0\epsilon_B+m_B^0 \epsilon_A}{M}\,,\label{eq:eMean}\\
\langle \omega\rangle&=\frac{m_A^0 \omega_A+m_B^0 \omega_B}{M}\,,\\
\langle\kappa\rangle&=\frac{m_A^0\kappa_B+m_B^0 \kappa_A}{M}\,,\\
\langle \xi\rangle&=\frac{m_A^0 \xi_A+m_B^0 \xi_B}{M}\,,\\
\langle\psi\rangle&=\frac{m_A^0\psi_B+m_B^0 \psi_A}{M}\,,
\end{align}
\end{subequations}
where Eqs.~\eqref{eq:bMean}-\eqref{eq:eMean} were already introduced in Ref.~\cite{Julie:2017pkb}, while the remaining quantities are new to this paper.  The results above are available online~\cite{FLJRepo}.

A few comments are in order.
The potentials \eqref{eq:finalEOBpotentials} are a beyond-GR extension
of the 3PN results of Damour, Jaranowski and Sch\"{a}fer~\cite{Damour:2000we}, which we recover in the GR limit detailed below Eqs.~\eqref{eq:combinations}. Indeed, in that limit $\langle\bar\beta\rangle$, $\bar\gamma_{AB}$ and the coefficients~\eqref{eq:coeffsEOB2PN} and \eqref{eq:coeffsEOB3PN} all vanish.
We observe that Newton's constant is now substituted by the effective gravitational coupling $G_{\rm AB}$ entering $u$ at all orders. Thus this effect can be absorbed in a redefinition of the total mass $M$.
When truncated to 2PN order, our potentials depend on five coefficients and reproduce those of Ref.~\cite{Julie:2017pkb} in ST theories. This can be checked using Eqs.~\eqref{eq:coeffsEOB2PN} and $B=(AD)^{-1}$.

The 3PN coefficients~\eqref{eq:coeffsEOB3PN} are the central new results of this paper. They show that among the eight coefficients in our ansatz at 3PN level [cf. Eq.~\eqref{eq:ADQ}], only five are nonzero.
The contribution from the nonlocal-in-time tail is driven by the constant $k_{\rm tail}$, which enters all coefficients in Eqs.~\eqref{eq:coeffsEOB3PN}.
This is necessary to include the non-GR tail beyond circular orbits, and at sixth order in the orbital eccentricity.
Note that the tail is fully responsible for the unique logarithmic correction $\bar a_{4,\ln}$ and the post-post-geodesic coefficient $\bar q_2$.
As for the ESGB corrections beyond ST, they are driven by $k_{\rm ESGB}$, and their inclusion is particularly simple: they only enter in $\delta \bar a_4$.

Contrary to the two-body Hamiltonian $\hat H$, the coefficients~\eqref{eq:coeffsEOB3PN} do not depend on $\ln_\pm=\ln(\hat r_A)\pm \ln(\hat r_B)$, where $\hat r_A$ and $\hat r_B$ are the regularization lengths mentioned in Sec.~\ref{sec:twoBodyH3PN}. As expected, they have been reabsorbed in a canonical transformation at 3PN level (see Appendix~\ref{app:generatingFunctionCoeffs}).

Finally, at the end of Sec.~\ref{sec:twoBodyH3PN} we split the two-body Hamiltonian into its local-in-time and nonlocal-in-time parts $\hat H^{\rm I}$ and $\hat H^{\rm II}$ by introducing an arbitrary constant $\hat s$ which propagated in both $A^{\rm I}$ and $A^{\rm II}$, cf. Eqs.~\eqref{eq:a4I} and \eqref{eq:a4II}. As expected, $\hat s$ cancels out from our final EOB potentials.

\section{The example of shift-symmetric ESGB gravity\label{sec:IV}}

\subsection{Hairy BH binaries\label{subsec:ESGBBBH}}

 The coefficients of the potentials \eqref{eq:finalEOBpotentials}, we recall,
are built out of 
the theory-dependent product $\ell^2 f'(\varphi_0)$ and of ten body-dependent parameters: the values of the masses $m_A(\varphi)$ and $m_B(\varphi)$ and their logarithmic derivatives \eqref{eq:sensis} evaluated at infinity (i.e., at $\varphi=\varphi_0$).

Now, these quantities can be calculated once the theory and the bodies are specified.
In ST theories, they were derived numerically for NSs and their scalarized counterparts
(e.g., see Refs.~\cite{Damour:1993hw,Zhao:2022vig} and references therein).
They were also calculated for BHs in ESGB models, both analytically in the small-$\ell$ limit, and numerically for nonperturbative solutions such as scalarized BHs
(cf. Refs.~\cite{Julie:2019sab,Julie:2022huo}).

Let us complete this paper with an explicit illustration. Consider a BH in the shift-symmetric theory $f(\varphi)=2\varphi$ and $\mathcal A(\varphi)=1$.
For simplicity, here we consider only terms at leading order in $\ell$, such that~\cite{Julie:2019sab}
\begin{align}
    \alpha_A^0&=-(\ell/m_A^0)^2+\mathcal O (\ell/m_A^0)^4 \,,\label{eq:alphaAsmallGB}
\end{align}
while $\beta_A^0$, ${\beta'}_A^0$ and ${\beta''}_A^0$ are at least of order $\mathcal O (\ell/m_A^0)^4$, and can thus be neglected in what follows.
The BH is fully described by the values of $\ell$ and $m_A^0$.
In the GR limit $\ell/m_A^0= 0$, the quantities above all vanish, because then the BH reduces to Schwarzschild and its mass $m_A(\varphi)$ is a constant (cf. Ref.~\cite{Julie:2019sab}).
We find it useful to introduce the dimensionless ratio
\begin{align}
 \hat \ell=\frac{\ell}{\mu}\,,
\end{align}
where $\mu$ is the binary's reduced mass defined in Eq.~\eqref{def:reducedMass}.
Then, for a binary BH system described by Eq.~\eqref{eq:alphaAsmallGB} and its $B$-counterpart, the coefficients of the effective potentials~\eqref{eq:finalEOBpotentials} boil down to simple functions of $\hat\ell$ and $\nu$ only:
\begin{subequations}
\begin{align}
    \bar\gamma_{AB}&=-2\hat\ell^4\nu^2+\mathcal O (\hat\ell^6)\,,\\
    \langle\bar\beta\rangle&=\mathcal O (\hat\ell^8)\,,
\end{align}\label{eq:coeffSS1PN}%
\end{subequations}
and
\begin{subequations}
\begin{align}
\delta \bar a_3&=\frac{1 }{3}\hat\ell^4\nu(3-\nu-16\nu^2)+\mathcal O (\hat\ell^6)\,,\label{eq:coeffSS2a}\\
 \delta\bar d_2&=\hat\ell^4\nu(1+3\nu-8\nu^2)+\mathcal O (\hat\ell^6)\,,
\end{align}
\end{subequations}
while at 3PN,
\begin{subequations}
\begin{align}
\delta\bar a_{4}&=\frac{1}{288}\hat\ell^4\nu\Big[(1-4\nu)(768 \gamma_{\rm E}+768 \ln 2+63 \pi ^2-1976)\nonumber\\
&\qquad+648+\nu^2(801 \pi ^2-26240)\Big]+\mathcal O (\hat\ell^6)\,,\label{eq:deltaA4shiftSym}\\
   \delta\bar d_3&=\frac{1}{3}\hat\ell^4\nu\Big[-2(1-4\nu)(21-32\ln 2)\nonumber\\
    &\qquad-3+40\nu-326\nu^2+48\nu^3\Big]+\mathcal O (\hat\ell^6)\,,\\
 \delta\bar q_1&=\frac{1}{9}\hat \ell^4\nu\Big[(1-4\nu)(-93-1753 \ln 2+729 \ln 6)\nonumber\\
    &\qquad -6\nu(1+23\nu-12\nu^2)\Big]+\mathcal O (\hat\ell^6)\,,\label{eq:deltaq1shiftSym}\\
    \bar q_2&=-\frac{1}{5}\hat\ell^4\nu(1-4\nu)(37-5707 \ln 2+2187 \ln 6)\nonumber\\
    &+\mathcal O (\hat\ell^6)\,,\label{eq:q2shiftSym}
\end{align}
\end{subequations}
with the logarithmic counterpart
\begin{align}
    \bar a_{4,\ln}&=\frac{4}{3}\hat\ell^4\nu(1-4\nu)+\mathcal O (\hat\ell^6)\,.\label{eq:coeffSS3PN}
\end{align}
At this order in $\hat\ell$, only the terms proportional to $\bar\gamma_{AB}$, $\delta_{A/B}$, $k_{\rm tail}$ and $k_{\rm ESGB}$ contribute to the beyond-GR coefficients above, which vanish in the GR limit $\hat \ell=0$.

Let us note two more useful limits. 

First, when $\nu=1/4$, the first lines of Eqs.~\eqref{eq:deltaA4shiftSym}-\eqref{eq:deltaq1shiftSym}, and also $\bar q_2$ and $\bar a_{4,\ln}$, are zero, because the tail corrections [driven by $k_{\rm tail}=-(2/3)\hat \ell^4\nu(1-4\nu)+\mathcal O (\hat\ell^6)$] vanish for symmetric binaries with constant scalar dipoles.
 
Second, in the extreme mass-ratio limit $\nu=0$, the effective potentials simplify to $A= 1-2u$, $D= 1$ and $Q= 0$.
Since moreover $H_{\rm EOB}-M=H_{\rm eff}-\mu$ in this limit, the two-body dynamics reduces to geodesics of the Schwarzschild metric, even when $\hat \ell\neq 0$.
The reason is simple: take, say, $m_A^0\gg m_B^0$. Since $\ell/m_B^0$ is kept fixed to small values in our approximation scheme \eqref{eq:alphaAsmallGB},  
$\ell/m_A^0=\mathcal O(\nu)$ vanishes, and $\hat \ell=\ell/m_B^0+\mathcal O(\nu)$.
This means that body $A$ reduces to a Schwarzschild spacetime with constant scalar field, cf. Eq.~\eqref{eq:alphaAsmallGB} and below.
We recover the conservative sector of the extreme mass-ratio analysis of Ref.~\cite{Maselli:2020zgv}.

\subsection{Orbital frequency at the ISCO}

We can now evaluate the beyond-GR modifications to the dynamics, focusing on circular orbits for simplicity.
 Consider the motion described by the effective Hamiltonian $\hat H_{\rm eff}$ given in Eq.~\eqref{eq:effHam}.
It does not depend on $\hat t$ nor on $\Phi$, and thus
\begin{subequations}
\begin{align}
    \hat H_{\rm eff}&=E\,,\\
    \hat P_\Phi&=J\,,
\end{align}\label{eq:systemEJ}%
\end{subequations}
are constants of motion.
When $\hat P_R=0$, we have from the system above that
\begin{align}
    E^2=W(\hat R)\label{eq:conditionCircular1}\quad \text{with}\quad W(\hat R)=A\left(1+\frac{J^2}{\hat R^2}\right)\,,
\end{align}
while the circularity of the orbit also requires $d\hat P_R/d\hat t=0$, that is
\begin{align}
    \frac{\partial \hat H_{\rm eff}}{\partial\hat R}=\frac{1}{2\hat H_{\rm eff}}\frac{\partial W(\hat R)}{\partial\hat R}=0\,.\label{eq:conditionCirc2}
\end{align}
The ISCO is characterized by a third (inflection point) condition, 
\begin{align}
    \frac{\partial^2 \hat H_{\rm eff}}{\partial\hat R^2}=\frac{1}{2\hat H_{\rm eff}}\frac{\partial^2 W(\hat R)}{\partial\hat R^2}=0\,.
\end{align}
Hence $E$ and $j=J/G_{AB}$ relate to $u$ [cf. Eq.~\eqref{def:u}] as
\begin{subequations}
\begin{align}
    j(u)&=\left[-\frac{A'}{(Au^2)'}\right]^{1/2}\,,\\
    E(u)&=A\left[\frac{2u}{(Au^2)'}\right]^{1/2}\,,
\end{align}\label{eq:JandEofU}%
\end{subequations}
where the primes denote derivatives with respect to $u$, while $u_{\rm ISCO}$ is the outermost root of
\begin{align}
    \frac{A''}{A'}=\frac{(Au^2)''}{(Au^2)'}\,.\label{eq:ISCOlocation}
\end{align}

Let us turn to the EOB Hamiltonian $\hat H_{\rm EOB}$ given in Eq.~\eqref{eq:defHeob}.
The associated Hamilton equations define a resummed two-body dynamics.
In this paper, we focus on the dimensionless orbital frequency $\hat\Omega=M\Omega=d\Phi/d\hat t$, which reads
\begin{align}
    \hat\Omega=\frac{\partial \hat H_{\rm EOB}}{\partial \hat H_{\rm eff}}\frac{\partial \hat H_{\rm eff}}{\partial \hat P_\Phi}=\frac{j u^2 A}{ G_{AB} E\sqrt{1+2\nu(E-1)}}\,,\label{eq:ISCOfrequency}
\end{align}
where $E(u)$ and $j(u)$ are given by Eqs.~\eqref{eq:JandEofU} on circular orbits.
The orbital frequency, which we shall evaluate at the ISCO, is thus fully fixed by the effective potential $A$~\cite{Buonanno:1998gg,Damour:2009sm}.
We follow Refs.~\cite{Damour:2000we,Damour:2013hea} and resum our 3PN result~\eqref{eq:AeobPotentialFinal} by means of the $(1,3)$-Pad\'e approximant
\begin{align}
   A_{\rm P}(u)=  \mathcal P^1_3[A(u)]\,.
\end{align}
This ensures that $A_{\rm P}(u)$ has a simple zero (by construction) and the presence of an ISCO, by continuity with the Schwarzschild metric recovered in the GR, test-mass limit. The Pad\'e resummation was adopted in several studies that calibrated GR-EOB waveforms to NR~\cite{Damour:2013hea,Rettegno:2019tzh}.
For further discussions on the effects of the Pad\'e resummation in the ST case at 2PN, see Ref.~\cite{Julie:2017pkb}.

\begin{figure*}[ht]
\includegraphics[width=.95\columnwidth]{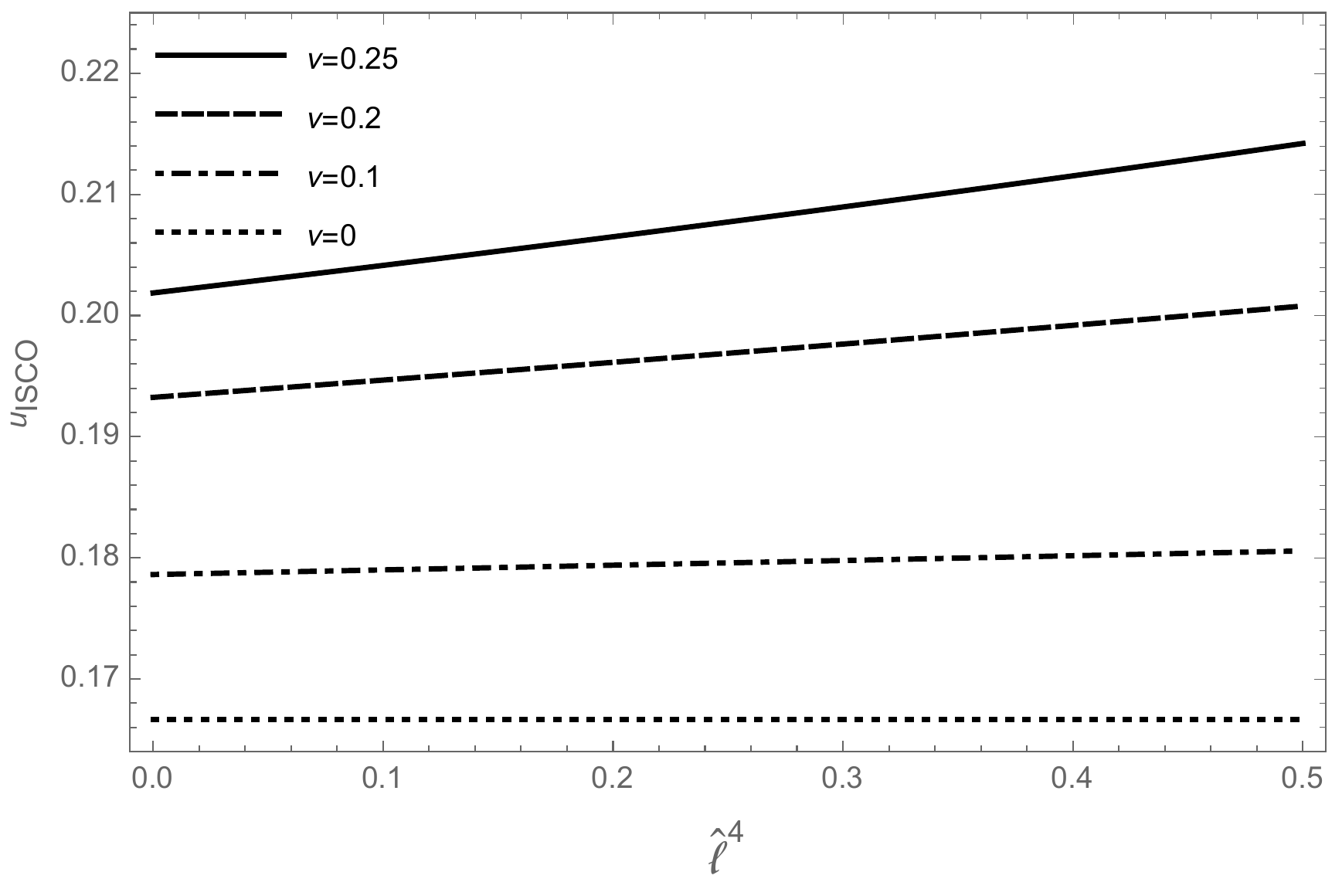}
\includegraphics[width=.959\columnwidth]{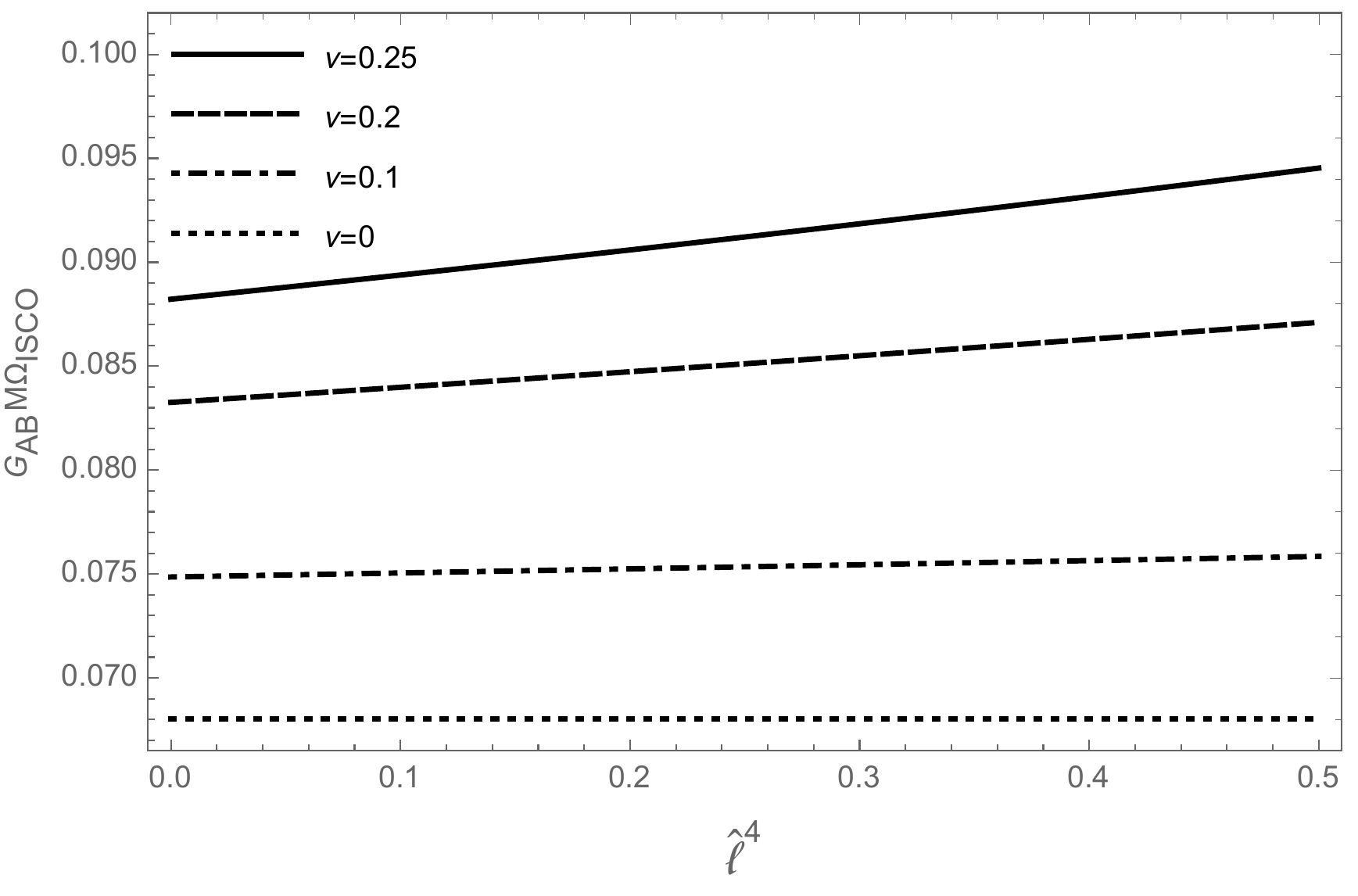}
\caption{Location $u$ and dimensionless orbital frequency $G_{AB}\hat\Omega=G_{AB}M\Omega$ at the ISCO of a BH binary in the shift-symmetric ESGB model $f(\varphi)=2\varphi$ and $\mathcal A=1$.
Here, $\nu=\{0,0.1,0.2,1/4\}$ is the symmetric mass ratio and $\hat\ell=\ell/\mu$ is the dimensionless Gauss-Bonnet coupling, with $\mu$ the reduced mass. GR is recovered when $\hat{\ell}=0$.
When $\nu=0$, the ISCO location and orbital frequency reduce to the Schwarzschild values, while they increase the most with $\hat\ell$ in the equal-mass case $\nu=1/4$.
The relative modification to the GR ISCO frequency 
then reaches the percent level when $\hat\ell=0.528$, that is $\ell/M=0.132$.}
\label{fig}
\end{figure*}

Figure~\ref{fig} shows the ISCO location $u_{\rm ISCO}$ and dimensionless frequency $G_{AB}\hat\Omega$ of a binary BH system in the shift-symmetric ESGB model discussed in Sec.~\ref{subsec:ESGBBBH}.
The beyond-GR coefficients of the potential $A$ are thus the functions of $\nu$ and $\hat\ell=\ell/\mu$ given in Eqs.~\eqref{eq:coeffSS1PN}-\eqref{eq:coeffSS3PN}, which we truncated at the leading order in $\hat\ell$ given there.
We recover GR when $\hat \ell=0$, and we consider four symmetric mass ratio values, $\nu=\{0,0.1,0.2,1/4\}$.

When $\nu=0$, we find that $u_{\rm ISCO}=1/6$ and $G_{AB}\hat\Omega=6^{-3/2}$ reduce to their Schwarzschild values for all $\hat \ell$,
consistently with the extreme mass-ratio limit described at the end of Sec.~\ref{subsec:ESGBBBH}.
However, when $\nu\neq 0$, both $u_{\rm ISCO}$ and $G_{AB}\hat\Omega$ increase with $\hat\ell$. In particular, the slope (or ``sensitivity'') of the ISCO frequency is maximal when $\nu=1/4$:
\begin{align}
\left.\frac{d(G_{AB}\hat\Omega)_{\rm ISCO}}{d(\hat\ell^4)}\right|_{\begin{subarray}{l}\hat\ell=0\\
    \nu=1/4\end{subarray}}\!\!=1.14\times 10^{-2}\,.
\end{align}
For such equal-mass binaries (with $\mu=M/4$), the relative modification to the GR ISCO frequency, $(G_{AB}\hat\Omega)_{\rm ISCO}/(G_{AB}\hat\Omega)_{\rm ISCO}^{\hat\ell=0}-1$,
then reaches the percent level when $\ell/M=0.132$.
For comparison, Ref.~\cite{Lyu:2022gdr} 
obtained one of the most stringent constraints to this day in shift-symmetric ESGB gravity, $\ell/M<0.344$, from the  BH-NS system GW200115 with total mass $M=7.1M_{\rm Sun}$. (Note that we translate between our conventions and those of Ref.~\cite{Lyu:2022gdr} 
by setting $\varphi=\sqrt{4\pi}\phi$ and $\ell^2=2\sqrt{4\pi}\alpha_{\rm GB}$.)

The ISCO analysis above motivates the obtention of full EOB waveforms, including the dissipative sector~\cite{Damour:1992we,Lang:2013fna,Shiralilou:2021mfl,Bernard:2022noq}, to be confronted to GW signals. This issue will be addressed in future work.

\section{Conclusions}
\label{sec:V}

In this paper we have extended the work of Refs.~\cite{Julie:2017pkb,Julie:2017ucp} and built an EOB Hamiltonian in ST and ESGB gravity at 3PN order.
Our main new results are:
\begin{enumerate}
\item An ordinary two-body Hamiltonian [Eq.~\eqref{eq:2bodyHamiltonian}] at 3PN and in ST-ESGB gravity;
\item The associated EOB Hamiltonian [Eq.~\eqref{eq:defHeob}] and its 3PN coefficients [Eq.~\eqref{eq:coeffsEOB3PN}], which account for the beyond-GR tail at sixth order in the eccentricity;
\item The application to hairy BH binaries in shift-symmetric ESGB gravity [Eqs.~\eqref{eq:coeffSS1PN}-\eqref{eq:coeffSS3PN}], and the first estimate of their ISCO frequency (Fig.~\ref{fig}).
\end{enumerate}

It is important that the EOB framework can be extended beyond GR.
Here, we have reduced the
3PN dynamics to the (nongeodesic) motion in a modification of the GR EOB metric, and accounted for the beyond-GR tail effects by adapting the 4PN methods of Ref.~\cite{Damour:2015isa}.

The EOB framework is also suitable to include other modified theories of gravity, such as Einstein-Maxwell-scalar models at 1PN \cite{Julie:2017rpw,Khalil:2018aaj,Julie:2018lfp}.
Our work can thus be regarded as another step toward the development of a parametrized EOB framework, by providing a ``dictionary'' between modified gravity theories and the values of the coefficients of the effective potentials~\eqref{eq:finalEOBpotentials}.
 In the future, the tools and methods we developed in this paper could be applied  to other models, such as disformal ST, massive gravity, or Horndeski theories (which also predict hairy BHs~\cite{VanAelst:2019kku}).

We have focused here on the conservative part of the dynamics. The corresponding EOB radiation-reaction force, to be inferred from already available energy fluxes~\cite{Damour:1992we,Lang:2013fna,Shiralilou:2021mfl,Bernard:2022noq}, and gravitational waveforms, will be the topic of future work.
Since NSs and BHs are, in general, spinning, it will also be important to extend the present work to include spin effects. For the PN analysis of spin-orbit effects in ST gravity some work has been done in Ref.~\cite{Brax:2021qqo}. As a first step, the beyond-GR EOB Hamiltonian derived here could be included in the state-of-the-art spinning EOB Hamiltonians in GR (see, e.g., Refs.~\cite{Rettegno:2019tzh,Khalil:2020mmr} and references therein), and then used to generate beyond-GR inspiral waveforms.

The EOB approach uses a resummation of the two-body dynamics that can be extended through the plunge of the two BHs, after which the waveform is matched to the merger-ringdown signal.
The latter should make use of the quasinormal mode spectrum of ESGB BHs, which has been computed up to second order in a slow-rotation expansion~\cite{Pierini:2021jxd,Pierini:2022eim}. The Pad\'e-resummed spectrum of Kerr BHs computed at the same order in the slow-rotation approximation is typically accurate at the percent level when evaluated at the dimensionless spins $\sim 0.7$ of interest for LIGO-Virgo-KAGRA observations~\cite{Pierini:2022eim}. Therefore it is reasonable to assume that deviations induced by beyond-GR terms could be testable at the same (percent) level of accuracy.
The quasinormal mode spectrum could be included in the EOB model using the parametrized spin expansion coefficient ({\sc ParSpec})  framework~\cite{Maselli:2019mjd}, which has been used to perform theory-specific tests of GR with ringdown signals using the {\sc pyRing}
code in Ref.~\cite{Carullo:2021dui}, and with EOB waveforms in Refs.~\cite{Silva:2022srr}.

For the case of binary BHs, once the EOB waveforms are completed with physically motivated ansatzes for the merger-ringdown in ESGB gravity, they could be compared with and informed by NR simulations (see, e.g., Ref.~\cite{Corman:2022xqg}). Developing precise and complete EOB-NR waveform models is crucial to obtain new experimental bounds on ESGB models, and more generally, on wider classes of modified gravity theories in the future.

\vspace{.2cm}

\noindent
{\bf Note added:} While this project was nearing completion, we became aware of an independent effort that recently appeared on the arXiv~\cite{Jain:2022nxs}. Their work focuses on the computation of the 3PN EOB Hamiltonian in ST theories, and restricts the inclusion of tail effects to circular orbits. This limit amounts to setting formally $k_{\rm ESGB}=0$, and $k_{\rm tail}=0$ in Eqs.~\eqref{Eq:dd3}-\eqref{Eq:q2}.

After both works appeared on the arXiv, we compared the results in the limit given above. We found that:\footnote{The cross-check of our results against Ref.~\cite{Jain:2022nxs} used
[arXiv:2211.15580v2].}
\begin{enumerate}
    \item Equation~(5.16) of Ref.~\cite{Jain:2022nxs} still differs from Eq.~\eqref{Eq:dq1} by an overall minus sign;
    \item \label{point2Comparison} The term proportional to $\langle\delta\rangle/\alpha_{AB}$ in Eq.~(5.14) of Ref.~\cite{Jain:2022nxs} differs from that of Eq.~\eqref{Eq:da4} by a factor $\bar\gamma_{AB}$. Note that $\alpha_{AB}=\tilde\alpha$ in our conventions.
\end{enumerate}
We explicitly checked that reexpanding our $H_{\rm EOB}$ at 3PN order yields Eqs.~(5.4)-(5.5) of Ref.~\cite{Bernard:2018ivi} on circular orbits. By contrast, we find that it does not if we replace Eq.~\eqref{Eq:da4} by Eq.~(5.14) of Ref.~\cite{Jain:2022nxs}.

After our work appeared on the arXiv, Ref.~\cite{Jain:2022nxs} was extended in Ref.~\cite{Jain:2023fvt} to include the tail effects up to $\mathcal O(e^4)$.
This limit amounts to setting formally $k_{\rm ESGB}=0$, and $k_{\rm tail}=0$ in Eq.~\eqref{Eq:q2}, since we recall that our work includes the tail effects up to $\mathcal O(e^6)$.
The tail contributions (4.12)-(4.17) in Ref.~\cite{Jain:2023fvt} all differ from ours by an overall factor $\mathcal A_0^2$. The latter indeed enters Eq.~\eqref{def:ktail}, and it originates from the translation of the Jordan-frame Lagrangian (A3) of Ref.~\cite{Bernard:2018ivi}, see Sec.~\ref{subsec:2-bodyLag3PN}.\footnote{The cross-check of our results against Ref.~\cite{Jain:2023fvt} used
[arXiv:2301.01070v1].}

\acknowledgments

We thank Laura Bernard for sharing a \textit{Mathematica} notebook containing the 3PN Lagrangian of Ref.~\cite{Bernard:2018ivi}. We are also grateful to Thibault Damour and Nathalie Deruelle for comments and discussions. F.L.J. acknowledges support  from the German Research Foundation (Deutsche Forschungsgemeinschaft, DFG, Project No. 386119226). E.B. is supported by NSF Grants No. AST-2006538, No. PHY-2207502, No. PHY-090003 and No. PHY-20043, and NASA Grants No. 19-ATP19-0051, No. 20-LPS20-0011 and No. 21-ATP21-0010.
The complete EOB potentials presented in Sec.~\ref{subseq:completePotentials} are gathered in a \textit{Mathematica} notebook available online~\cite{FLJRepo}.

\appendix

\section{Einstein and Jordan frames\label{app:EinsteinVsJordan}}

In Refs.~\cite{Mirshekari:2013vb,Bernard:2018hta,Bernard:2018ivi}, the ST two-body Lagrangian was computed up to 3PN order by adopting the Jordan-frame formulation of the theory (we use tildes for clarity):
\begin{align}
I_{\rm ST} =  \int \frac{d^{4}x  \sqrt{-\tilde g}}{16 \pi}  \left(\phi \tilde R - \frac{\omega(\phi)}{\phi}(\partial\phi)^2\right)+I_{\rm m}[\Psi,\tilde g_{\mu\nu}]\,,\label{eq:JFaction1}
\end{align}
where $(\partial\phi)^2=\tilde g^{\mu\nu} \partial_{\mu}\phi \partial_{\nu}\phi$, and where $\omega(\phi)$ is a function defining the theory. As for compact bodies, they were described by performing the substitution
\begin{align}
I_{\rm m}\to  I_{\rm m}^{\rm pp}=-\sum_A\int \tilde m_A(\phi)d\tilde s_A\,,\label{eq:JFaction2}
\end{align}
with $d\tilde s_A=\sqrt{-\tilde g_{\mu\nu}dx_A^\mu dx_A^\nu}$.
In the present paper, we describe ST theories by the Einstein-frame action, which (setting to zero the GB coupling) reads:
\begin{align}
I_{\rm ST} =  \int \frac{d^{4}x  \sqrt{-g}}{16 \pi}  \left(R - 2(\partial\varphi)^2\right)+I_{\rm m}[\Psi,\mathcal A^2g_{\mu\nu}]\,,\label{eq:EFaction1}
\end{align}
where $(\partial\varphi)^2=g^{\mu\nu} \partial_{\mu}\varphi \partial_{\nu}\varphi$, and where we account for compact bodies by the substitution
\begin{align}
I_{\rm m}\to I_{\rm m}^{\rm pp}=-\sum_A\int m_A(\varphi)ds_A\,,\label{eq:EFaction2}
\end{align}
with $ds_A=\sqrt{-g_{\mu\nu}dx_A^\mu dx_A^\nu}$.
The actions \eqref{eq:JFaction1}-\eqref{eq:JFaction2} and \eqref{eq:EFaction1}-\eqref{eq:EFaction2} are identical, modulo boundary terms, via the redefinitions:
\begin{subequations}
\begin{align}
\tilde g_{\mu\nu}&=\mathcal A^2 g_{\mu\nu}\,,\\
3+2\omega(\phi)&=\left(\frac{d\ln\mathcal A}{d\varphi}\right)^{-2},\\
m_A(\varphi)&=\mathcal A(\varphi) \tilde m_A(\varphi)\,,\label{eq:conversionJfEfmasses}%
\end{align}\label{eq:conversionJfEf}%
\end{subequations}
where $\varphi(\phi)$ is obtained by inverting $\mathcal A(\varphi)=1/\sqrt{\phi}$.
Let us also introduce the notation:
\begin{subequations}
\begin{align}
\alpha_0&=\frac{d\ln \mathcal A}{d\varphi}(\varphi_0)\,,\label{eq:sensisUniversellesAlpha}\\
\beta_0&=\frac{d\alpha}{d\varphi}(\varphi_0)\,,\\
{\beta'}_0&=\frac{d\beta}{d\varphi}(\varphi_0)\,,\\
{\beta''}_0&=\frac{d{\beta'}}{d\varphi}(\varphi_0)\,,
\end{align}\label{eq:sensisUniverselles}%
\end{subequations}
where the subscript $0$ denotes a quantity evaluated at infinity, $\varphi(\phi_0)=\varphi_0$.
The quantities above can be obtained by inserting Eq.~\eqref{eq:conversionJfEfmasses} into Eqs.~\eqref{eq:sensis} and taking the limit $\tilde m_A(\phi)=\rm const.$ In this limit, body $A$ is said to have negligible self-gravity, and its motion reduces to geodesics of $\tilde g_{\mu\nu}$, cf. Eq.~\eqref{eq:JFaction2}.

By using Eqs.~\eqref{eq:conversionJfEf} and below, we can translate the parameters of Refs.~\cite{Mirshekari:2013vb,Bernard:2018hta,Bernard:2018ivi} into our conventions. The results are gathered in Table~\ref{table:JFvsEFparameters}.

\begin{widetext}
\begin{center}
\begin{table}[h]
\begin{tabular}{|c |l|}
  \hline
    Refs.~\cite{Mirshekari:2013vb,Bernard:2018hta,Bernard:2018ivi} & Refs.~\cite{Damour:1992we,Damour:1995kt,Julie:2019sab} and this paper \\
  \hline \hline
Theory dependent &  \\
   $\tilde G$ & $\mathcal A_0^2 (1+\alpha_0^2)$\\
  $\zeta$ & $\frac{\alpha_0^2}{1+\alpha_0^2}$ \\ 
  $\lambda_1$ & $\frac{\beta_0}{2(1+\alpha_0^2)}$ \\
  $\lambda_2$ & $\frac{1}{4(1+\alpha_0^2)^2}(-\beta'_0\alpha_0-2\beta_0\alpha_0^2+4\beta_0^2)$ \\ 
  $\lambda_3$ & $\frac{1}{8(1+\alpha_0^2)^3}(6\beta'_0\alpha_0^3-24\beta_0^2\alpha_0^2+8\beta_0\alpha_0^4-13\beta'_0\beta_0\alpha_0+24\beta_0^3+{\beta''}_0\alpha_0^2)$ \\ 
  \hline
    Body dependent &  \\
  $m_A$ & $m_A^0/\mathcal A_0$ \\
  $s_A$ & $\frac{1}{2}(1-\alpha_A^0/\alpha_0)$ \\
  $s_A'$ & $\frac{1}{4\alpha_0}(\beta_A^0/\alpha_0-\alpha_A^0\beta_0/\alpha_0^2)$ \\
  $s_A''$ & $\frac{1}{8\alpha_0^5}(3\beta_A^0\alpha_0\beta_0-3\alpha_A^0\beta_0^2-{\beta'}_A^0\alpha_0^2+\alpha_A^0\beta'_0\alpha_0)$ \\
  $s_A'''$ & $\frac{1}{16\alpha_0^7}(15\beta_A^0\alpha_0\beta_0^2-15\alpha_A^0\beta_0^3-5{\beta'}_A^0\beta_0\alpha_0^2+10\alpha_A^0\beta_0'\beta_0\alpha_0-{\beta'}_A^0\alpha_0^2\beta_0$
  $-4\beta_A^0\alpha_0^2\beta_0'+{\beta''}_A^0\alpha_0^3-\alpha_A^0{\beta''}_A^0\alpha_0^2)$\\
  \hline
   0PN &  \\
  $\tilde G\tilde\alpha$ & $\mathcal A_0^2(1+\alpha_A^0\alpha_B^0)=\mathcal A_0^2G_{AB}$ \\
  \hline
   1PN &  \\
  $\bar\gamma$ & $\frac{-2\alpha_A^0\alpha_B^0}{1+\alpha_A^0\alpha_B^0}=\bar\gamma_{AB}$ \\
  $\bar\beta_A$ & $\frac{\beta_A^0(\alpha_B^0)^2}{2(1+\alpha_A^0\alpha_B^0)^2}=\bar\beta_A$ \\
  \hline
   2PN &  \\
  $\bar\delta_A$ & $\frac{(\alpha_A^0)^2}{(1+\alpha_A^0\alpha_B^0)^2}=\delta_A$ \\
  $\bar\chi_A$ & $\frac{-{\beta'}_A^0(\alpha_B^0)^3}{4(1+\alpha_A^0\alpha_B^0)^3}=-\epsilon_A/4$ \\
    $\bar\beta_A\bar\beta_B/\bar\gamma$ & $\frac{-{\beta}_A^0{\beta}_B^0\alpha_A^0\alpha_B^0}{8(1+\alpha_A^0\alpha_B^0)^3}=-\zeta_{AB}/8$ \\
  \hline
   3PN &  \\
  $\bar\kappa_A$ & $\frac{(\alpha_B^0)^4{\beta''_A}^0}{8(1+\alpha_A^0\alpha_B^0)^4}=\kappa_A$ \\
  $\bar\beta_A\bar\delta_A/\bar\gamma$ & $\frac{-\alpha_A^0\alpha_B^0\beta_A^0}{4(1+\alpha_A^0\alpha_B^0)^3}=-\psi_A/4$ \\
  $\bar\beta_A\bar\chi_B/\bar\gamma$ & $\frac{(\alpha_A^0)^2\alpha_B^0\beta_A^0{\beta'}_B^0}{16(1+\alpha_A^0\alpha_B^0)^4}=\xi_A/16$ \\
$\bar\beta_A(\bar\beta_B)^2/\bar\gamma^2$   & $\frac{(\alpha_A^0)^2\beta_A^0(\beta_B^0)^2}{32(1+\alpha_A^0\alpha_B^0)^4}=\omega_A/32$ \\
  \hline
\end{tabular}
\caption{Translation of the parameters from Refs.~\cite{Mirshekari:2013vb,Bernard:2018hta,Bernard:2018ivi}. Their $\alpha$ is renamed here as $\tilde\alpha$ to avoid confusion with ours.}
\label{table:JFvsEFparameters}
\end{table}
\end{center}
\end{widetext}

\section{Two-body Lagrangian at 3PN order\label{app:3PNlagrangian}}

The contributions to the two-body, harmonic Lagrangian~\eqref{eq:twoBodyLstructure} are, up to 2PN:

\begin{widetext}
\begin{subequations}
\begin{align}
L_{0\rm PN}&=\frac{1}{2}m_A^0 \mathbf{v}_A^2+\frac{1}{2}m_B^0 \mathbf{v}_B^2+\frac{G_{AB}m_A^0m_B^0}{r}\,,\\
L_{1\rm PN}&=\frac{1}{8}m_A^0\mathbf{v}_A^4+\frac{1}{8}m_B^0\mathbf{v}_B^4+\frac{G_{AB}m_A^0m_B^0}{r}\left[\frac{3}{2}(\mathbf{v}_A^2+\mathbf{v}_B^2)-\frac{7}{2}(\mathbf v_A\cdot \mathbf v_B)-\frac{1}{2}(\mathbf n\cdot \mathbf v_A)(\mathbf n\cdot \mathbf v_B)+\bar\gamma_{AB}(\mathbf v_A-\mathbf v_B)^2\right]\nonumber\\
&-\frac{G_{AB}^2m_A^0m_B^0}{2r^2}\left[m_A^0(1+2\bar\beta_B)+m_B^0(1+2\bar\beta_A)\right]\,,\\
L_{2\rm PN}&=\frac{1}{16}m_A^0\mathbf{v}_A^6\nonumber\\
&+\frac{G_{AB}m_A^0m_B^0}{r}\left[\frac{1}{8}(7+4\bar{\gamma}_{AB})\left(\mathbf v_A^4-\mathbf  v_A^2(\mathbf n\cdot \mathbf v_B)^2\right)-(2+\bar{\gamma}_{AB})\mathbf v_A^2(\mathbf v_A\cdot \mathbf v_B)+\frac{1}{8}(\mathbf v_A\cdot\mathbf v_B)^2\right.\nonumber\\
&\qquad\left.+\frac{1}{16}(15+8\bar{\gamma}_{AB})\mathbf v_A^2 \mathbf v_B^2+\frac{3}{16}(\mathbf n\cdot\mathbf v_A)^2(\mathbf n\cdot\mathbf v_B)^2+\frac{1}{4}(3+2\bar{\gamma}_{AB})\mathbf v_A\cdot\mathbf v_B(\mathbf n\cdot\mathbf v_A)(\mathbf n\cdot\mathbf v_B)\right]\nonumber\\
&+\frac{G_{AB}^2m_B^0(m_A^0)^2}{8r^2}\bigg[\left(2+12\bar{\gamma}_{AB}+7\bar{\gamma}_{AB}^2+8\bar{\beta}_B-4\delta_A\right)\mathbf v_A^2+\left(14+20\bar{\gamma}_{AB}+7\bar{\gamma}_{AB}^2+4\bar{\beta}_B-4\delta_A\right)\mathbf v_B^2\nonumber\\
&\qquad -2\left(7+16\bar{\gamma}_{AB}+7\bar{\gamma}_{AB}^2+4\bar{\beta}_B-4\delta_A\right)\mathbf v_A\cdot\mathbf v_B-2\left(14+12\bar{\gamma}_{AB}+\bar{\gamma}_{AB}^2-8\bar{\beta}_B+4\delta_A\right)(\mathbf n\cdot\mathbf v_A)(\mathbf n\cdot\mathbf v_B)\nonumber\\
&\qquad +\left(28+20\bar{\gamma}_{AB}+\bar{\gamma}_{AB}^2-8\bar{\beta}_B+4\delta_A\right)(\mathbf n\cdot\mathbf v_A)^2+\left(4+4\bar{\gamma}_{AB}+\bar{\gamma}_{AB}^2+4\delta_A\right)(\mathbf n\cdot\mathbf v_B)^2\bigg]\nonumber\\
&\left.+\frac{G_{AB}^3(m_A^0)^3m_B^0}{2r^3}\left[1+\frac{2}{3}\bar{\gamma}_{AB}+\frac{1}{6}\bar{\gamma}_{AB}^2+2\bar{\beta}_B+\frac{2}{3}\delta_A+\frac{1}{3}\epsilon_B\right]+\frac{G_{AB}^3(m_A^0)^2(m_B^0)^2}{8r^3}\bigg[19+8\bar{\gamma}_{AB}+8(\bar{\beta}_A+\bar{\beta}_B)+4\zeta_{AB}\bigg]\right.\nonumber\\
&-\frac{1}{8}G_{AB}m_A^0m_B^0\bigg[2(7+4\bar{\gamma}_{AB})(\mathbf v_B\cdot\mathbf a_A)(\mathbf n\cdot\mathbf v_B)+(\mathbf n\cdot\mathbf a_A)(\mathbf n\cdot\mathbf v_B)^2-(7+4\bar{\gamma}_{AB})(\mathbf n\cdot\mathbf a_A)\mathbf v_B^2\bigg]\nonumber\\
&+(A\leftrightarrow B)\,.
\end{align}\label{eq:2bodyL2PN}%
\end{subequations}
The contributions $L_{3\rm PN}^{(i)}$ at 3PN level~\eqref{eq:3PNlagrangianStructure} are respectively proportional to $G_{AB}^i$ and read:
\begin{subequations}
\begin{align}
L_{3\rm PN}^{(0)}&=\frac{5}{128}m_A^0\mathbf v_A^8+\frac{5}{128}m_B^0\mathbf v_B^8\,,\\
L_{3\rm PN}^{(1)}&=\frac{G_{AB} m_A^0 m_B^0}{32 r}\bigg[
-4 \mathbf v_A^4 (\mathbf n\cdot\mathbf v_B)^2 \left(12 \bar{\gamma }_{A B}+23\right)+\mathbf v_A^4( \mathbf n\cdot\mathbf v_A)( \mathbf n\cdot\mathbf v_B) \left(20 \bar{\gamma }_{A B}+42\right)-8 \mathbf v_A^2 \mathbf v_B^2 (\mathbf n\cdot\mathbf v_A)^2 \left(19 \bar{\gamma }_{A B}+39\right)\nonumber\\
&\qquad+\mathbf v_A^2 \mathbf v_B^2( \mathbf n\cdot\mathbf v_A)( \mathbf n\cdot\mathbf v_B) \left(140 \bar{\gamma }_{A B}+283\right)-156 \mathbf v_A^2 (\mathbf n\cdot\mathbf v_A) (\mathbf n\cdot\mathbf v_B)^3 \left(\bar{\gamma }_{A B}+2\right)+144 \mathbf v_A^2 (\mathbf n\cdot\mathbf v_A)^2 (\mathbf n\cdot\mathbf v_B)^2 \left(\bar{\gamma }_{A B}+2\right)\nonumber\\
&\qquad-6 \mathbf v_A^2 (\mathbf n\cdot\mathbf v_A)^3 (\mathbf n\cdot\mathbf v_B) \left(2 \bar{\gamma }_{A B}+5\right)-16 \mathbf v_A^2 (\mathbf n\cdot\mathbf v_A)( \mathbf n\cdot\mathbf v_B) \mathbf v_A\cdot\mathbf v_B \left(18 \bar{\gamma }_{A B}+35\right)+8 (\mathbf n\cdot\mathbf v_A)^4 \mathbf v_A\cdot\mathbf v_B \left(3 \bar{\gamma }_{A B}+5\right)\nonumber\\
&\qquad-2 \mathbf v_A^2 (\mathbf n\cdot\mathbf v_A)^2 \mathbf v_A\cdot\mathbf v_B \left(26 \bar{\gamma }_{A B}+45\right) +4 \mathbf v_A^2 (\mathbf n\cdot\mathbf v_B)^2 \mathbf v_A\cdot\mathbf v_B \left(65 \bar{\gamma }_{A B}+128\right)+4 \mathbf v_B^2 (\mathbf n\cdot\mathbf v_A)^4 \left(13 \bar{\gamma }_{A B}+27\right)\nonumber\\
&\qquad-10 (\mathbf n\cdot\mathbf v_A)^4 (\mathbf n\cdot\mathbf v_B)^2 \left(8 \bar{\gamma }_{A B}+15\right)-12 (\mathbf n\cdot\mathbf v_A)^2 (\mathbf v_A\cdot\mathbf v_B)^2 \left(8 \bar{\gamma }_{A B}+15\right)+5 (\mathbf n\cdot\mathbf v_A)^3 (\mathbf n\cdot\mathbf v_B)^3 \left(16 \bar{\gamma }_{A B}+29\right)\nonumber\\
&\qquad+16 (\mathbf n\cdot\mathbf v_A)^3( \mathbf n\cdot\mathbf v_B) \mathbf v_A\cdot\mathbf v_B \left(17 \bar{\gamma }_{A B}+32\right)+2( \mathbf n\cdot\mathbf v_A)(\mathbf n\cdot\mathbf v_B) (\mathbf v_A\cdot\mathbf v_B)^2 \left(108 \bar{\gamma }_{A B}+197\right)\nonumber\\
&\qquad-3 (\mathbf n\cdot\mathbf v_A)^2 (\mathbf n\cdot\mathbf v_B)^2 \mathbf v_A\cdot\mathbf v_B \left(108 \bar{\gamma }_{A B}+199\right)+12 \mathbf v_A^6 \bar{\gamma }_{A B}+120 \mathbf v_A^4 \mathbf v_B^2 \bar{\gamma }_{A B}-4 \mathbf v_A^4 \mathbf v_A\cdot\mathbf v_B \bar{\gamma }_{A B}\nonumber\\
&\qquad-192 \mathbf v_A^2 \mathbf v_B^2 \mathbf v_A\cdot\mathbf v_B \bar{\gamma }_{A B}+96 \mathbf v_A^2 (\mathbf v_A\cdot\mathbf v_B)^2 \bar{\gamma }_{A B}-32 (\mathbf v_A\cdot\mathbf v_B)^3 \bar{\gamma }_{A B}+22 \mathbf v_A^6+240 \mathbf v_A^4 \mathbf v_B^2-10 \mathbf v_A^4 \mathbf v_A\cdot\mathbf v_B\nonumber\\
&\qquad-387 \mathbf v_A^2 \mathbf v_B^2 \mathbf v_A\cdot\mathbf v_B+188 \mathbf v_A^2 (\mathbf v_A\cdot\mathbf v_B)^2-54 (\mathbf v_A\cdot\mathbf v_B)^3\bigg]\nonumber\\
+&\frac{1}{48}G_{AB} m_A^0 m_B^0
\bigg[
 48 \mathbf v_A^2 \mathbf v_B^2 (\mathbf n \cdot \mathbf a_A) \left(3 \bar{\gamma }_{A B}+\frac{97}{16}\right)-42 \mathbf v_A^2 (\mathbf n \cdot \mathbf a_A) (\mathbf n \cdot \mathbf v_B)^2 \left(\bar{\gamma }_{A B}+2\right)+120 \mathbf v_A^2 \mathbf v_B \cdot \mathbf a_A (\mathbf n \cdot \mathbf v_B) \left(\bar{\gamma }_{A B}+2\right)\nonumber\\
 &\qquad+48 \mathbf v_A^2( \mathbf n \cdot \mathbf a_A) \mathbf v_A \cdot \mathbf v_B \left(\bar{\gamma }_{A B}+\frac{15}{8}\right)+6 \mathbf v_A^2 (\mathbf n \cdot \mathbf a_A) (\mathbf n \cdot \mathbf v_A)( \mathbf n \cdot \mathbf v_B) \left(2 \bar{\gamma }_{A B}+5\right)-12 \mathbf v_A^2 \mathbf v_A \cdot \mathbf a_A (\mathbf n \cdot \mathbf v_B) \left(6 \bar{\gamma }_{A B}+11\right)\nonumber\\
 &\qquad+6 \mathbf v_A^2 \mathbf v_B \cdot \mathbf a_A( \mathbf n \cdot \mathbf v_A) \left(8 \bar{\gamma }_{A B}+15\right)+48 \mathbf v_B^2 \mathbf v_A \cdot \mathbf a_A (\mathbf n \cdot \mathbf v_A) \left(6 \bar{\gamma }_{A B}+\frac{97}{8}\right)-6 \mathbf v_B^2( \mathbf n \cdot \mathbf a_A) (\mathbf n \cdot \mathbf v_A)^2 \left(13 \bar{\gamma }_{A B}+27\right)\nonumber\\
 &\qquad-84 \mathbf v_A \cdot \mathbf a_A( \mathbf n \cdot \mathbf v_A) (\mathbf n \cdot \mathbf v_B)^2 \left(\bar{\gamma }_{A B}+2\right)+240 \mathbf v_A \cdot \mathbf a_A( \mathbf n \cdot \mathbf v_B) \mathbf v_A \cdot \mathbf v_B \left(\bar{\gamma }_{A B}+2\right)+12 (\mathbf n \cdot \mathbf a_A) (\mathbf v_A \cdot \mathbf v_B)^2 \left(4 \bar{\gamma }_{A B}+7\right)\nonumber\\
 &\qquad-12( \mathbf n \cdot \mathbf a_A )(\mathbf n \cdot \mathbf v_A)^2 \mathbf v_A \cdot \mathbf v_B \left(3 \bar{\gamma }_{A B}+5\right)-4 \mathbf v_B \cdot \mathbf a_A (\mathbf n \cdot \mathbf v_A)^3 \left(3 \bar{\gamma }_{A B}+5\right)+6 \mathbf v_A \cdot \mathbf a_A (\mathbf n \cdot \mathbf v_A)^2 (\mathbf n \cdot \mathbf v_B) \left(2 \bar{\gamma }_{A B}+5\right)\nonumber\\
 &\qquad+24 \mathbf v_B \cdot \mathbf a_A (\mathbf n \cdot \mathbf v_A) \mathbf v_A \cdot \mathbf v_B \left(4 \bar{\gamma }_{A B}+7\right)+9( \mathbf n \cdot \mathbf a_A) (\mathbf n \cdot \mathbf v_A)^2 (\mathbf n \cdot \mathbf v_B)^2 \left(8 \bar{\gamma }_{A B}+15\right)-6 \mathbf v_A^4 (\mathbf n \cdot \mathbf a_B) \left(6 \bar{\gamma }_{A B}+11\right)\nonumber\\
 &\qquad-12( \mathbf n \cdot \mathbf a_A) (\mathbf n \cdot \mathbf v_A) (\mathbf n \cdot \mathbf v_B) \mathbf v_A \cdot \mathbf v_B \left(22 \bar{\gamma }_{A B}+41\right)-6 \mathbf v_B \cdot \mathbf a_A (\mathbf n \cdot \mathbf v_A)^2 (\mathbf n \cdot \mathbf v_B )\left(22 \bar{\gamma }_{A B}+41\right)\nonumber\\
 &\qquad+6 \mathbf v_A^2 (\mathbf n \cdot \mathbf a_B) (\mathbf n \cdot \mathbf v_A)^2 \left(2 \bar{\gamma }_{A B}+5\right)+12 \mathbf v_A^2 \mathbf v_A \cdot \mathbf a_B (\mathbf n \cdot \mathbf v_A) \left(8 \bar{\gamma }_{A B}+15\right)-72 \mathbf v_A^2 (\mathbf n \cdot \mathbf a_B)( \mathbf n \cdot \mathbf v_A)( \mathbf n \cdot \mathbf v_B) \left(\bar{\gamma }_{A B}+2\right)\nonumber\\
 &\qquad+144 \mathbf v_A^2 \mathbf v_A \cdot \mathbf a_B (\mathbf n \cdot \mathbf v_B) \left(\bar{\gamma }_{A B}+2\right)+144 \mathbf v_A^2 (\mathbf n \cdot \mathbf a_B) \mathbf v_A \cdot \mathbf v_B \left(\bar{\gamma }_{A B}+2\right)+12 \mathbf v_A^2 \mathbf v_B \cdot \mathbf a_B( \mathbf n \cdot \mathbf v_A) \left(20 \bar{\gamma }_{A B}+41\right)\nonumber\\
 &\qquad-8 \mathbf v_A \cdot \mathbf a_B (\mathbf n \cdot \mathbf v_A)^3 \left(3 \bar{\gamma }_{A B}+5\right)+12( \mathbf n \cdot \mathbf a_B )(\mathbf n \cdot \mathbf v_A)^3( \mathbf n \cdot \mathbf v_B) \left(4 \bar{\gamma }_{A B}+7\right)-8 \mathbf v_B \cdot \mathbf a_B (\mathbf n \cdot \mathbf v_A)^3 \left(5 \bar{\gamma }_{A B}+11\right)\nonumber\\
 &\qquad-24 \mathbf v_A \cdot \mathbf a_B (\mathbf n \cdot \mathbf v_A)^2( \mathbf n \cdot \mathbf v_B) \left(6 \bar{\gamma }_{A B}+11\right)-24( \mathbf n \cdot \mathbf a_B) (\mathbf n \cdot \mathbf v_A)^2 \mathbf v_A \cdot \mathbf v_B \left(6 \bar{\gamma }_{A B}+11\right)-3 (\mathbf n \cdot \mathbf a_B) (\mathbf n \cdot \mathbf v_A)^4\nonumber\\
 &\qquad+12 \mathbf v_A \cdot \mathbf a_B (\mathbf n \cdot \mathbf v_A) \mathbf v_A \cdot \mathbf v_B \left(8 \bar{\gamma }_{A B}+13\right)+12 \mathbf v_A \cdot \mathbf a_A (\mathbf n \cdot \mathbf v_A )\mathbf v_A \cdot \mathbf v_B \left(8 \bar{\gamma }_{A B}+15\right)
\bigg]\nonumber\\
&+(A\leftrightarrow B)\,,\\
L_{3\rm PN}^{(2)}&=-\frac{G_{AB}^2 m_A^0 m_B^0}{144\, r^2}\bigg[
3 \mathbf v_A^2 (\mathbf n \cdot \mathbf v_A)^2 \big(m^0_A \left(146 \bar{\gamma }_{A B}^2+564 \bar{\gamma }_{A B}-48 \bar{\beta }_B+8 \delta _A+490\right)-3 m^0_B (\left(\bar{\gamma }_{A B}+2\right)^2+4 \delta _B)\big)\nonumber\\
&\qquad +6 \mathbf v_A^2 (\mathbf n \cdot \mathbf v_A) (\mathbf n \cdot \mathbf v_B) \big(2 m^0_A \left(-73 \bar{\gamma }_{A B}^2-246 \bar{\gamma }_{A B}+24 \bar{\beta }_B-4 \delta _A-179\right)+3 m^0_B (\left(\bar{\gamma }_{A B}+2\right)^2+4 \delta _B)\big)\nonumber\\
&\qquad +6 \mathbf v_A^2 (\mathbf n \cdot \mathbf v_B)^2 m^0_A \left(98 \bar{\gamma }_{A B}^2+312 \bar{\gamma }_{A B}+8 \delta _A+235\right)+4 (\mathbf n \cdot \mathbf v_A)^4 m^0_A \left(5 \bar{\gamma }_{A B}^2-24 \bar{\gamma }_{A B}+36 \bar{\beta }_B+20 \delta _A-26\right)\nonumber\\
&\qquad +3 \mathbf v_B^2 (\mathbf n \cdot \mathbf v_A)^2 m^0_A \left(37 \bar{\gamma }_{A B}^2+96 \bar{\gamma }_{A B}-24 \bar{\beta }_B-44 \delta _A+14\right)+12 (\mathbf n \cdot \mathbf v_A)^2 (\mathbf n \cdot \mathbf v_B)^2 m^0_A \big(5 \bar{\gamma }_{A B}^2+24 \bar{\gamma }_{A B}+48 \bar{\beta }_B\nonumber\\
&\qquad +20 \delta _A+70\big) -8 (\mathbf n \cdot \mathbf v_A)^3 (\mathbf n \cdot \mathbf v_B) m^0_A \left(10 \bar{\gamma }_{A B}^2+24 \bar{\gamma }_{A B}+72 \bar{\beta }_B+40 \delta _A+83\right) +24( \mathbf n \cdot \mathbf v_A )(\mathbf n \cdot \mathbf v_B) \mathbf v_A \cdot \mathbf v_B m^0_A \big(97\nonumber\\
&\qquad +26 \bar{\gamma }_{A B}^2+111 \bar{\gamma }_{A B}-24 \bar{\beta }_B+8 \delta _A\big)+6 (\mathbf n \cdot \mathbf v_A)^2 \mathbf v_A \cdot \mathbf v_B m^0_A \left(-149 \bar{\gamma }_{A B}^2-600 \bar{\gamma }_{A B}+72 \bar{\beta }_B-20 \delta _A-529\right)\nonumber\\
&\qquad -3 \mathbf v_A^4 \left(3 m^0_B \left(2 \bar{\beta }_A+15 \bar{\gamma }_{A B}^2+52 \bar{\gamma }_{A B}-4 \delta _B+45\right)+m^0_A \left(122 \bar{\gamma }_{A B}^2+432 \bar{\gamma }_{A B}+8 \delta _A+373\right)\right)\nonumber\\
&\qquad +3 \mathbf v_A^2 \mathbf v_B^2 m^0_A \left(-167 \bar{\gamma }_{A B}^2-564 \bar{\gamma }_{A B}+4 \delta _A-463\right)+6 (\mathbf v_A \cdot \mathbf v_B)^2 m^0_A \left(-125 \bar{\gamma }_{A B}^2-492 \bar{\gamma }_{A B}+24 \bar{\beta }_B-20 \delta _A-463\right)\nonumber\\
&\qquad +6 \mathbf v_A^2 \mathbf v_A \cdot \mathbf v_B \left(m^0_A \left(223 \bar{\gamma }_{A B}^2+816 \bar{\gamma }_{A B}-24 \bar{\beta }_B+28 \delta _A+719\right)+3 m^0_B \left(23 \bar{\gamma }_{A B}^2+84 \bar{\gamma }_{A B}-4 \delta _B+76\right)\right)
\bigg]\nonumber\\
&-\frac{G_{AB}^2 m_A^0 m_B^0}{144\, r}\bigg[
3 \mathbf v_A^2 (\mathbf n \cdot \mathbf a_B) \left(4 m^0_B \left(6 \bar{\beta }_A+25 \bar{\gamma }_{A B}^2+87 \bar{\gamma }_{A B}+4 \delta _B+80\right)+m^0_A \left(98 \bar{\gamma }_{A B}^2+312 \bar{\gamma }_{A B}+8 \delta _A+235\right)\right)\nonumber\\
&\qquad
-9 \mathbf v_A^2( \mathbf n \cdot \mathbf a_A) m^0_A \left(52 \bar{\gamma }_{A B}^2+196 \bar{\gamma }_{A B}+16 \delta _A+185\right)+18 (\mathbf n \cdot \mathbf a_A) \mathbf v_A \cdot \mathbf v_B m^0_A \left(52 \bar{\gamma }_{A B}^2 +196 \bar{\gamma }_{A B}+16 \delta _A+185\right)\nonumber\\
&\qquad -6 \mathbf v_A \cdot \mathbf a_B( \mathbf n \cdot \mathbf v_A) \left(m^0_A \left(-56 \bar{\gamma }_{A B}^2-240 \bar{\gamma }_{A B}+24 \bar{\beta }_B-32 \delta _A-235\right)+m^0_B \left(98 \bar{\gamma }_{A B}^2+312 \bar{\gamma }_{A B}+8 \delta _B+235\right)\right)\nonumber\\
&\qquad +6 (\mathbf n \cdot \mathbf a_B) (\mathbf n \cdot \mathbf v_A)^2 \left(m^0_A \left(\bar{\gamma }_{A B}^2+6 \bar{\gamma }_{A B}+24 \bar{\beta }_B+4 \delta _A+29\right)-2 m^0_B \left(2 \bar{\gamma }_{A B}^2+21 \bar{\gamma }_{A B}+8 \delta _B+34\right)\right)
\bigg]\nonumber\\
&+(A\leftrightarrow B)\,,\\
L_{3\rm PN}^{(3)}&=\frac{G_{AB}^3 m^0_A m^0_B}{24\, \tilde\alpha \, r^3 \left(\bar{\gamma }_{A B}+2\right)}
\bigg[(m^0_A)^2
 \left(-3( \mathbf n \cdot \mathbf v_A)( \mathbf n \cdot \mathbf v_B)+3 (\mathbf n \cdot \mathbf v_A)^2-\mathbf v_A^2+\mathbf v_A \cdot \mathbf v_B\right) \big(11 \bar{\gamma }_{A B} \left(\bar{\gamma }_{A B}+2\right)^2-4 \delta _A \left(\bar{\gamma }_{A B}+10\right)\big)
 \bigg]\nonumber\\
& -\frac{G_{AB}^3 m_A^0 m_B^0}{2304\, r^3}
\bigg[
-576(m_A^0)^2\mathbf v_A^2 \ln(r/r_A) \left(4 \delta _A-11 \left(\bar{\gamma }_{A B}+2\right)^2\right)  +1728 (m_A^0)^2 (\mathbf n \cdot \mathbf v_A)^2 \ln(r/r_A) \left(4 \delta _A-11 \left(\bar{\gamma }_{A B}+2\right)^2\right) \nonumber\\
&\qquad -1728(m_A^0)^2 (\mathbf n \cdot \mathbf v_A)( \mathbf n \cdot \mathbf v_B) \ln(r/r_A) \left(4 \delta _A-11 \left(\bar{\gamma }_{A B}+2\right)^2\right) +576(m_A^0)^2 \mathbf v_A \cdot \mathbf v_B \ln(r/r_A) \left(4 \delta _A-11 \left(\bar{\gamma }_{A B}+2\right)^2\right) \nonumber\\
&\qquad -\mathbf v_A^2\bigg(32(m_A^0)^2 \left(69 \bar{\gamma }_{A B}^3+558 \bar{\gamma }_{A B}^2+1472 \bar{\gamma }_{A B}-18 \epsilon _B+54 \bar{\beta }_B \left(4 \bar{\gamma }_{A B}+5\right)-12 \delta _A \left(5 \bar{\gamma }_{A B}+2\right)+1232\right) \nonumber\\
&\qquad \left.+m^0_A m^0_B\left(-63 \pi ^2 \bar{\gamma }_{A B}^3+2304 \bar{\gamma }_{A B}^3+180 \pi ^2 \bar{\gamma }_{A B}^2+3136 \bar{\gamma }_{A B}^2+1350 \pi ^2 \bar{\gamma }_{A B}-126 \pi ^2 \delta _A \bar{\gamma }_{A B}-126 \pi ^2 \delta _B \bar{\gamma }_{A B}-6656 \bar{\gamma }_{A B}\right.\right.\nonumber\\
&\qquad \left.\left.  +1476 \pi ^2 -252 \pi ^2 \delta _A+8704 \delta _A-252 \pi ^2 \delta _B+3840 \delta _B-2304 \zeta _{A B}+2304 \psi _A+2304 \psi _B+4608 \bar{\beta }_B \left(\bar{\gamma }_{A B}+1\right)\right.\right.\nonumber\\
&\qquad \left.+576 \bar{\beta }_A \left(16 \bar{\gamma }_{A B}+21\right)-9760\right) +192(m_B^0)^2 \left(6 \bar{\gamma }_{A B}^3+25 \bar{\gamma }_{A B}^2+34 \bar{\gamma }_{A B}-\epsilon _A-4 \delta _B \left(2 \bar{\gamma }_{A B}+3\right)+6 \bar{\beta }_A \left(2 \bar{\gamma }_{A B}+3\right)+15\right) \bigg) \nonumber\\
&\qquad+\mathbf v_A \cdot \mathbf v_B  \bigg(32(m^0_A)^2 \left(105 \bar{\gamma }_{A B}^3+711 \bar{\gamma }_{A B}^2+1688 \bar{\gamma }_{A B}-18 \epsilon _B-12 \delta _A \left(9 \bar{\gamma }_{A B}+7\right)+18 \bar{\beta }_B \left(16 \bar{\gamma }_{A B}+23\right)+1340\right) \nonumber\\
&\qquad +m^0_A m^0_B\left(-63 \pi ^2 \bar{\gamma }_{A B}^3+2304 \bar{\gamma }_{A B}^3+180 \pi ^2 \bar{\gamma }_{A B}^2+3136 \bar{\gamma }_{A B}^2+1350 \pi ^2 \bar{\gamma }_{A B}-5504 \bar{\gamma }_{A B}+1476 \pi ^2-1728 \zeta _{A B}+4608 \psi _A\right.\nonumber\\
&\qquad \left.+1728 \bar{\beta }_A \left(8 \bar{\gamma }_{A B}+11\right)-28 \delta _A \left(9 \pi ^2 \left(\bar{\gamma }_{A B}+2\right)-448\right)-7024\right)\bigg)\nonumber\\
&\qquad -3( \mathbf n \cdot \mathbf v_A)( \mathbf n \cdot \mathbf v_B )\bigg(32 (m^0_A)^2 \left(21 \bar{\gamma }_{A B}^3+405 \bar{\gamma }_{A B}^2+1400 \bar{\gamma }_{A B}-18 \epsilon _B-60 \delta _A \left(\bar{\gamma }_{A B}+1\right)+18 \bar{\beta }_B \left(8 \bar{\gamma }_{A B}+7\right)+1316\right)\nonumber\\
&\qquad \left. + m^0_A m^0_B\left(-63 \pi ^2 \bar{\gamma }_{A B}^3+768 \bar{\gamma }_{A B}^3+180 \pi ^2 \bar{\gamma }_{A B}^2-3008 \bar{\gamma }_{A B}^2+1350 \pi ^2 \bar{\gamma }_{A B}-15296 \bar{\gamma }_{A B}+1476 \pi ^2-1344 \zeta _{A B}+4608 \psi _A\right.\right.\nonumber\\
&\qquad \left.+192 \bar{\beta }_A \left(56 \bar{\gamma }_{A B}+75\right)-28 \delta _A \left(9 \pi ^2 \left(\bar{\gamma }_{A B}+2\right)-448\right)-14224\right)\bigg)\nonumber\\
&\qquad +3 (\mathbf n \cdot \mathbf v_A)^2 \bigg(32(m_A^0)^2 \left(27 \bar{\gamma }_{A B}^3+441 \bar{\gamma }_{A B}^2+1472 \bar{\gamma }_{A B}-12 \epsilon _B-12 \delta _A \left(3 \bar{\gamma }_{A B}+1\right)+18 \bar{\beta }_B \left(8 \bar{\gamma }_{A B}+9\right)+1370\right) \nonumber\\
&\qquad+m^0_A m^0_B\left(-63 \pi ^2 \bar{\gamma }_{A B}^3+768 \bar{\gamma }_{A B}^3+180 \pi ^2 \bar{\gamma }_{A B}^2-3008 \bar{\gamma }_{A B}^2+1350 \pi ^2 \bar{\gamma }_{A B}-126 \pi ^2 \delta _A \bar{\gamma }_{A B}-126 \pi ^2 \delta _B \bar{\gamma }_{A B}-14528 \bar{\gamma }_{A B}\right.\nonumber\\
&\qquad +1476 \pi ^2-252 \pi ^2 \delta _A+8704 \delta _A-252 \pi ^2 \delta _B+3840 \delta _B-768 \zeta _{A B}+2304 \psi _A+2304 \psi _B+1536 \bar{\beta }_B \left(2 \bar{\gamma }_{A B}+3\right)\nonumber\\
&\qquad\left. +192 \bar{\beta }_A \left(40 \bar{\gamma }_{A B}+63\right)-12256\right) -96(m_B^0)^2 \left(2 \bar{\gamma }_{A B}+3\right) \big(\left(\bar{\gamma }_{A B}+2\right)^2+4 \delta _B\big) \bigg)
\bigg]\nonumber\\
&+(A\leftrightarrow B)\,,\label{eq:2bodyL3PN3}\\
L_{3\rm PN}^{(4)}&=-\frac{G_{A B}^4 (m^0_A)^3 (m^0_B)^2}{12 \tilde\alpha\,  r^4\left(\bar{\gamma }_{A B}+2\right)} 
\bigg[11 \bar{\gamma }_{A B} \left(\bar{\gamma }_{A B}+2\right)^2-4 \delta _A \left(\bar{\gamma }_{A B}-5\right)\bigg]\nonumber\\
&-\frac{G_{AB}^4 (m_A^0)^3 m_B^0}{144\, r^4}
\bigg[
36 m^0_B \ln(r/r_A) \left(4 \delta _A-11 \left(\bar{\gamma }_{A B}+2\right)^2\right)\nonumber\\
&\qquad +6 m^0_A \left(4 \bar{\beta }_B \left(\bar{\gamma }_{A B}^2+4 \bar{\gamma }_{A B}+4 \delta _A+7\right)+2 \bar{\gamma }_{A B}^2+8 \bar{\gamma }_{A B}+12 \bar{\beta }_B^2+8 \delta _A+8 \kappa _B+4 \epsilon _B+9\right)\nonumber\\
&\qquad+m^0_B \left(24 \bar{\beta }_A \left(\bar{\gamma }_{A B}^2+4 \bar{\gamma }_{A B}+12 \bar{\beta }_B+4 \delta _A+16\right)+36 \bar{\beta }_B \left(32 \bar{\gamma }_{A B}+79\right)+12 \delta _A \bar{\gamma }_{A B}-33 \bar{\gamma }_{A B}^3+624 \bar{\gamma }_{A B}^2+3340 \bar{\gamma }_{A B}\right.\nonumber\\
&\qquad \left.+288 \bar{\beta }_B^2+288 \zeta _{A B}+136 \delta _A+72 \xi _A+72 \omega_A+96 \psi _A+200 \delta _B+72 \epsilon _B+4008\right)
\bigg]\nonumber\\
&+(A\leftrightarrow B)\,.\label{eq:2bodyL3PN4}
\end{align}\label{eq:2bodyL3PN}%
\end{subequations}
\end{widetext}

\section{Contact transformations\label{app:contactTransfo}}

The order-reduced two-body Lagrangian \eqref{eq:lagRedTwoBody} is obtained by replacing the accelerations by their on-shell expressions at 1PN:
\begin{widetext}
\begin{align}
(a_F)_A^{i} &= \frac{G_{AB}m_B^0}{r^2}\bigg[\,n^i\bigg(\!-1+(5+2\bar\beta_B+2\bar\gamma_{AB})\frac{G_{AB}m_A^0}{r}+(4+2\bar\beta_A+2\bar\gamma_{AB})\frac{G_{AB}m_B^0}{r}+\frac{3}{2}(\mathbf n\cdot \mathbf v_B)^2\\
&-(1+\bar\gamma_{AB})\mathbf v_A^2-(2+\bar\gamma_{AB})(\mathbf v_B^2-2\mathbf v_A\cdot\mathbf v_B)\bigg)+(\mathbf v_A^i-\mathbf v_B^i)\big((4+2\bar\gamma_{AB})(\mathbf n\cdot\mathbf v_A)-(3+2\bar\gamma_{AB})(\mathbf n\cdot\mathbf v_B)\big)\bigg]\,,\nonumber
\end{align}
and $(A\leftrightarrow B)$.

The six-by-six Hessian matrix associated with $F=L_{0\rm PN}+L_{1\rm PN}$ defined in Eq.~\eqref{eq:Hessian1PNdef} is
\begin{align}
(H_F)^{Ci}_{Dj}=m_C^0\delta_{CD}&\left[\delta_{ij}\left(1+\frac{1}{2}\mathbf v_C^2+(3+2\bar\gamma_{AB})\frac{G_{AB}(M-m_C^0)}{r}\right)+v_C^i v_C^j\right]\nonumber\\
&-\frac{1}{2}\frac{G_{AB}m_A^0m_B^0}{r}\left(\delta_{AC}\delta_{BD}+\delta_{BC}\delta_{AD}\right)\left(n^i n^j+(7+4\bar\gamma_{AB})\delta_{ij}\right)\,,
\end{align}
with inverse
\begin{align}
(H^{-1}_F)^{Ek}_{Fl}=\frac{1}{m_E}\delta_{EF}&\left[\delta_{kl}\left(1-\frac{1}{2}\mathbf v_E^2-(3+2\bar\gamma_{AB})\frac{G_{AB}(M-m_E^0)}{r}\right)+v_E^k v_E^l\right]\nonumber\\
&+\frac{1}{2}\frac{G_{AB}}{r}\left(\delta_{AE}\delta_{BF}+\delta_{BE}\delta_{AF}\right)\left(n^k n^l+(7+4\bar\gamma_{AB})\delta_{kl}\right)\,.
\end{align}
The contact transformation defined by Eqs.~\eqref{eq:Hessian1PNdefComplete} then reads $\delta \mathbf x_A=\delta \mathbf x_A^{2\rm PN}+\delta \mathbf x_A^{3\rm PN}$, with
\begin{subequations}
\begin{align}
\delta \mathbf x_A^{2\rm PN}&=-\frac{1}{8}G_{AB}m_B^0\left(2(7+4\bar\gamma_{AB})(\mathbf n\cdot\mathbf v_B) \mathbf v_B+(\mathbf n\cdot\mathbf v_B)^2\mathbf n-(7+4\bar\gamma_{AB}) \mathbf v_B^2 \mathbf n\right)\,,\\
\delta \mathbf x_A^{3\rm PN}&=\frac{G_{AB}^2 m_B^0}{48 r}\Big[
2 \mathbf n (\mathbf n\cdot\mathbf v_B)^2 \left(m^0_B \left(24 \bar{\beta }_A+\bar{\gamma }_{A B}^2+12 \bar{\gamma }_{A B}+4 \delta _B+38\right)-2 m^0_A \left(2 \bar{\gamma }_{A B}^2+21 \bar{\gamma }_{A B}+8 \delta _A+34\right)\right)\nonumber\\
&\qquad-2 \mathbf v_B (\mathbf n\cdot\mathbf v_B) \left(m^0_B \left(24 \bar{\beta }_A-104 \bar{\gamma }_{A B}^2-396 \bar{\gamma }_{A B}-32 \delta _B-361\right)+m^0_A \left(98 \bar{\gamma }_{A B}^2+312 \bar{\gamma }_{A B}+8 \delta _A+235\right)\right)\nonumber\\
&\qquad+6 \mathbf n (\mathbf n\cdot\mathbf v_A)^2 m^0_A \left(6 \bar{\gamma }_{AB}+11\right)+6 \mathbf v_A (\mathbf n\cdot\mathbf v_A) m^0_A \left(4 \bar{\gamma }_{A B}+7\right)^2+3 \mathbf n \mathbf v_A^2 m^0_A \left(36 \bar{\gamma }_{A B}^2+136 \bar{\gamma }_{A B}+16 \delta _A+129\right)\nonumber\\
&\qquad+\mathbf n \mathbf v_B^2 m^0_B \left(50 \bar{\gamma }_{A B}^2+156 \bar{\gamma }_{A B}+8 \delta _B+109\right)+4 \mathbf n \mathbf v_B^2 m^0_A \left(25 \bar{\gamma }_{A B}^2+87 \bar{\gamma }_{A B}+6 \bar{\beta }_B+4 \delta _A+80\right)\nonumber\\
&\qquad-6 \mathbf n (\mathbf v_A\cdot\mathbf v_B) m^0_A \left(52 \bar{\gamma }_{A B}^2+196 \bar{\gamma }_{A B}+16 \delta _A+185\right)
\Big] \nonumber \\
&+\frac{G_{AB} m_B^0}{48}\Big[
-12 \mathbf n (\mathbf n\cdot \mathbf v_A) (\mathbf n\cdot \mathbf v_B)^3 \left(4 \bar{\gamma }_{A B}+7\right)+24 \mathbf n (\mathbf n\cdot \mathbf v_B)^2 \mathbf v_A\cdot\mathbf v_B \left(6 \bar{\gamma }_{A B}+11\right)+9 \mathbf n (\mathbf n\cdot \mathbf v_A)^2 (\mathbf n\cdot \mathbf v_B)^2 \left(8 \bar{\gamma }_{A B}+15\right)\nonumber\\
&\qquad-3 \mathbf n \mathbf v_A^2 (\mathbf n\cdot \mathbf v_B)^2 \left(14 \bar{\gamma }_{A B}+27\right)+72 \mathbf n \mathbf v_B^2 (\mathbf n\cdot \mathbf v_A)(\mathbf n\cdot \mathbf v_B) \left(\bar{\gamma }_{A B}+2\right)+6 \mathbf n \mathbf v_A^2 (\mathbf n\cdot \mathbf v_A)( \mathbf n\cdot \mathbf v_B) \left(2 \bar{\gamma }_{A B}+5\right)\nonumber\\
&\qquad-12 \mathbf n (\mathbf n\cdot \mathbf v_A) (\mathbf n\cdot \mathbf v_B )\mathbf v_A\cdot\mathbf v_B \left(22 \bar{\gamma }_{A B}+41\right)-12 \mathbf n (\mathbf n\cdot \mathbf v_A)^2 \mathbf v_A\cdot\mathbf v_B \left(3 \bar{\gamma }_{A B}+5\right)-6 \mathbf n \mathbf v_B^2 (\mathbf n\cdot \mathbf v_A)^2 \left(13 \bar{\gamma }_{A B}+27\right)\nonumber\\
&\qquad -6 \mathbf n \mathbf v_B^2 (\mathbf n\cdot \mathbf v_B)^2 \left(2 \bar{\gamma }_{A B}+5\right)-6 \mathbf v_A (\mathbf n\cdot \mathbf v_A) (\mathbf n\cdot \mathbf v_B)^2 \left(14 \bar{\gamma }_{A B}+27\right)+288 \mathbf v_A (\mathbf n\cdot \mathbf v_B )\mathbf v_A\cdot\mathbf v_B \bar{\gamma }_{A B}\nonumber\\
&\qquad +6 \mathbf v_A (\mathbf n\cdot \mathbf v_A)^2 (\mathbf n\cdot \mathbf v_B) \left(2 \bar{\gamma }_{A B}+5\right)-12 \mathbf v_A^2 \mathbf v_A (\mathbf n\cdot \mathbf v_B) \left(6 \bar{\gamma }_{A B}+11\right)+12 \mathbf v_A (\mathbf n\cdot \mathbf v_A) \mathbf v_A\cdot\mathbf v_B \left(8 \bar{\gamma }_{A B}+15\right)\nonumber\\
&\qquad+12 \mathbf v_A \mathbf v_B^2 (\mathbf n\cdot \mathbf v_A) \left(22 \bar{\gamma }_{A B}+45\right)+24 \mathbf v_B (\mathbf n\cdot \mathbf v_A) (\mathbf n\cdot \mathbf v_B)^2 \left(6 \bar{\gamma }_{A B}+11\right)-96 \mathbf v_B (\mathbf n\cdot \mathbf v_B) \mathbf v_A\cdot\mathbf v_B \bar{\gamma }_{A B}\nonumber\\
&\qquad-6 \mathbf v_B (\mathbf n\cdot \mathbf v_A)^2 (\mathbf n\cdot \mathbf v_B) \left(22 \bar{\gamma }_{A B}+41\right)+6 \mathbf v_A^2 \mathbf v_B (\mathbf n\cdot \mathbf v_B) \left(24 \bar{\gamma }_{A B}+47\right)-144 \mathbf v_B^2 \mathbf v_B (\mathbf n\cdot \mathbf v_A) \left(\bar{\gamma }_{A B}+2\right)\nonumber\\
&\qquad +24 \mathbf v_B (\mathbf n\cdot \mathbf v_A) \mathbf v_A\cdot\mathbf v_B \left(4 \bar{\gamma }_{A B}+7\right)-4 \mathbf v_B (\mathbf n\cdot \mathbf v_A)^3 \left(3 \bar{\gamma }_{A B}+5\right)+6 \mathbf v_A^2 \mathbf v_B (\mathbf n\cdot \mathbf v_A) \left(8 \bar{\gamma }_{A B}+15\right)\nonumber\\
&\qquad+8 \mathbf v_A (\mathbf n\cdot \mathbf v_B)^3 \left(5 \bar{\gamma }_{A B}+11\right)-240 \mathbf v_A \mathbf v_B^2 (\mathbf n\cdot \mathbf v_B) \bar{\gamma }_{A B}+8 \mathbf v_B (\mathbf n\cdot \mathbf v_B)^3 \left(3 \bar{\gamma }_{A B}+5\right)-96 \mathbf v_B^2 \mathbf v_B( \mathbf n\cdot \mathbf v_B) \bar{\gamma }_{A B}\nonumber\\
&\qquad-144 \mathbf n \mathbf v_B^2 \mathbf v_A\cdot\mathbf v_B \left(\bar{\gamma }_{A B}+2\right)+12 \mathbf n (\mathbf v_A\cdot\mathbf v_B)^2 \left(4 \bar{\gamma }_{A B}+7\right)+6 \mathbf n \mathbf v_A^2 \mathbf v_A\cdot\mathbf v_B \left(8 \bar{\gamma }_{A B}+15\right)+6 \mathbf n \mathbf v_A^2 \mathbf v_B^2 \left(22 \bar{\gamma }_{A B}+45\right)\nonumber\\
&\qquad+6 \mathbf n (\mathbf v_B)^4 \left(6 \bar{\gamma }_{A B}+11\right)+3 \mathbf n (\mathbf n\cdot \mathbf v_B)^4+564 \mathbf v_A (\mathbf n\cdot \mathbf v_B) \mathbf v_A\cdot\mathbf v_B-156 \mathbf v_B( \mathbf n\cdot \mathbf v_B) \mathbf v_A\cdot\mathbf v_B-492 \mathbf v_A \mathbf v_B^2 (\mathbf n\cdot \mathbf v_B)\nonumber\\
&\qquad-180 \mathbf v_B^2 \mathbf v_B (\mathbf n\cdot \mathbf v_B)\Big]\nonumber\\
&-\frac{4M\mathcal A_0^2}{3}(G_{AB}m_A^0m_B^0)\alpha_A^0(\alpha_A^0-\alpha_B^0)\,\underset{2r}{\rm PF}\!\int_{\Bbb R}\frac{d\tau}{|\tau|}\left.\left(\frac{\mathbf n}{r^2}\right)\right|_{t+\tau}\,,
\end{align}\label{eq:appendixContactExplicit}
\end{subequations}
and $(A\leftrightarrow B)$.
\end{widetext}

\section{Two-body Hamiltonian at 3PN order\label{app:twoBodHamiltonian}}

The contributions to the center-of-mass frame, two-body Hamiltonian~\eqref{eq:2bodyHamiltonian} are, up to 2PN:
\begin{widetext}
\begin{subequations}
\begin{align}
\hat H_{0\rm PN}&=\frac{\hat p^2}{2}-\frac{G_{AB}}{\hat r}\,,\label{eq:2bodyHamiltonian0PN}\\
\hat H_{1\rm PN}&=-\frac{1}{8}(1-3\nu)\hat p^4-\frac{G_{AB}}{2\hat r}\left[\nu\, \hat p_r^2+(3+2\bar\gamma_{AB}+\nu)\hat p^2\right]+\frac{G_{AB}^2}{2\hat  r^2}\left[1+\bar\beta_+-m_-\bar\beta_-\right]\,,\\
\hat H_{2\rm PN}&=\frac{1}{16}(1-5\nu+5\nu^2)\hat p^6+\frac{G_{AB}}{8\hat r}\left[\left(5-22\nu-3\nu^2+4\bar\gamma_{AB}(1-4\nu)\right)\hat p^4+2(1-\nu)\nu\hat p^2\hat p_r^2-3\nu^2\hat p_r^4\right]\nonumber\\
&+\frac{G_{AB}^2}{8\hat r^2}\bigg[\left(22+2\delta_++28\bar\gamma_{AB}+9\bar\gamma_{AB}^2+58\nu+36\bar\gamma_{AB}\nu-2\delta_+(1+\nu)+2m_-(\bar\beta_-+\delta_--\bar\beta_-\nu)\right)\hat p^2\nonumber\\
&\qquad-\left(2\delta_++\bar\gamma_{AB}^2+4\bar\gamma_{AB}(1+6\nu)+4(1+8\nu-3\bar\beta_+\nu)+2m_-(\delta_-+2\bar\beta_-\nu)\right)\hat p_r^2\bigg]\nonumber\\
&-\frac{G_{AB}^3}{12\hat r^3}\bigg[6+\epsilon_++4\bar\gamma_{AB}+\bar\gamma_{AB}^2+45\nu-2\epsilon_+\nu+16\bar\gamma_{AB}\nu-2\bar\gamma_{AB}^2\nu+6\bar\beta_+(1+2\nu)+\delta_+(2-4\nu)+12\zeta_{AB}\nu\nonumber\\
&\qquad-m_-(6\bar\beta_--2\delta_-+\epsilon_-)\bigg]\,.
\end{align}
\end{subequations}
The contributions $\hat H_{3\rm PN}^{(i)}$ at 3PN level~\eqref{eq:2bodyHamiltonian3PN} are respectively proportional to $G_{AB}^i$ and read:
\begin{subequations}
\begin{align}
\hat H_{\rm 3PN}^{(0)}&=\frac{5}{128}(-1+7\nu-14\nu^2+7\nu^3)\hat p^8\,,\\
\hat H_{\rm 3PN}^{(1)}&=\frac{G_{AB}}{16\,\hat r}\bigg[-\left(2 \bar\gamma_{AB}  (93 \nu ^2-23 \nu +3)+5 \nu ^3+331 \nu ^2-64 \nu +7\right)\hat p^6\nonumber\\
&\qquad +\left(-27+4 \bar\gamma_{AB}  (89 \nu -4)-3 \nu ^2+676 \nu \right)\nu\,\hat p^4\hat p_r^2-\left(-5+3 \nu ^2+468 \nu+6 \bar\gamma_{AB}  (43 \nu -1) \right)\nu\, \hat p^2\hat p_r^4\nonumber\\
&\qquad+5(15+8 \bar\gamma_{AB} -\nu )\nu ^2\hat p_r^6\bigg]\,,\\
\hat H_{\rm 3PN}^{(2)}&=\frac{G_{AB}^2}{48\,\hat r^2}\bigg[
\Big(-3 \bar\beta_+ \left(\nu ^2+3 \nu -3\right)-308 \bar\gamma_{AB} ^2 \nu ^2+199 \bar\gamma_{AB} ^2 \nu -51 \bar\gamma_{AB} ^2-348 \bar\gamma_{AB}  \nu ^2+402 \bar\gamma_{AB}  \nu -132 \bar\gamma_{AB}\nonumber\\
&\qquad -2 \delta_+ \left(20 \nu ^2-7 \nu +3\right)+395 \nu ^2+164 \nu -87-m_-(\delta_-(6-34\nu)+\bar\beta_-(9\nu^2-39\nu+9))\Big)\hat p^4\nonumber\\
&\qquad+\Big(
\left(596 \nu ^2+137 \nu +3\right) \bar{\gamma }_{AB}^2+6 \left(504 \nu ^2+71 \nu +2\right) \bar{\gamma }_{AB}-36 \nu ^2 \bar{\beta }_+-72 \nu  \bar{\beta }_++2 \delta _+ \left(20 \nu ^2-7 \nu +3\right)\nonumber\\
&\qquad+3556 \nu ^2+211 \nu +12-2m_- \left(6 \nu  (\nu +2) \bar{\beta }_-+\delta _- (17 \nu -3)\right)
\Big)\hat p^2\hat p_r^2\nonumber\\
&\qquad+\left(\frac{1}{3}\nu  \left(80 \nu  \bar{\gamma }_{AB}^2-504 \nu  \bar{\gamma }_{AB}+20 \bar{\gamma }_{AB}^2+228 \bar{\gamma }_{AB}+72 (5 \nu +1) \bar{\beta }_++40 \delta _+ (4 \nu +1)-581 \nu +454\right)\right.\nonumber\\
&\qquad\left.-\frac{8}{3}\nu  m_- \left(9 (\nu -1) \bar{\beta }_-+5 \delta _-\right)\right)\hat p_r^4
\bigg]\,,\\
\hat H_{\rm 3PN}^{(3)}&=\frac{G_{AB}^3}{\hat r^3}\bigg[
\bigg(\nu\frac{  11 \bar{\gamma }_{AB} \left(\bar{\gamma }_{AB}+2\right)^2-2 \delta _+ \left(\bar{\gamma }_{AB}+10\right)}{24 \tilde\alpha  \left(\bar{\gamma }_{AB}+2\right)}-\frac{\delta _- \nu  m_- \left(\bar{\gamma }_{AB}+10\right)}{12 \tilde\alpha  \left(\bar{\gamma }_{AB}+2\right)}-\frac{1}{8} \bar{\beta }_+ \left(20 \nu  \bar{\gamma }_{AB}-4 \bar{\gamma }_{AB}+4 \nu ^2+17 \nu -6\right)\nonumber\\
&\qquad+\frac{7}{128} \pi ^2 \delta _+ \nu  \bar{\gamma }_{AB}-\frac{7}{12} \delta _+ \nu  \bar{\gamma }_{AB}-\frac{2}{3} \delta _+ \bar{\gamma }_{AB}+\frac{13}{2} \nu ^2 \bar{\gamma }_{AB}^2+\frac{58}{3} \nu ^2 \bar{\gamma }_{AB}+\frac{7}{256} \pi ^2 \nu  \bar{\gamma }_{AB}^3-\frac{11}{24} \nu  \bar{\gamma }_{AB}^3-\frac{5}{64} \pi ^2 \nu  \bar{\gamma }_{AB}^2\nonumber\\
&\qquad-\frac{923}{72} \nu  \bar{\gamma }_{AB}^2-\frac{75}{128} \pi ^2 \nu  \bar{\gamma }_{AB}-\frac{1421 \nu  \bar{\gamma }_{AB}}{36}-\bar{\gamma }_{AB}^3-\frac{29 \bar{\gamma }_{AB}^2}{6}-\frac{47 \bar{\gamma }_{AB}}{6}-\frac{1}{4} \nu  \ln(\hat r) (2 \delta _+-11 \left(\bar{\gamma }_{AB}+2\right)^2)\nonumber\\
&\qquad -\frac{11}{8} \ln_+ \nu  \bar{\gamma }_{AB}^2-\frac{11}{2} \ln_+ \nu  \bar{\gamma }_{AB}-\frac{\nu ^2 \zeta _{AB}}{2}+\nu  \zeta _{AB}+\delta _+ \nu ^2+\frac{7}{64} \pi ^2 \delta _+ \nu -\frac{143 \delta _+ \nu }{36}-\delta _+ +\frac{1}{4} \delta _+ \ln_+ \nu -\frac{11 \ln_+ \nu }{2}\nonumber\\
&\qquad+\frac{157 \nu ^2}{12}-\nu  \psi _+-\frac{1531 \nu }{48}-\frac{41 \pi ^2 \nu }{64}+\frac{\nu ^2 \epsilon _+}{12}-\frac{\nu  \epsilon _+}{24}+\frac{\epsilon _+}{24}-\frac{17}{4}+\frac{1}{4} \delta _- \ln_- \nu\nonumber\\
&\qquad-\frac{m_-}{72}\Big(-9 \bar{\beta }_- \left(28 \nu  \bar{\gamma }_{AB}-4 \bar{\gamma }_{AB}+37 \nu -6\right)-6 \delta _- \nu  \bar{\gamma }_{AB}+48 \delta _- \bar{\gamma }_{AB}+99 \ln_- \nu  \bar{\gamma }_{AB}^2+396 \ln_- \nu  \bar{\gamma }_{AB}+14 \delta _- \nu \nonumber\\
&\qquad +72 \delta _-+36 \delta _- \nu  \ln(\hat r)-18 \delta _+ \ln_- \nu +396 \ln_- \nu -18 \delta _- \ln_+ \nu +3 \nu  \epsilon _-+3 \epsilon _-\Big)\bigg)\hat p^2\nonumber\\
&\qquad+\bigg(\nu\frac{ 2 \delta _+ \left(\bar{\gamma }_{AB}+10\right)-11 \bar{\gamma }_{AB} \left(\bar{\gamma }_{AB}+2\right)^2}{8 \tilde\alpha  \left(\bar{\gamma }_{AB}+2\right)}+\frac{\delta _- \nu  m_- \left(\bar{\gamma }_{AB}+10\right)}{4 \tilde\alpha  \left(\bar{\gamma }_{AB}+2\right)}+5 \nu  \bar{\beta }_+ \bar{\gamma }_{AB}+\frac{27}{2} \nu ^2 \bar{\gamma }_{AB}^2+53 \nu ^2 \bar{\gamma }_{AB}\nonumber\\
&\qquad-\frac{1}{256} 21 \pi ^2 \nu  \bar{\gamma }_{AB}^3+\frac{23}{8} \nu  \bar{\gamma }_{AB}^3+\frac{15}{64} \pi ^2 \nu  \bar{\gamma }_{AB}^2+\frac{779}{24} \nu  \bar{\gamma }_{AB}^2+\frac{225}{128} \pi ^2 \nu  \bar{\gamma }_{AB}+\frac{1103 \nu  \bar{\gamma }_{AB}}{12}+\frac{\bar{\gamma }_{AB}^3}{4}+\frac{11 \bar{\gamma }_{AB}^2}{8}+\frac{5 \bar{\gamma }_{AB}}{2}\nonumber\\
&\qquad+\frac{3}{4} \nu  \ln(\hat r) (2 \delta _+-11 \left(\bar{\gamma }_{AB}+2\right)^2)+\frac{1}{128} \delta _+\left(\left(96-21 \pi ^2\right) \nu +64\right) \bar{\gamma }_{AB}-\frac{1}{192}\delta _+ \left(144 \ln_++63 \pi ^2-1904\right) \nu \nonumber\\
&\qquad +3 \delta _+\nu ^2+\frac{3}{4}\delta _++\frac{33}{8} \ln_+ \nu  \bar{\gamma }_{AB}^2+\frac{33}{2} \ln_+ \nu  \bar{\gamma }_{AB}-3 \nu ^2 \bar{\beta }_++\frac{47 \nu  \bar{\beta }_+}{8}-\frac{3 \nu ^2 \zeta _{AB}}{2}-\nu  \zeta _{AB} +\frac{33 \ln_+ \nu }{2}+\frac{197 \nu ^2}{4}\nonumber\\
&\qquad +3 \nu  \psi _+ +\frac{123 \pi ^2 \nu }{64}+\frac{309 \nu }{4}+\frac{\nu ^2 \epsilon _+}{4}-\frac{3 \nu  \epsilon _+}{8}+\frac{3}{2}-\frac{3}{4} \delta _- \ln_- \nu+\frac{m_- }{24}\Big(-3 \nu  \bar{\beta }_- \left(40 \bar{\gamma }_{AB}+47\right)\nonumber\\
&\qquad+9 \nu  (\ln_- (11 \left(\bar{\gamma }_{AB}+2\right)^2-2 \delta _+)+\epsilon _-)-2 \delta _- \left(3 (\nu -2) \bar{\gamma }_{AB}+9\ln_+ \nu +35 \nu -9\right)+36 \delta _- \nu  \ln(\hat r)\Big)\bigg)\hat p_r^2
\bigg]\,,\\
\hat H_{\rm 3PN}^{(4)}&=\frac{G_{AB}^4}{144\,\hat r^4}\bigg[
12\nu\frac{ 11 \bar{\gamma }_{AB} \left(\bar{\gamma }_{AB}+2\right)^2-2 \delta _+ \left(\bar{\gamma }_{AB}-5\right)}{ \tilde\alpha  \left(\bar{\gamma }_{AB}+2\right)}
-\frac{24\delta _- \nu  m_- \left(\bar{\gamma }_{AB}-5\right)}{ \tilde\alpha  \left(\bar{\gamma }_{AB}+2\right)}
-24 \nu  \bar{\beta }_+ \bar{\gamma }_{AB}^2+480 \nu  \bar{\beta }_+ \bar{\gamma }_{AB}+12 \bar{\beta }_+ \bar{\gamma }_{AB}^2\nonumber\\
&\qquad+48 \bar{\beta }_+ \bar{\gamma }_{AB}+6 \delta _+ \nu  \bar{\gamma }_{AB}-33 \nu  \bar{\gamma }_{AB}^3+588 \nu  \bar{\gamma }_{AB}^2+3196 \nu  \bar{\gamma }_{AB}+12 \bar{\gamma }_{AB}^2+48 \bar{\gamma }_{AB}+36 \nu \ln(\hat r) (2 \delta _+-11 \left(\bar{\gamma }_{AB}+2\right)^2)\nonumber\\
&\qquad+198 \ln_+ \nu  \bar{\gamma }_{AB}^2+792\ln_+ \nu  \bar{\gamma }_{AB}+24 \delta _- (4 \nu -1) \bar{\beta }_--48 \delta _+ \nu  \bar{\beta }_++24 \delta _+ \bar{\beta }_++(18-54 \nu ) \bar{\beta }_-^2+90 \nu  \bar{\beta }_+^2+1362 \nu  \bar{\beta }_+\nonumber\\
&\qquad+18 \bar{\beta }_+^2+84 \bar{\beta }_++288 \nu  \zeta _{AB}+96 \delta _+ \nu +24 \delta _+-72 \kappa _+ \nu +24 \kappa _+-36 \delta _+\ln_+ \nu +792\ln_+ \nu +36 \nu  \xi _+\nonumber\\
&\qquad +48 \nu  \psi _++3846 \nu +36 \nu  w_++12 \epsilon _++54-36 \delta _-\ln_- \nu 
+2m_- \Big(3 \delta _- \nu  \bar{\gamma }_{AB}+99\ln_- \nu  \bar{\gamma }_{AB}^2+396\ln_- \nu  \bar{\gamma }_{AB}\nonumber\\
&\qquad -3 \bar{\beta }_- \left(-4 \nu  \bar{\gamma }_{AB}^2+80 \nu  \bar{\gamma }_{AB}+2 \bar{\gamma }_{AB}^2+8 \bar{\gamma }_{AB}+18 \nu  \bar{\beta }_++6 \bar{\beta }_+-8 \delta _+ \nu +4 \delta _++191 \nu +14\right)+12 \delta _- \bar{\beta }_+-28 \delta _- \nu\nonumber\\
&\qquad +12 \delta _-+36 \delta _- \nu \ln(\hat r)+12 \kappa _- \nu -12 \kappa _--18 \delta _+\ln_- \nu +396\ln_- \nu -18 \delta _-\ln_+ \nu +18 \nu  \xi _-+24 \nu  \psi _-\nonumber\\
&\qquad+18 \nu  w_--12 \nu  \epsilon _--6 \epsilon _-\Big)
\bigg]\,.
\end{align}
\end{subequations}
\end{widetext}

\section{Generating functions \label{app:generatingFunctionCoeffs}}

In the GR limit described below Eqs.~\eqref{eq:combinations}, the part $\hat H^{\rm I}$ of the two-body Hamiltonian is equal to the 3PN (ADM) Hamiltonian of Ref.~\cite{Damour:2014jta} modulo the canonical transformation built in Sec.~\ref{sec:local}. Its nonzero coefficients yield a coordinate change at 2PN and 3PN, and they read, respectively:
\begin{subequations}
\begin{align}
\gamma_{101}&=\frac{\nu}{4}\,,\\
 \gamma_{002}&=-\frac{1}{4}-3\nu\,,
 \end{align}
 \end{subequations}
\vspace*{-0.5cm}
\begin{subequations}
\begin{align}
\gamma_{201}&=\frac{\nu}{4}(-5+68\nu)\,,\\
 \gamma_{111}&=\frac{\nu}{48}(5-396\nu)\,,\\
\gamma_{021}&=\frac{15\nu^2}{16}\,,\\
\gamma_{102}&=\frac{\nu}{24}(92+491\nu)\,,\\
\gamma_{012}&=\frac{\nu}{144}(91+124\nu)\,,\\
\gamma_{003}&=\frac{\nu}{32}\big[3(156+7\pi^2)
+176(\ln_++m_-\ln_-)\big]\,,\\
\gamma_{003}^{\ln}&=-11\nu\,,
\end{align}
\end{subequations}
where $\gamma_{003}^{\ln}$ allows us to eliminate the $\ln\hat r$-dependent terms.

In ST-ESGB gravity, the Hamiltonian can be identified to its EOB counterpart $\hat H_{\rm EOB}^{\rm I}$ modulo a coordinate change at 1PN, 2PN and 3PN. The nonzero coefficients of the canonical transformation are now, respectively:
\begin{subequations}
\begin{align}
\gamma_{100}&=-\frac{\nu}{2}\,,\\
\gamma_{001}&=\frac{G_{AB}}{2}(2+2\bar\gamma_{AB}+\nu)\,,
\end{align}
\end{subequations}
\vspace*{-0.5cm}
\begin{subequations}
\begin{align}
\gamma_{200}&=\frac{\nu}{8}(1-\nu)\,,\\
\gamma_{020}&=\frac{\nu^2}{2}\,,\\
\gamma_{101}&=\frac{G_{AB}\,\nu}{8}(12+8\bar\gamma_{AB}-\nu)\,,\\
\gamma_{011}&=-\frac{G_{AB}\,\nu^2}{8}\,,\\
\gamma_{002}&=\frac{G_{A B}^2}{8} \big(-24 \nu  \bar{\gamma }_{A B}+3 \bar{\gamma }_{A B}^2+4 \bar{\gamma }_{A B}+4 (\nu +1) \bar{\beta }_+\nonumber\\
&-4 m_- \bar{\beta }_--2 \delta _++2 \nu ^2-2 \delta _- m_--38 \nu \big)\,,
\end{align}
\end{subequations}
\vspace{-1cm}
\begin{widetext}
\begin{subequations}
\begin{align}
\gamma_{300}&=-\frac{\nu}{16}(1-3\nu+\nu^2)\,,\\
\gamma_{210}&=\frac{\nu^2}{24}(3-4\nu)\,,\\
\gamma_{120}&=\frac{\nu^2}{12}(7\nu-6)\,,\\
\gamma_{030}&=-\frac{2 \nu ^3}{3}\,,\\
\gamma_{201}&=-\frac{G_{A B} \,\nu}{16}   \left(-2 (77 \nu -9) \bar{\gamma }_{A B}+\nu ^2-292 \nu +29\right)\,,\\
\gamma_{111}&=-\frac{  G_{A B}\,\nu}{48}  \left(2 (97 \nu -3) \bar{\gamma }_{A B}+4 \nu ^2+360 \nu -5\right)\,,\\
\gamma_{021}&=\frac{G_{A B} \nu ^2 }{16} \left(-24 \bar{\gamma }_{A B}+5 \nu -33\right)\,,\\
\gamma_{102}&=\frac{G_{A B}^2\,\nu}{48}   \big(308 \nu  \bar{\gamma }_{A B}^2+1044 \nu  \bar{\gamma }_{A B}+11 \bar{\gamma }_{A B}^2+60 \bar{\gamma }_{A B}-12 (\nu -1) \bar{\beta }_+-30 m_- \bar{\beta }_-+\delta _+ (40 \nu -2)\nonumber\\
&-22 \delta _- m_-+850 \nu +55\big)\,,\\
\gamma_{012}&=\frac{G_{A B}^2\,\nu}{72}   \left(10 \nu  \bar{\gamma }_{A B}^2+150 \nu  \bar{\gamma }_{A B}+16 \bar{\gamma }_{A B}^2+18 \bar{\gamma }_{A B}+42 \bar{\beta }_+-24 m_- \bar{\beta }_-+4 \delta _+ (5 \nu +2)+3 \nu ^2-2 \delta _- m_-+266 \nu +17\right)\,,\\
\gamma_{003}&=\frac{G_{A B}^3}{2304}\bigg[\frac{ 96 \nu \big((2 \delta _++2\delta_-m_-) \left(\bar{\gamma }_{A B}+10\right)-11 \bar{\gamma }_{A B} \left(\bar{\gamma }_{A B}+2\right)^2\big)}{ \tilde\alpha \left(\bar{\gamma }_{A B}+2\right)}+288 \bar{\beta }_+ \left(4 (7 \nu +1) \bar{\gamma }_{A B}-7 \nu ^2+19 \nu +4\right)\nonumber\\
&\qquad+192 \nu ^2 \bar{\gamma }_{A B}^2+1920 \nu ^2 \bar{\gamma }_{A B}-63 \pi ^2 \nu  \bar{\gamma }_{A B}^3+1056 \nu  \bar{\gamma }_{A B}^3+180 \pi ^2 \nu  \bar{\gamma }_{A B}^2+5872 \nu  \bar{\gamma }_{A B}^2+1350 \pi ^2 \nu  \bar{\gamma }_{A B}+20672 \nu  \bar{\gamma }_{A B}\nonumber\\
&\qquad+288 \bar{\gamma }_{A B}^3+672 \bar{\gamma }_{A B}^2+384 \bar{\gamma }_{A B}+3168 \ln_- m_- \nu  \bar{\gamma }_{A B}^2+12672 \ln_- m_- \nu  \bar{\gamma }_{A B}+3168\ln_+ \nu  \bar{\gamma }_{A B}^2+12672 \ln_+ \nu  \bar{\gamma }_{A B}\nonumber\\
&\qquad-2 \delta _+ \left(21 \left(3 \pi ^2-32\right) \nu  \bar{\gamma }_{A B}+96 \bar{\gamma }_{A B}+2 \nu  \left(144 \ln_- m_-+144 \ln_++63 \pi ^2-2264\right)-192 \nu ^2+96\right)-6912 m_- \nu  \bar{\beta }_- \bar{\gamma }_{A B}\nonumber\\
&\qquad-1152 m_- \bar{\beta }_- \bar{\gamma }_{A B}-192 \delta _- m_- \nu  \bar{\gamma }_{A B}-192 \delta _- m_- \bar{\gamma }_{A B}+288 m_- \nu ^2 \bar{\beta }_--8640 m_- \nu  \bar{\beta }_--1152 m_- \bar{\beta }_--1152 \nu ^2 \zeta _{A B}\nonumber\\
&\qquad -3456 \nu  \zeta _{A B}-576 \delta _- \ln_- \nu +12672 \ln_- m_- \nu -576 \delta _- \ln_+ m_- \nu +12672 \ln_+ \nu +288 \nu ^3-992 \delta _- m_- \nu-192 \delta _- m_-\nonumber\\
&\qquad  +864 \nu ^2+2304 \nu  \psi _++288 m_- \nu  \epsilon _-+192 m_- \epsilon _-+1476 \pi ^2 \nu +18816 \nu +192 \nu ^2 \epsilon _++96 \nu  \epsilon _+-192 \epsilon _+\bigg]\,,\\
\gamma_{003}^{\ln}&=\frac{G_{A B}^3\,\nu}{4}  \left(-11 \bar{\gamma }_{A B}^2-44 \bar{\gamma }_{A B}+2 \delta _++2 \delta _- m_--44\right)\,.
\end{align}
\end{subequations}
\end{widetext}

\bibliography{FLJbib}

\end{document}